\documentclass{aa}  
\usepackage{graphicx}
\usepackage{txfonts}
\usepackage{xcolor}
\usepackage{multirow}
\usepackage{subcaption}
\usepackage{mhchem}
\usepackage{hyperref}
\hypersetup{colorlinks=true,citecolor=blue}

\titlerunning{A broad exploration of climate and observability of close-in rocky exoplanets}
\authorrunning{M. Houelle et al.}

\begin{document} 

\title{A broad exploration of climate and observability of close-in rocky exoplanets: applications to Ross~128~b}
   
\author{Mathilde Houelle\inst{1,2} \and
        Emeline Bolmont\inst{1,2} \and
        Christophe Lovis\inst{1,2} \and
        Martin Turbet\inst{3,4} \and
        Guillaume Chaverot\inst{2,5} \and
        Alexandre Mechineau\inst{4} \and
        Jérémy Leconte\inst{4}}
 
\institute{Observatoire de Genève, Université de Genève, Chemin Pegasi 51, 1290 Versoix, Switzerland 
           \and
           Centre pour la Vie dans l’Univers, Université de Genève, 1211 Geneva, Switzerland \\
           \email{mathilde.houelle@unige.ch}
           \and 
           Laboratoire de Météorologie Dynamique/IPSL, CNRS, Sorbonne Université, \'{E}cole Normale Supérieure, PSL Research University, \'{E}cole Polytechnique, Institut Poytechnique de Paris, 75005 Paris, France
           \and 
           Laboratoire d’Astrophysique de Bordeaux, Univ. Bordeaux, CNRS, B18N, allée Geoffroy Saint-Hilaire, 33615 Pessac, France
           \and 
           Univ. Grenoble Alpes, CNRS, IPAG, 38000 Grenoble, France}
    

\abstract{
{The upcoming instruments VLT/RISTRETTO, ELT/ANDES and ELT/PCS will soon enable atmospheric characterization of the expanding population of non-transiting, small, rocky exoplanets orbiting closer than the inner edge of the Habitable Zone around M~dwarf stars. This advancement is made possible thanks to the combination of high-contrast imaging and high-resolution spectroscopy in reflected light. A critical parameter for reflected-light observability is the wavelength-dependent reflectivity, which is strongly influenced by climate, surface and atmospheric properties.}
{The objective of this work is to refine the predictions of the spectral reflectivity for this population of planets. Using Ross~128~b as a prototype, we aim to provide physically consistent geometric albedo estimates within the RISTRETTO and ANDES spectral ranges across diverse atmospheric scenarios.}
{Using the Generic Planetary Climate Model, we perform 3D climate simulations of Ross~128~b under varying atmospheric compositions, surface pressures, water inventories and spin-orbit resonances to investigate the potential climate regimes of the planet. Then, we compute synthetic reflectance spectra with the radiative transfer code Pytmosph3R to assess spectral signatures for each scenario and discuss the planet detectability and instrument capacity for constraining the climate.}
{The results show that hazeless rocky planets receiving stellar irradiations similar to Ross~128~b exhibit rather low reflectivity, with geometric albedos ranging from 0.07 to 0.2 in the RISTRETTO bandpass and from 0 to 0.14 in the ANDES bandpass across all simulated scenarios. This low reflectivity can result from the lack of clouds or surface ice deposits on the dayside or from a strong atmospheric absorption due to high water vapor concentrations, depending on the parameter configurations. These climate features are characteristic of the climate moist bistability in close-in planets with low water reservoirs investigated in previous studies. Our results suggest that arbitrary albedo assumptions, such as the commonly adopted 0.3 Earth-like value, can overestimate the reflectivity for this specific planet population when assessing detection limits.}
{This highlights the importance of using more accurate climate models, as done in this study, to improve the reflectivity predictions, better assess the planet detectability and optimize the preparation of future observations with next-generation high-resolution spectrographs.}}


\maketitle

\section{Introduction}
\label{sec:intro}

The atmospheric characterization of small, rocky exoplanets is now possible through instruments such as the James Webb Space Telescope (JWST). Since 2022, this space-based mission has enabled the first observations of rocky planets, including those within the TRAPPIST-1 system \citep{gillon2016, gillon2017}, using transmission spectroscopy \citep{piaulet_Ghorayeb2025, espinoza2025, glidden2025} and thermal emission measurements \citep{greene2023, zieba2023, gillon2025}. Statistical analyses from detection missions like Kepler reveal that small exoplanets with radii between 1 and 4~$R_\oplus$ are particularly prevalent, with over 50\% of Sun-like stars hosting at least one planet within this size range \citep{batalha2013, petigura2013, fulton2018, he2019}. Moreover, due to observational biases, many super-Earth and Earth-mass planets are detected in short-period orbits around M~dwarf stars \citep{dressing2015, mignon2025}. This results in a considerable population of close-in, rocky exoplanets that are relatively highly irradiated and likely tidally locked.

However, several factors limit the atmospheric characterization of rocky planets around M~dwarf stars with transmission and emission spectroscopy.
First, stellar contamination (especially the transit light source effect with the presence of spots and faculae in the stellar photosphere) can impact the observed transmission spectrum and limit reliable detections of atmospheric signatures \citep[e.g.,][]{rackham_transit_2019}, as observed for TRAPPIST-1~b \citep{lim2023}. 
Second, our limited understanding of stellar spectra constrains atmospheric retrievals based on emission spectroscopy \citep{fauchez2025}. 
Finally, both techniques of transmission and emission spectroscopy are currently restricted to transiting exoplanets.

The combination of high-contrast imaging and high-resolution spectroscopy in reflected light \citep{snellen2015, lovis2017} is an alternative technique expected to enable atmospheric characterization of nearby non-transiting exoplanets previously inaccessible with existing methods.
This new approach will be employed by RISTRETTO \citep{lovis2022, lovis2024, bugatti2025} on the Very Large Telescope (VLT) from 2030, pathfinder instrument for ANDES \citep{marconi2022, palle2025} and PCS \citep{kasper2021}, planned on the forthcoming Extremely Large Telescope (ELT) from 2035 and 2040, respectively.
RISTRETTO (high-resolution integral-field spectrograph for the tomography of resolved exoplanets through timely observations) is a pioneering instrument integrating extreme adaptive optics with high-resolution spectroscopy in the visible spectral range. Among other science cases, the spectrograph will particularly focus on  Proxima~Centauri~b, the closest known exoplanet to Earth \citep{anglada_escude2016}.
Proxima~b will also be observed with ANDES (armazones high dispersion echelle spectrograph), along with other promising nearby targets with smaller angular separations, referred to as the golden sample \citep{palle2025}. 
ANDES and PCS (planetary camera and spectrograph) are expected to reach unprecedented angular resolutions with an inner working angle of 20~mas (allowing the observation of a larger number of targets), achieve higher spectral resolutions ($R=100\,000$), significantly decrease integration times and broaden the spectral coverage across the entire visible and near-infrared. 

\begin{figure}[!htb]
    \centering
    \includegraphics[width=\hsize]{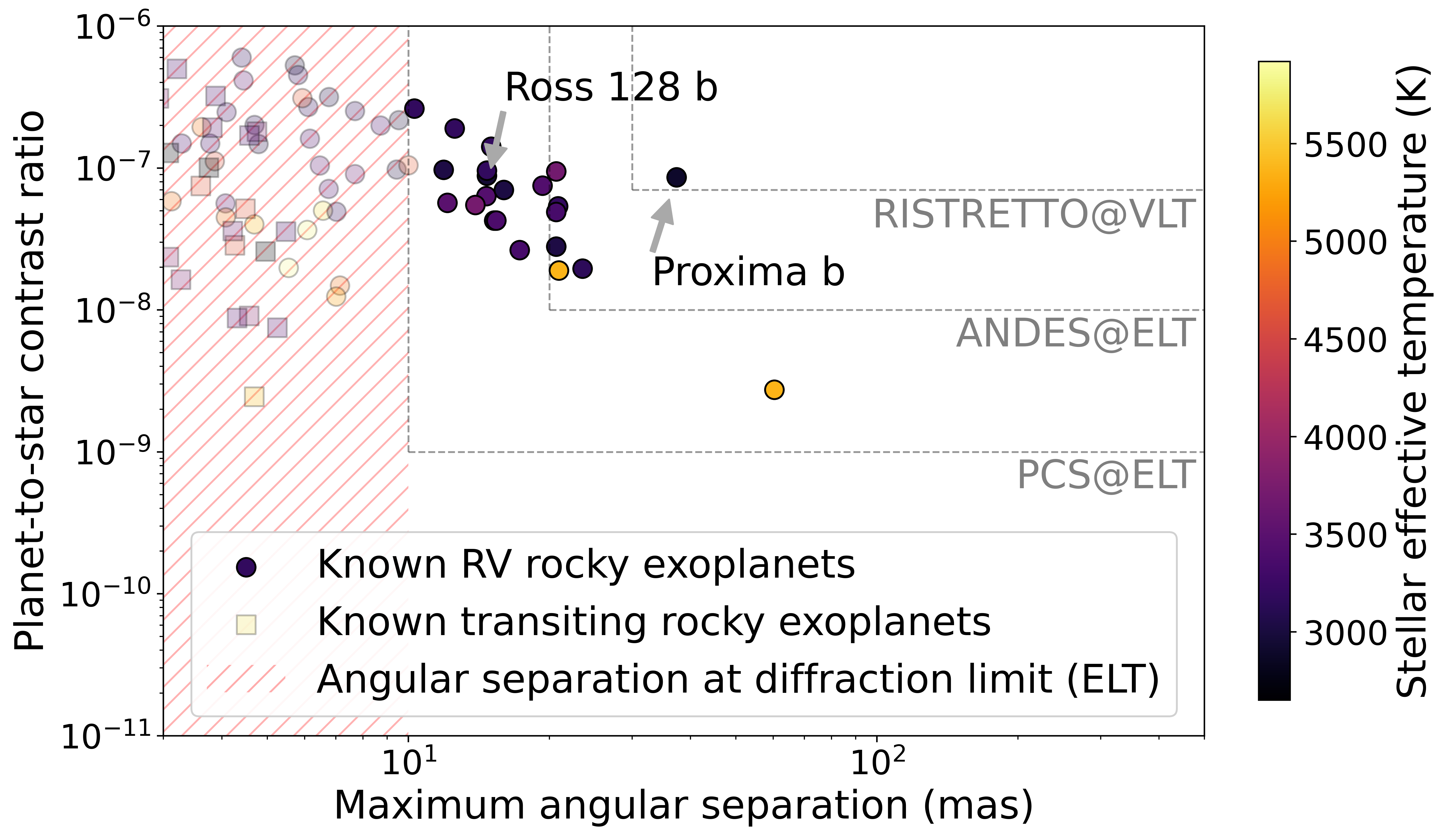}
    \caption{Estimated detectability of the currently known (RV and transiting) rocky exoplanets, with estimated masses below 5~$M_\oplus$ and radii below 1.7~$R_\oplus$. The masses are estimated assuming a most probable orbital inclination of 60$^\circ$ and the radii are computed using the empirical mass-radius relationship from \cite{parc2024}. The contrast ratios are estimated assuming a geometric albedo of 0.3 and planets are assumed at maximum elongation. The maximum angular separations come from NASA Exoplanet Archive. The ELT diffraction limit and the approximate detection limits of RISTRETTO, ANDES and PCS are depicted in the figure.}
    \label{fig:contrast_vs_sep}
\end{figure}

An important initial step involves assessing the detectability of the targets using current and upcoming observational capabilities. Two key aspects to consider are (1)~the planet-to-star contrast ratio (i.e., ratio of the reflected flux from the planet to the stellar flux) and (2)~the angular separation from the host star. Figure~\ref{fig:contrast_vs_sep} illustrates the estimated detectability of the currently known rocky exoplanets discovered through radial velocity (RV) and transit methods. 
The diffraction limit of the ELT, defined by the 39-metre-diameter primary mirror, limits the detectability of planets with angular separations smaller than 10~mas.
The contrast ratios are estimated assuming a geometric albedo of 0.3 for each planet.

The geometric albedo is a key parameter influencing target detectability.
However, despite the expected diversity among the targets, studies generally rely on unique and arbitrary albedo assumptions to estimate contrast ratios and required exposure times, commonly using the 0.3 Earth-like albedo assumption \citep{Madden2018} such as in Figure~\ref{fig:contrast_vs_sep}.
In this study, we aim to provide more robust and physically consistent estimates of the planet spectral reflectivity by taking atmospheric and surface properties into account. Given that the planetary albedo is strongly influenced by the climate, the most effective approach is to perform climate simulations to explore the potential climate regimes. Moreover, the climate is governed by a broader set of parameters beyond stellar irradiation and orbital distance, notably including the atmospheric composition, which currently remains largely unknown. Therefore, such simulations need to be conducted across a wide range of orbital and atmospheric configurations, to refine the predictions and better prepare the future observations by providing a wide grid of potential atmospheric scenarios.

Historical methods for climate studies used 1D single-column climate models \citep{kasting1988}. Although these models provide reasonable estimates for planets with fast rotations and dense atmospheres, they are limited when applied to planets with synchronous or slow rotations \citep[e.g.,][]{yang2013}, such as expected for those orbiting close to their host stars, or even planets with Earth-like rotations \citep[e.g.,][]{leconte2013nat}. This is because they are unable to consistently represent the three-dimensional processes of the atmosphere, including global circulation patterns and cloud distribution. For albedo estimations, clouds especially are a first-order factor influencing the results.

In such cases, 3D global climate models (GCMs) are essential for accurately simulating energy redistribution, temperature variations and mean planetary temperatures, for both Earth-like rotations \citep[e.g.,][]{leconte2013nat}, and slow or synchronous rotations \citep[e.g.,][]{yang2013, yang2014, way2016, kopparapu2017, turbet2021}. Incorporating this complexity is vital for deriving physically grounded albedo values across various planetary configurations, and also assessing the potential detectability of spectral features for each scenario, as done in \cite{leconte2013aa, turbet2016, wolf2019, fauchez2025b}. 

GCMs have been used to investigate the climate and potential habitability of Proxima~b \citep[e.g.,][]{turbet2016, Boutle2017, del_genio2019, Yang2020}. According to the current definitions of the Habitable Zone (HZ), which is defined as the circumstellar region where surface water could be sustained in its liquid form \citep{kasting1993, kopparapu2013, kopparapu2014, kopparapu2016, kopparapu2017}, Proxima~b is most likely located within the HZ of its host star. This assessment is based on the measured stellar effective temperature, stellar luminosity and orbital distance of Proxima~b, that we summarized in Table~\ref{tab:targets}, but is also supported by the GCM studies. In comparison, most other planets within the golden sample of ANDES are located closer to their host star than the inner boundary of the HZ, according to the definitions of \citet{kopparapu2017} and \citet{turbet2023} for synchronously rotating planets. More accurate studies of these planets with GCMs can help to better understand their climate and atmospheric processes, which can also inform the criteria defining the inner edge of the HZ. 

This study focuses on the ANDES target Ross~128~b \citep{bonfils2018}, which is the second closest known rocky exoplanet to Earth, discovered via the RV method. With a stellar irradiation of 1.5~$S_\oplus$ (see Table~\ref{tab:targets}), Ross~128~b is closer to its star than the inner edge of the HZ as defined by \citet{kopparapu2017}, which corresponds to $\sim$1.25~$S_\oplus$ for the effective temperature of Ross~128. Moreover, with a maximum angular separation of 14.7~mas, Ross~128~b is beyond the detection limits of RISTRETTO and ANDES, as shown in Figure~\ref{fig:contrast_vs_sep}. However, it remains a potential target for PCS, which aims to operate also in the visible and may achieve the required contrast and separation for a direct characterization. 
Since Ross~128~b is the most irradiated object in the ANDES golden sample and has the smallest angular separation, we consider this target as a representative case study for both currently and potentially future confirmed exoplanets within the population of small, rocky exoplanets located closer than the inner edge of the HZ, including the remaining targets of the golden sample such as GJ~273~b \citep{astudillo_defru2017} or Wolf~1061~c \citep{wright2016}.

\begin{table}[h!]
    \caption{Stellar effective temperature, stellar luminosity, semi-major axis, stellar irradiation and maximum angular separation values of Proxima~b and Ross~128~b.} 
    \label{tab:targets}    
    \centering                 
    \begin{tabular}{c c c}       
        \hline\hline        
                          & Proxima b$^{(1)}$ & Ross 128 b$^{(2)}$ \\
        \hline  
        $T_{\rm eff}$ (K) & 2980$^{(3)}$      & 3192$^{(4)}$       \\
        $L$ ($L_\odot$)   & 0.00151$^{(3)}$   & 0.00362$^{(4)}$    \\
        $a$ (AU)          & 0.04848$^{(5)}$   & 0.0496$^{(2)}$     \\
        $S$ ($S_\oplus$)  & 0.64              & 1.5                \\
        $\theta$ (mas)    & 37.3              & 14.7               \\
        \hline                              
    \end{tabular}
    \tablefoot{The angular separations come from \cite{palle2025}.}
    \tablebib{
        (1) \citet{anglada_escude2016};
        (2) \citet{bonfils2018};
        (3) \citet{ribas2017};
        (4) \citet{mann2015};
        (5) \citet{suarez_mascareno2025}.}
\end{table}

The objective of this study is to provide a more robust and accurate assessment of the spectral reflectivity of small, rocky exoplanets located closer than the inner edge of the Habitable Zone of M~dwarf stars. Applying our methodology (detailed in Section~\ref{sec:method}) on Ross~128~b, we first investigate in Section~\ref{sec:climate_results} the potential climate regimes of the planet using a GCM, by exploring the impact of a broad range of parameters (orbital configurations, atmospheric compositions, surface pressures and water inventories) on the climate. Then, we provide in Section~\ref{sec:observability_results} new geometric albedo estimates and analyze potential observable signatures for each scenario. 
The updated geometric albedo estimates can improve the accuracy of target detectability and better inform the future reflected-light observations with next-generation ground-based high-resolution spectrographs such as RISTRETTO, ANDES and PCS. 

\section{Method}
\label{sec:method}

\subsection{Modeling the climate by varying multiple critical planetary parameters}
\label{sub:method:climate}

The climate simulations are performed using the Generic Planetary Climate Model (Generic-PCM, previously known as LMD-Generic), which is a 3D GCM developed to simulate a diverse range of planetary atmospheres, readily applicable to exoplanets \citep{wordsworth2010, wordsworth2011, selsis2011, leconte2013aa, bolmont2016, turbet2016, turbet2018, fauchez2019}. Based on fundamental hydrodynamic and thermodynamic equations, the model solves the primitive equations of geophysical fluid dynamics using the 3D dynamical core of the LMDz 3D General Circulation Model \citep{hourdin2006}, originally developed to study the climate of the Earth. 

We conducted the GCM simulations using the physical parameters of the Ross~128 system, detailed in Table~\ref{tab:parameters}. The star is an M~dwarf with an effective temperature of 3192~K and a luminosity of 0.00362~$L_\odot$ \citep{mann2015}. The stellar spectral data used for the simulations were computed using the BT-Settl model grid of theoretical spectra \citep{rajpurohit2013}, considering $T_{\rm eff}=3200$~K, $\log g=5$~dex and $\text{[Fe/H]}=0$~dex for the stellar effective temperature, surface gravity and metallicity, respectively \citep{mann2015, souto2018}. The planet rotates around the star in 9.86~days with a semi-major axis of 0.0496~AU \citep{bonfils2018}. The measured stellar luminosity and semi-major axis give a received stellar irradiation of about 2009~W\,m$^{-2}$, which is $\sim$1.5~times the solar flux received by the Earth. The projected mass $M_{\rm p}\,\sin i$ is 1.35~$M_\oplus$ \citep{bonfils2018}. The planetary mass considered for the simulations (1.56~$M_\oplus$) is calculated assuming that the orbital inclination of the system is 60$^\circ$, which is the median value for a random distribution of orbital inclinations. The planetary radius (1.13~$R_\oplus$) and gravity field (12~m\,s$^{-2}$) are derived assuming that the planet has an Earth-like rocky composition \citep{souto2018, zeng2019, parc2024}. 
Given the uncertainties regarding planetary surface properties and internal composition, we adopt the simplest assumption of a homogenous, rock surface, as used in \citet{turbet2016}.
Each simulation is initialized with a flat surface, with a surface albedo of 0.2 and a thermal inertia of 1000~J\,m$^{-2}$\,K$^{-1}$\,s$^{-1/2}$. 
This surface albedo assumption is consistent with Earth-based studies \citep{pierrehumbert2010, Roccetti2024, Roccetti2025}, and observations of airless solar system bodies \citep{Buratti1996, Mallama2017} and rocky exoplanets \citep{greene2023, zieba2023, Zieba2026}.
Moreover, darkening processes such as space weathering \citep{Cassidy1975, Pieters2016} support the assumption of a low-albedo planetary surface.

The spatial grid used for the simulations has a resolution of $64\times48\times30$ in longitude, latitude and altitude, respectively.
An initial temperature profile of 300~K is uniformly assigned across all atmospheric grid columns. 
We fix the mean atmospheric surface pressure at 0.1, 1 or 10~bars, depending on the simulation setup (presented below).

\subsubsection{Orbital configurations}

Given that Ross~128~b orbits close to its host star, we initially assumed a circular orbit ($e=0$), a synchronous rotation (1:1 spin-orbit resonance) and zero obliquity.
This is consistent with the final stage of a two-body tidal evolution, which tends to (1)~synchronize the rotation and orbital periods, (2)~align the planet spin axis perpendicularly to the orbit and (3)~damp the eccentricity to zero. 

However, this state can be reached over timescales potentially longer than the lifetime of the system. Moreover, non-zero eccentricities can facilitate capture into higher order resonances, which can delay the evolution towards the synchronous state, and induce high obliquities instead of damping to zero \citep{valente2022}. For the specific case of Ross~128~b, \citet{valente2022} showed that eccentricities of 0.09 and 0.16 can respectively lead to capture into 5:2 and 7:2 spin-orbit resonances and obliquities of 79$^\circ$ and 66$^\circ$. These high obliquity configurations can be maintained over several gigayears (10~Gyr), potentially exceeding the estimated age of Ross~128, which is of the order of a few gigayears \citep{newton2016}.

To analyze the impact of the synchronous and asynchronous rotations on the planetary climate and their implications for observability, we conducted the simulations for two different sets of spin-orbit resonances, eccentricities and obliquities (see also Table~\ref{tab:parameters}):
\begin{itemize}
    \item 1:1 spin-orbit resonance, 0 eccentricity, 0$^\circ$ obliquity
    \item 5:2 spin-orbit resonance, 0.09 eccentricity, 79$^\circ$ obliquity
\end{itemize}
The orbital parameters significantly influence the planetary climate, as the stellar irradiation distribution across the planetary surface is governed by the rotation period and obliquity \citep{Milankovitch1941}. Therefore, the climate regime of a synchronously rotating planet, characterized by permanent dayside and nightside, substantially differs from a planet with an asynchronous rotation. Moreover, planets with high eccentricities induce important temporal variations in the stellar irradiation, thereby limiting the assumption of the mean flux approximation, although this approximation should hold for a planet like Ross~128~b \citep{bolmont2016}. These orbital parameter variations are critical for assessing the planetary climate and habitability.

\begin{table}[h!]
    \caption{Stellar and planetary characteristics of the Ross~128 system considered for the simulations.}
    \label{tab:parameters}      
    \centering                         
    \begin{tabular}{c c}        
        \hline\hline
        \multicolumn{2}{c}{Stellar parameters$^{(1,2)}$}          \\
        \hline
        Effective temperature, $T_{\rm eff}$ (K) & 3192           \\
        Luminosity, $L$ ($L_\odot$)              & 0.00362        \\
        Mass, $M_\star$ ($M_\odot$)              & 0.168          \\
        Radius, $R_\star$ ($R_\odot$)            & 0.1967         \\
        Surface gravity, $\log g$                & 5              \\
        Metallicity, [Fe/H]                      & 0              \\
        \hline
        \multicolumn{2}{c}{Planetary parameters}                  \\
        \hline
        Orbital period, $P$ (days)               & 9.86$^{(3)}$   \\
        Semi-major axis, $a$ (AU)                & 0.0496$^{(3)}$ \\
        Stellar irradiation, $S$ (W\,m$^{-2}$)   & 2009           \\
        Projected mass, $M_{\rm p}\,\sin i$ ($M_\oplus$) & 1.35$^{(3)}$ \\
        Mass, $M_{\rm p}$ ($M_\oplus$)     & 1.56\tablefootmark{*}    \\
        Radius, $R_{\rm p}$ ($R_\oplus$)   & 1.13\tablefootmark{*}    \\
        Gravity, $g_{\rm p}$ (m\,s$^{-2}$) & 12\tablefootmark{*}      \\
        Spin-orbit resonance              & 1:1/5:2\tablefootmark{**} \\
        Obliquity ($^\circ$)               & 0/79\tablefootmark{**}   \\
        Eccentricity, $e$                  & 0/0.09\tablefootmark{**} \\
        Rotation rate, $\Omega$ (rad\,s$^{-1}$) & $7.4\times10^{-6}$/$1.8\times10^{-5}$ \\
        \hline
        \multicolumn{2}{c}{Assumed surface parameters}             \\
        \hline  
        Type                                                & Rock \\
        Albedo, $A_{\rm surf}$                              & 0.2  \\
        Thermal inertia (J\,m$^{-2}$\,K$^{-1}$\,s$^{-1/2}$) & 1000 \\
        Topography                                          & Flat \\
        \hline 
    \end{tabular}
    \tablefoot{
        \tablefoottext{*}{The planetary mass, radius and gravity are calculated under the assumptions of an orbital inclination $i=60^\circ$ and a bulk density consistent with an Earth-like rocky composition.} 
        \tablefoottext{**}{Two different orbital configurations are considered for this study. The asynchronous case (5:2) and its related eccentricity and obliquity come from \cite{valente2022}.}}
    \tablebib{
        (1)~\citet{mann2015};
        (2)~\citet{souto2018};
        (3)~\citet{bonfils2018}.}
\end{table}

\subsubsection{Atmospheric compositions}

In our simulations, we consider atmospheres dominated by \ce{N2} and \ce{CO2}.
The choice of \ce{N2}-dominated atmospheres is based on the prevalence of \ce{N2} as a common constituent in secondary atmospheres \citep{forget2014, maurice2024} and motivated by examples in the solar system, such as the Earth and Titan. 
The choice of a \ce{CO2}-dominated atmosphere is supported by planetary formation models showing that atmospheres of low-mass rocky exoplanets are likely to evolve as \ce{CO2}-rich environments \citep{Hu2012dec, bower2025}. Also, analogous cases of planets with \ce{CO2}-dominated atmospheres exist in the solar system, such as Venus and Mars. 

More specifically, we consider three different mixtures of atmospheric gases: 
\begin{itemize}
    \item \ce{H2O + N2}
    \item \ce{H2O + N2 + 376ppm CO2} (Earth-like composition)
    \item \ce{H2O + CO2}
\end{itemize}
The different atmospheric setups used in the simulations are summarized in Table~\ref{tab:atmospheres}.
The radiative transfer is computed using correlated-k tables from \cite{chaverot2022} for the mixture of \ce{H2O + N2}, \cite{leconte2013nat} for the Earth-like composition and \cite{turbet2019} for the mixture of \ce{H2O + CO2}. The correlated-k tables include spectral bands for the thermal emission (IR) and visible wavelengths. For the \ce{H2O + N2} mixture, the spectral resolution is 58 for the IR and 28 for the visible, whereas for both Earth-like and \ce{H2O + CO2} mixtures, the IR and visible resolutions are 58 and 28, respectively. The opacity due to \ce{N2}-\ce{N2} collision-induced absorption (CIA) is accounted and come from the HITRAN CIA database \citep{karman2019}. The simulations include also the \ce{H2O}-\ce{H2O} and \ce{H2O}-air continua \citep{mlawer2012}. 
The correlated-k table for \ce{H2O + CO2} also includes the opacity due to the \ce{CO2}-\ce{CO2} continua.

Water is treated as a condensable gas, thus the water vapor pressure varies during the simulations. Water can exist in its vapor and condensed phases, in the atmosphere (gas), in clouds and on the surface (liquid and ice). The transport, pressure variations and phase transitions of water are computed following the water cycle scheme described in \citet{leconte2013nat}. For the cloud formation, water (liquid or ice, depending on the temperature) can condense on cloud condensation nuclei (CCN). A detailed description of the cloud formation scheme can be found in \citet{leconte2013nat}. In our \ce{CO2}-dominated atmospheres, we also allow \ce{CO2} condensing into ice at the surface and on CCNs. For all cloud particles (liquid water, water ice and \ce{CO2} ice), we keep the number of CCN per unit mass of moist air fixed at $10^5$~kg$^{-1}$, and the CCN radii variable. 

\subsubsection{Water inventories}

In the model, the initial global water amount can be numerically finite (limited water amount on the surface and/or in the atmosphere) or infinite (ocean with or without heat transport).
The presence of a large water reservoir such as an ocean on a planet receiving a stellar irradiation similar to Ross~128~b would lead the climate to the runaway greenhouse state \citep{kopparapu2017, turbet2023}.  
The runaway greenhouse is a state where all available water is vaporized in the atmosphere and the lower atmosphere becomes optically thick, leading to extremely high surface temperatures.
This will be further discussed in Section~\ref{sec:discussion}.
Thus, here we initialized the simulations with limited water reservoirs (initially as water vapor in the atmosphere), with varying global water amounts for each atmospheric setup (including dry conditions). For the setups under dry conditions, the correlated-k tables used for the radiative transfer computations include a tiny amount of water, which has a negligible impact on the GCM results, but may affect the computed spectra.

\begin{table}[h!]
    \caption{Setups of atmospheric gas compositions and surface pressures considered in the simulations.} 
    \label{tab:atmospheres}    
    \centering                 
    \begin{tabular}{cccc}       
        \hline\hline 
        Case\tablefootmark{*} & \ce{N2} pressure & \ce{CO2} pressure & \ce{H2O} pressure\tablefootmark{**} \\
        \hline
        1 & 0.1 bar & -           & variable \\
        2 & 1 bar   & -           & variable \\
        3 & 10 bars & -           & variable \\
        4 & 1 bar   & \ce{376ppm} & variable \\
        5 & -       & 0.1 bar     & variable \\
        6 & -       & 1 bar       & variable \\
        7 & -       & 10 bars     & variable \\
        \hline                              
    \end{tabular}
    \tablefoot{
        \tablefoottext{*}{Most cases involve multiple global water inventory scenarios, including dry scenarios. For Case~2 specifically, we define three sub-cases characterized by distinct water contents, denoted as Cases~2A, 2B and~2C, with specific values to be detailed in Section~\ref{sec:climate_results}.} 
        \tablefoottext{**}{As water is treated as a condensable gas in each gas mixture, the water vapor pressure can vary during the simulations.}}
\end{table}

\subsection{Computing the reflectance spectra}
\label{sub:method:pytmo}

\subsubsection{Description of Pytmosph3R}

Pytmosph3R \citep{Caldas2019, Falco2022} allows processing Generic-PCM simulation outputs to generate synthetic observables, including transmission and emission spectra, at a different resolution than the simulation. In this study, the code is used to produce phase curves that take into account the surface albedo and the incident stellar flux. The phase curves are computed using high-resolution opacity data binned down at a lower spectral resolution, which remains higher than the resolution of the correlated-k tables used in the Generic-PCM. The spectral resolutions of the correlated-k tables adapted for Pytmosph3R are $R=300$ for the \ce{H2O + N2} gas mixture, and $R=500$ for the Earth-like and \ce{H2O + CO2} tables. The \ce{H2O}-\ce{H2O} and \ce{H2O}-\ce{N2} continua and CIA spectral data are also included. 
Pytmosph3R computes the emission spectrum by seeing the 3D~atmosphere as a 2D~grid of independent atmospheric columns, using the latitude-longitude discretization of the Generic-PCM outputs. By extension, the phase curves are a collection of emission spectra at different orbital phases.
The spectrum of each atmospheric column is computed using exo\_k \citep{Leconte2021}, a library allowing to solve the radiative transfer equations for a 1D~atmospheric column, based on \cite{Toon1989} and using a two-stream approximation \citep{Meador1980}. An effective angle $\mu_0$ is used, which represents how the line of sight of the observer probes the atmosphere. The algorithm from \citep{Toon1989} allows to specify a source term to treat the stellar flux as a collimated beam with an independent effective angle $\mu_\star$. The proper $\mu_\star$ and $\mu_0$ are obtained from the observing geometry (from the observer and star positions and the latitude/longitude coordinates of the columns). The surface albedo is also specified. The spectrum seen by the observer is then computed by performing a weighted sum of the spectra of each column. The weights take into account the surface of the columns and their effective surface for the observer. This method allows taking into account the geometric effects of the stellar illumination and the surface albedo. 

\subsubsection{Computing the geometric albedo}

\begin{figure}[!htb]
    \centering
    \begin{subfigure}{\hsize}
        \centering
        \includegraphics[width=\hsize, trim=4cm 2.5cm 3cm 1cm, clip]{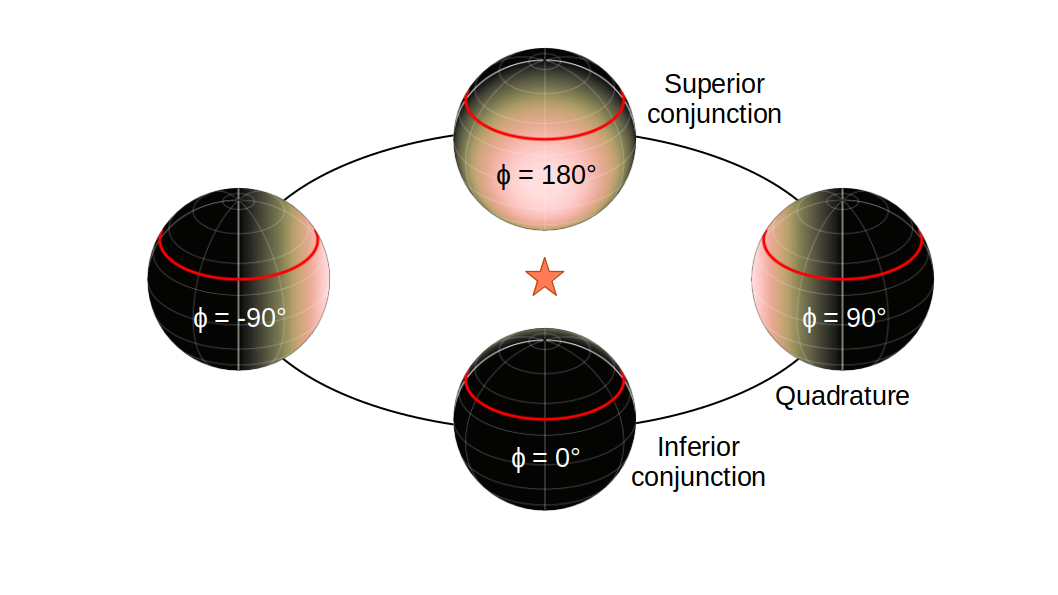} 
        \label{fig:phases}
    \end{subfigure}
    \begin{subfigure}{\hsize}
        \centering
        \includegraphics[width=\hsize, trim=0cm 2cm 7cm 2cm, clip]{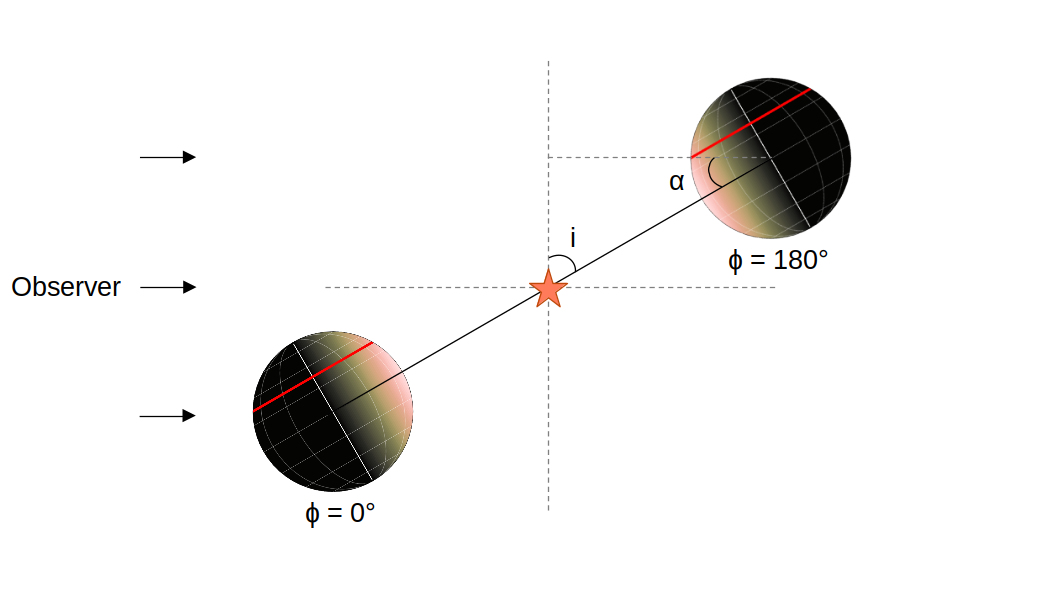}
        \label{fig:angles}
    \end{subfigure}
    \caption{\textit{Top:} Orbital phases $\phi$ as seen from the observer for a planet with an orbital inclination $i$ of $60^\circ$.
    \textit{Bottom:} Illustration of the orbital phase $\phi$, orbital inclination $i$ and phase angle $\alpha$ (for the same inclination $i=60^\circ$).
    The red line indicates the observed latitude (30$^\circ$) corresponding to an inclination $i$ of 60$^\circ$.}
    \label{fig:geometry}
\end{figure}

The wavelength-dependent planet-to-star contrast ratio at phase angle $\alpha$ is given by
\begin{equation}
    C(\lambda,\,\alpha) = A_{\rm g}(\lambda) \ g(\alpha) \ \left( \frac{R_{\rm p}}{a} \right)^{2},
    \label{equ:contrast}
\end{equation}
where $C(\lambda,\,\alpha)$ is the ratio of the reflected spectrum at phase angle $\alpha$ to the stellar spectrum, $A_{\rm g}(\lambda)$ is the geometric albedo spectrum, $g(\alpha)$ is the phase function defining the scattering properties of the planet (from the surface and the atmosphere), $R_{\rm p}$ is the planet radius and $a$ is the orbital semi-major axis. The phase angle $\alpha$ is defined as $\cos \alpha = -\sin i \cos \phi$, with $i$ the orbital inclination angle and $\phi$ the orbital phase, defined as zero at inferior conjunction, as shown in Figure~\ref{fig:geometry}. For $i=60^\circ$, the phase angle of a planet at orbital phase $\phi=180^\circ$ is $\alpha=30^\circ$.

The contrast ratio is computed with Pytmosph3R for all simulated scenarios. Then, we derive the geometric albedo spectra, defined as the ratio of the planet reflectivity at specific wavelengths under full illumination to that of an idealized fully reflecting Lambertian disk of equal radius. 
We use the phase function of a Lambertian spherical planet, which is given by
\begin{equation}
    g(\alpha) = \frac{\sin \alpha + (\pi - \alpha) \cos \alpha}{\pi}.
    \label{equ:lambert}
\end{equation}
This is a valid approximation for planets dominated by isotropic scattering (like Rayleigh scattering). For a perfectly reflecting Lambertian spherical surface, $A_{\rm g}=2/3$.

This study focuses on the spectral ranges pertinent to RISTRETTO and ANDES. RISTRETTO will operate within the visible ($0.62-0.84$~$\mu$m), whereas the single conjugate adaptive optics (SCAO) of ANDES will cover the Y, J and H near-infrared spectral bands, defined in Table \ref{tab:bands}.
We will compute the mean geometric albedo over the RISTRETTO and YJH spectral bands for all scenarios considered in this study.

\begin{table}[h!]
    \caption{Spectral ranges of RISTRETTO and the YJH bands of ANDES considered for mean geometric albedo computations.}
    \label{tab:bands}      
    \centering                         
    \begin{tabular}{c c}        
        \hline\hline                
        Bandpass  & Spectral range ($\mu$m) \\    
        \hline                        
        RISTRETTO & $0.62-0.84$             \\
        Y band    & $0.95-1.13$             \\
        J band    & $1.12-1.36$             \\
        H band    & $1.41-1.80$             \\
        \hline                                  
    \end{tabular}
\end{table}

\section{Potential climate regimes of Ross~128~b}
\label{sec:climate_results}

This section focuses on the interpretation of the different climate regimes obtained from the GCM simulations of Ross~128~b. We begin by defining three scenarios (designated as Cases~2A, 2B and 2C), all simulating a planet with a synchronous rotation, an atmosphere of \ce{N2} and a mean surface pressure of 1~bar (corresponding to Case~2 as specified in Table~\ref{tab:atmospheres}), but with three distinct initial global water inventories. 
In this study, we express the total planetary water inventory in global equivalent layer (GEL), which represents the depth of a hypothetical global surface water layer assuming that all available water is in liquid form. For reference, Venus has an estimated water content of $\sim$2~cm~GEL, while the total water content of the Earth is around 3~km~GEL. In this work, we adopt an order of magnitude closer to Venus.
The total water inventories in our three cases are approximately:
\begin{itemize}
    \item 6~cm~GEL (Case~2A)
    \item 8.3~cm~GEL (Case~2B)
    \item 83.3~cm~GEL (Case~2C)
\end{itemize}

We first focus on Case~2B, where the atmospheric circulation regime, and water distribution and evolution are discussed in detail in Section~\ref{sub:climate_results:refcase}. From this case (arbitrarily designated as the baseline reference) we explore the impact of varying different parameters on the climate of Ross~128~b. Section~\ref{sub:climate_results:water_variation} discusses the effect of varying the initial water inventory on the climate by comparing Cases~2A, 2B and~2C. From this analysis, we will define the climate regimes of the three cases and will refer to them in the rest of this paper.
Then, Section~\ref{sub:climate_results:atmosphere_variation} investigates the influence of the atmospheric composition on the climate; Section~\ref{sub:climate_results:pressure_variation} analyzes the effects of the atmospheric surface pressure; and Section~\ref{sub:climate_results:rotation_variation} compares scenarios with synchronous and asynchronous rotations. 
Finally, an overview of the results is presented in Section~\ref{sub:climate_results:overview} and summarized in Figure~\ref{fig:tsurf_summary}.

\subsection{Description of the reference case}
\label{sub:climate_results:refcase}

\subsubsection{Circulation regime}
\label{sub:climate_results:refcase:circulation}

\begin{figure*}
    \centering
    \includegraphics[width=\hsize]{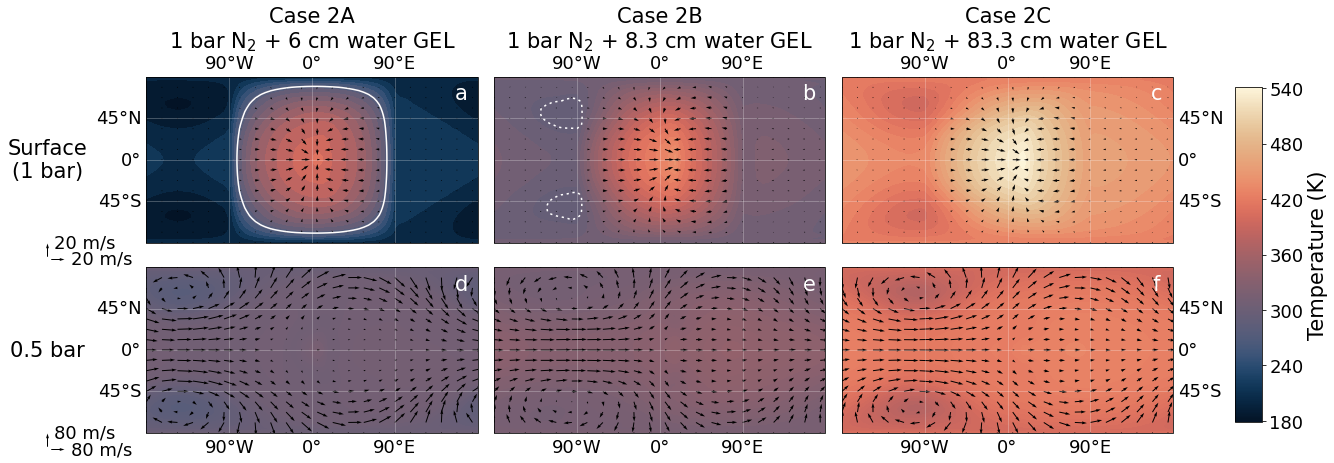}
    \caption{Latitude-longitude maps of temperature and winds averaged over $\sim$50 orbits (corresponding to $\sim$500 Earth days) for Cases 2A, 2B and 2C (specified in Table~\ref{tab:atmospheres}) and at different pressure levels. The zonal and meridional components of the winds are represented by the arrows. Panels a, b and c show maps at the surface and panels d, e and f show maps at a mean pressure level of 0.5~bar. 
    The solid white line contour corresponds to the 273.15~K isotherm below which water freezes and the dotted white line contour corresponds to the location of surface liquid water.}
    \label{fig:maps_T_winds}
\end{figure*}

In this section, we discuss the atmospheric circulation regime for Case~2B. Figure~\ref{fig:maps_T_winds} presents the maps of temperature and winds at the surface and at a higher atmospheric level corresponding to a pressure of 0.5~bar. The maps corresponding to Case~2B are shown in panels b and e; therefore, this section focuses on these panels.

At the surface, Figure~\ref{fig:maps_T_winds} (panel b) shows a clear temperature dichotomy between the dayside and the nightside. This pattern is characteristic of a planet in synchronous rotation, with permanent hot dayside and cold nightside. This results in a high day/night thermal contrast of $\sim$152~K, inducing a convergence of surface winds at the hottest point of the surface and an upwelling towards higher altitudes. 
On the nightside near the western terminator and at symmetric positions around latitudes (approximately $-45^\circ$ and $+45^\circ$), the surface map exhibits two cold regions with temperatures low enough to enable atmospheric water vapor to condense into liquid phase at the surface.

At higher altitudes, Figure~\ref{fig:maps_T_winds} (panel e) reveals a more homogeneous temperature distribution, with a temperature contrast of $\sim$32~K, and again two colder regions at mid-latitudes on the nightside in both hemispheres. The map also displays antisymmetric cyclonic vortex structures over the colder regions, and strong eastward equatorial winds (up to $\sim$90~m\,s$^{-1}$), leading to a slight eastward shift of the hottest point from the substellar point. 

These circulation features \citep[also described in][]{leconte2013aa} are characteristic of synchronously rotating rocky planets with relatively fast rotation rate, as shown in previous GCM studies of tidally locked exoplanets \citep[e.g.,][]{Koll2016, Yang2019, Hammond2020}. 
The climate and atmospheric circulation of such planets are driven by the fixed day/night radiative heating contrast, which generates the formation of stationary, atmospheric planetary waves (Kelvin and Rossby waves), as described by the Matsuno-Gill response pattern \citep{matsuno1966, gill1980, showman2011, showman2013}. 
In the case of fast rotating planets (thus with strong Coriolis force), the Kelvin waves can induce transport of eastward momentum from mid-latitudes toward the equator, leading to the formation of a super-rotating equatorial jet (as observed in our simulations, see Figure~\ref{fig:maps_T_winds}, panel~e). 
Forming on the western side of the substellar point and propagating westward, the Rossby waves can create cyclonic gyres at mid-latitudes both in the northern and southern hemispheres (Figure~\ref{fig:maps_T_winds}, panel e), referred to as Rossby-wave gyres. 

The circulation regime of tidally locked planets can be determined by the ratio of the equatorial Rossby deformation radius $L_{\rm Ro}$ (i.e., characteristic length scale at which the Coriolis force, and thus the planet rotation, starts to significantly affect atmospheric dynamics) and the planetary radius $R_{\rm p}$. 
Therefore, Case~2B is consistent with the Matsuno-Gill circulation regime expected for a synchronously rotating rocky planet with $L_{\rm Ro} \approx R_{\rm p}$, which is characterized by the presence of a planet-wide equatorial jet and Rossby-wave gyres.
We checked and quantified the effect of the rotation rate of Ross~128~b on its atmospheric circulation by calculating the equatorial Rossby deformation length, defined as $L_{\rm Ro} / R_{\rm p}$. The calculations are detailed in Appendix~\ref{appendix:rossby}, where we find that $L_{\rm Ro} / R_{\rm p} \approx 1$ for \ce{N2} and \ce{CO2}-dominated atmospheres and for a diverse range of temperatures, confirming our conclusions.

However, given that the radius of Ross~128~b is not directly measured as the planet was discovered via the RV method, we calculated the equatorial Rossby deformation length with the assumed value of $R_{\rm p}$ in Table~\ref{tab:parameters}. 
Assuming an Earth-like rocky composition, the actual planetary radius may vary depending on the true orbital inclination of the system, which could impact the atmospheric circulation regime at a given rotation rate. 
If $R_{\rm p}$ exceeds $L_{\rm Ro}$, the atmospheric dynamics may favor the formation of a more confined equatorial jet (similar to the narrow bands of Jupiter), rather than a planet-wide jet as illustrated in Figure~\ref{fig:maps_T_winds} (panel e). 
In contrast, if the planet is too small compared to $L_{\rm Ro}$, it may be unable to contain Rossby waves, thereby preventing the formation of a super-rotating jet. 
Using the minimum mass of Ross~128~b derived from the RV measurements and assuming an Earth-like rocky composition, we can estimate a lower limit for $R_{\rm p}$ at $\sim$1.1~$R_\oplus$ via the mass-radius relationship from \citet{parc2024}. This value yields similar conclusions regarding the circulation regime as the previous calculations.
Therefore, Ross~128~b is more likely to exhibit a super-rotating atmospheric circulation regime (with the presence of a planet-wide equatorial jet and Rossby-wave gyres), assuming a synchronous rotation, for \ce{N2} and \ce{CO2} atmospheres.

\subsubsection{Atmospheric water vapor, surface liquid water and cloud distribution}
\label{sub:climate_results:refcase:water}

\begin{figure*}
    \centering
    \includegraphics[width=0.9\textwidth, clip]{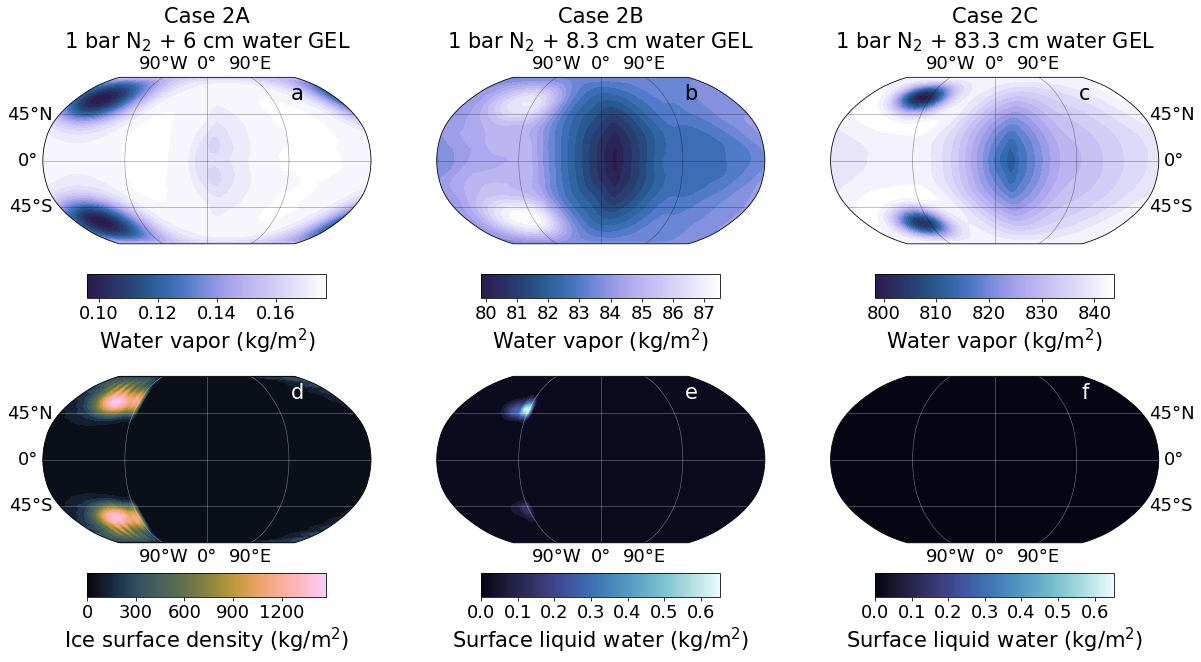}
    \caption{Latitude-longitude maps of column-integrated water vapor (panels a, b and c), water ice surface density (panel d) and surface liquid water (panels e and f) for Cases~2A, 2B and~2C. The variables are averaged over $\sim$50 orbits. Each map uses a different color map scaling.}
    \label{fig:maps_water}
\end{figure*}

\begin{figure*}
    \centering
    \includegraphics[width=0.95\textwidth, clip]{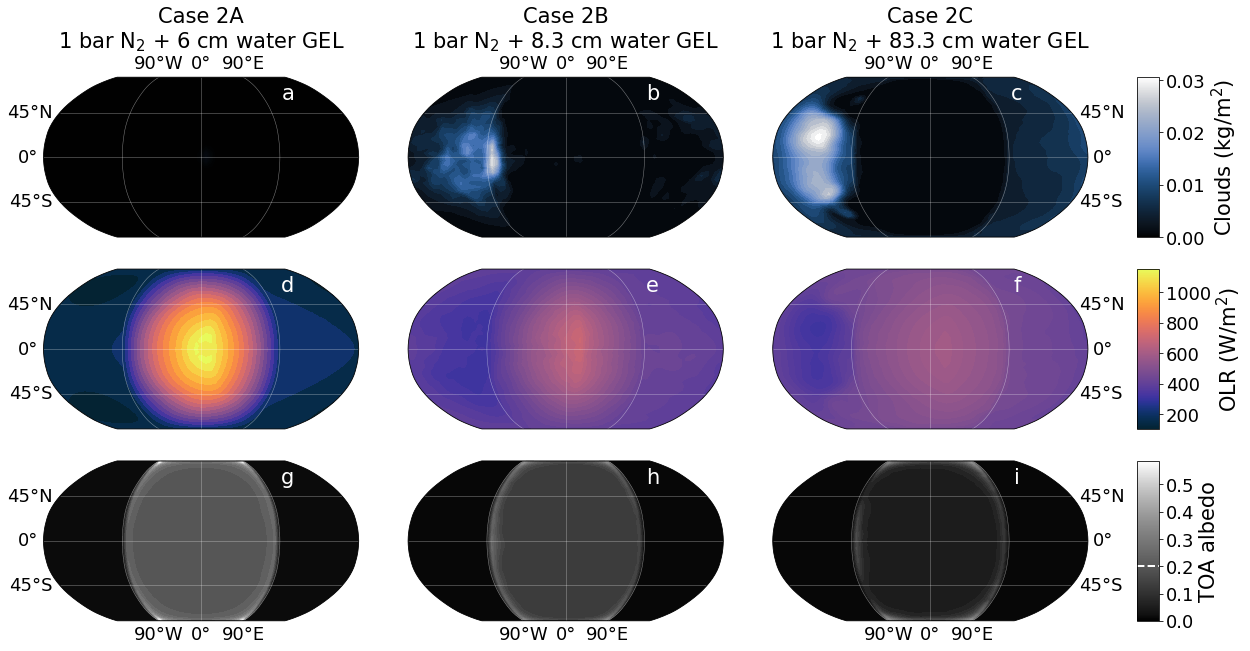}
    \caption{Latitude-longitude maps of column-integrated water clouds (panels a, b and c), outgoing longwave radiation (OLR, panels d, e and f) and albedo at the top of the atmosphere (TOA albedo, panels g, h and i) for Cases~2A, 2B and~2C. The variables are averaged over $\sim$50 orbits. The dashed white line on the albedo color bar corresponds to the surface albedo.}
    \label{fig:maps_T_olr_clouds_alb}
\end{figure*}

In this section, we discuss the distribution of the global water content in Case~2B. Figure~\ref{fig:maps_water} shows the column-integrated water vapor concentration, the liquid water content at the surface and the ice surface density. Figure~\ref{fig:maps_T_olr_clouds_alb} presents the maps of column-integrated water clouds, outgoing longwave radiation (OLR, i.e., thermal emission at the top of the atmosphere) and albedo at the top of the atmosphere (TOA albedo). Here we focus on the maps corresponding to Case~2B shown in panels b and e of Figure~\ref{fig:maps_water} (for water vapor and surface liquid water, respectively), and in panels b, e and h of Figure~\ref{fig:maps_T_olr_clouds_alb} (for clouds, OLR and TOA albedo, respectively).

Figure~\ref{fig:maps_water} (panel b) shows relatively homogenous water vapor concentrations (with a difference of $\sim$8~kg\,m$^{-2}$ between the minimum and maximum values) and an average water vapor mixing ratio of 83~kg\,m$^{-2}$.
The water vapor concentrations are lower above the heated region on the dayside.
This is caused by the atmospheric circulation, which transports water vapor vertically (with the ascending branches of the Hadley cells) and induces divergence from the heat source at higher altitudes, leading to the decrease of the water vapor concentrations.
The regions showing the highest water vapor concentrations are located on the nightside near the western terminator at mid-latitudes in both hemispheres, corresponding to the locations of the Rossby-wave gyres (as discussed in Section~\ref{sub:climate_results:refcase:circulation}).
These regions have surface temperatures low enough (see Figure~\ref{fig:maps_T_winds}) to allow condensation of atmospheric water vapor into liquid phase at the surface, as illustrated on the surface liquid water map (panel e).

Figure~\ref{fig:maps_T_olr_clouds_alb} (panel b) shows that in Case~2B, water clouds form preferentially on the nightside and near the western terminator.
These cloud patterns are consistent with the descriptions of \citet{Kodama2019}, \citet{turbet2021} and \citet{chaverot2023} for land planets or water-rich atmospheres.
For planets with water-rich atmospheres but lacking surface liquid water oceans, the high absorption of incoming stellar radiation (ISR) by atmospheric water vapor increases dayside temperatures. This temperature increase reduces moist convection above the substellar point, thereby inhibiting cloud formation on the dayside.
Moreover, the atmospheric circulation enables heat transport from the dayside to the nightside. The cooling of the warm and moist air parcels reaching the nightside leads to cloud formation on the nightside \citep{turbet2021, chaverot2023}.
This is consistent with Case~2B, characterized by a water-rich atmosphere, high dayside temperatures and the absence of surface liquid water on the dayside, which reduces moist convection in this region and prevents the formation of dayside clouds.

Nightside clouds have a net warming effect because of their efficient greenhouse effect \citep{turbet2021, chaverot2023}. They absorb the thermal emission from the lower atmospheric layers and the planetary surface, which reduces the OLR above the clouds. 
Figure~\ref{fig:maps_T_olr_clouds_alb} (panel e) shows this decrease in OLR on the nightside, with a pattern corresponding to the shape of the cloud distribution.
Furthermore, the TOA albedo map (panel h) shows a slight increase in albedo near the western terminator, which could be attributed to the clouds located on the western terminator (with a small fraction on the dayside, see panel b).
This point will be further discussed in Section~\ref{sec:observability_results}. 

\subsubsection{Evolution of the global water content}

\begin{figure}[!htb]
   \centering
   \includegraphics[width=\hsize]{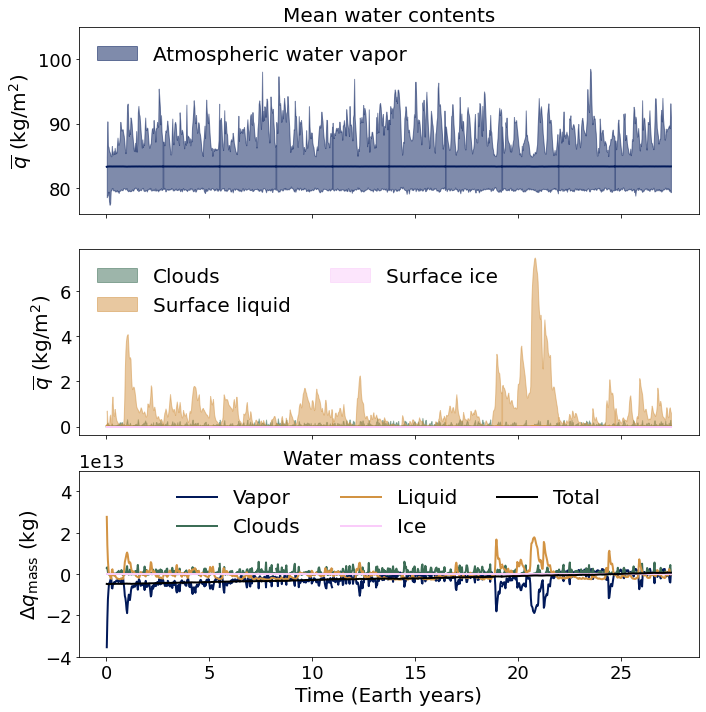}
    \caption{Temporal evolutions of the column-integrated water vapor (top panel), water clouds, surface liquid water and ice surface density (middle panel) for Case~2B. The mean values are represented by the solid lines, while the minimum and maximum values are defined by the filled areas in the first two panels. The lower panel represents the deviations of the total water mass content for each quantity from its respective initial value.}
    \label{fig:water_evolution}
\end{figure}

Figure~\ref{fig:water_evolution} shows that the global contents of atmospheric water vapor and surface liquid water exhibit a variability over time, characterized by fluctuations between higher and lower liquid water content levels at the surface, approaching near-zero values\footnote{The difference of total water mass ($\Delta \mathrm{q}_{\rm mass,~total}$) is expected to be at zero. However, this is not what is shown in Figure~\ref{fig:water_evolution}, indicating that the global water content of the system is not conserved. This requires further investigation. However, the difference corresponds only to $10^{-3}$\%, thus this should not have an impact on the simulation.}. 
We observe an anti-correlation between water vapor and the combination of clouds and surface liquid water, with the lowest surface liquid water levels corresponding to the highest atmospheric water vapor concentrations. This behavior is potentially indicative of a transient and unstable climate regime. We tested the robustness of this state by performing a simulation with a similar yet slightly different global water content corresponding to 8~cm~GEL, initially condensed at the surface. Despite these different initial conditions, we find a similar outcome where water alternates between condensation in the cold points at the surface and evaporation back in the atmosphere (see Figure~\ref{fig:water_evolution_appendix} in Appendix~\ref{appendix:water}). 
As we expect this state to exist only within a narrow range of water content, we explore in the next section the impact of varying the global water content on the climate. 

\subsection{Variation of the water inventory}
\label{sub:climate_results:water_variation}

In this section, we vary the global water content of the system to estimate its impact on the climate of Ross~128~b. The subsequent simulations correspond to Cases~2A, 2B and~2C, where the mean atmospheric surface pressure and gas composition are fixed at 1~bar of \ce{N2} atmosphere, as specified in Table~\ref{tab:atmospheres}. We vary the initial global water content (considered here as the only free parameter) from 6~cm~GEL (Case~2A) to 83.3~cm~GEL (Case~2C)\footnote{We also performed a simulation for a dry case, which is an idealized scenario useful to show the potential climate regimes and atmospheric circulation patterns on tidally locked rocky planets. This dry case exhibits the lowest mean surface temperature ($\sim$250~K) and the highest day/night temperature contrast ($\sim$255~K). Since Case~2A (6~cm~GEL) shows comparable results, the climate regimes of these two scenarios will be considered equivalent in this study.}. The results are shown in Figures~\ref{fig:maps_T_winds}, \ref{fig:maps_T_olr_clouds_alb} and~\ref{fig:maps_water}.

\subsubsection{Impact on the circulation regime}

The atmospheric circulation patterns illustrated in Figure~\ref{fig:maps_T_winds} show similar characteristics across all cases. All surface maps show a convergence of surface winds towards the hottest point (panels a, b and c). At higher altitudes, each case exhibits the presence of an eastward equatorial jet and Rossby-wave gyres (panels d, e and f).
This suggests that variations of the initial water reservoir for the specific case of Ross~128~b with an \ce{N2} atmosphere may not affect the circulation regime described in Section~\ref{sub:climate_results:refcase:circulation}.

\subsubsection{Impact on the climate regime}

Figure~\ref{fig:maps_T_winds} shows that, in all cases, surface temperatures tend to increase with increasing water content, with mean values of approximately 268~K, 337~K and 456~K for Case~2A (panel a), Case~2B (panel b) and Case~2C (panel c), respectively.
This corresponds to an increase of nearly 200~K from the driest to the most water-rich scenario. Moreover, the day/night temperature contrast decreases as water content increases. Specifically, the temperature difference between the hottest point on the dayside and the coldest regions on the nightside decreases from around 234~K for Case~2A to 152~K for Case~2B, and 144~K for Case~2C. This indicates a reduction of nearly 100~K in the thermal contrast between the two extreme scenarios. 

Case~2A exhibits nightside temperatures below the freezing point of water at 1~bar of pressure. This enables water condensation into ice on the two coldest points of the nightside surface, as illustrated on the ice surface density map in Figure~\ref{fig:maps_water} (panel d). This collapse of water into the surface cold traps leads to the depletion of the atmospheric water vapor, as seen in Figure~\ref{fig:maps_water} (panel a), which shows a mean water vapor mixing ratio of 0.2~kg\,m$^{-2}$.
The climate regime of the driest scenario represents a stable equilibrium characterized by dominant surface ice accumulation on the nightside. This state is triggered when the nightside surface temperatures drop below the water freezing point, causing the atmospheric water vapor to condense into ice at the surface. In this regime, the greenhouse warming effect of atmospheric water vapor is effectively overcome by the surface condensation process. This climate state is referred to as the collapsed state in \cite{leconte2013aa}, indicating that water exists predominantly in solid form at the surface, with atmospheric water vapor largely depleted and replaced by surface ice deposits.

On the other hand, in the case of the most water-rich scenario represented by Case~2C, the climate reaches another stable equilibrium where surface temperatures are sufficiently high to maintain water completely vaporized in the atmosphere. This results in a more efficient greenhouse effect leading to an increase of the global surface temperatures. This trend occurs because water vapor facilitates more efficient heat redistribution through the atmosphere. This leads to a reduction of the day/night temperature contrast and an offset of the hottest point at the surface from the substellar point. This state is referred to as the runaway state in \cite{leconte2013aa}. 
Other cases with higher water vapor contents would lead to higher surface temperatures.

These results are characteristic of the moist bistability described in \cite{leconte2013aa}. They show the existence of two stable climate regimes for rocky planets orbiting closer to their stars than the inner edge of the HZ. In this study, we will consider three representative climate states: 
(1) the collapsed state, illustrated in Case~2A, where all available water is trapped as nightside surface ice deposits,
(2) the transient state, described in Case~2B in Section~\ref{sub:climate_results:refcase}, where water is mostly vaporized in the atmosphere but surface liquid water is sustained in the two nightside cold regions and 
(3) the runaway state, represented by Case~2C, where all available water is vaporized in the atmosphere. 

\subsubsection{Impact on the water vapor distribution}

Figures~\ref{fig:maps_water} (panels a, b and c) shows that Cases~2A, 2B and 2C correspond to distinct regimes of water vapor concentration, characterized by different mean mixing ratios: 0.2~kg\,m$^{-2}$ in Case~2A (indicative of a water vapor-depleted atmosphere), 83~kg\,m$^{-2}$ in Case~2B and 833~kg\,m$^{-2}$ in Case~2C (both representative of states in which water is globally vaporized in the atmosphere).
As the global water content increases, the water vapor concentrations increase and the atmospheric heat transport becomes more efficient.
This induces the eastward shifts of (1)~the hottest point from the substellar point at the surface and (2)~the Rossby-wave gyres (and the corresponding cold points at the surface) toward the western terminator (Figure~\ref{fig:maps_T_winds}).

Although the column-integrated water vapor concentrations are spatially homogeneous in all cases, with relative differences between minimum and maximum values of 46\%, 9\% and 5\% for Cases~2A, 2B and~2C, respectively, the water vapor spatial distribution seems to vary with the global water content, and thus the prevailing climate regime. 
First, all cases show lower water vapor concentrations above the dayside, due to the ascending branches of the Hadley cells transporting water vapor away from the heated regions (as discussed in Section~\ref{sub:climate_results:refcase:water}).
Second, both stable equilibrium states (Cases~2A and~2C) show lower water vapor concentrations at the Rossby-wave gyre regions. These regions experience subsiding air motion (from the descending branches of the circulation cells) and divergence near the surface, which effectively transport water vapor away from the gyre regions, inducing this decrease of water vapor concentrations. 
In contrast, the transient state (Case~2B) differs by presenting higher concentrations in these regions, which coincide with the locations of the colder surface regions where liquid water is maintained.
Liquid water can evaporate from the surface back to the atmosphere, and this local evaporation (which tends to increase water vapor concentrations in these regions) may predominate over the large-scale water vapor transport (which tends to reduce water vapor concentrations in these regions).
This may result in higher water vapor concentrations in the Rossby-wave gyre regions, as observed in Case~2B.

\subsubsection{Impact on the cloud distribution}

Figure~\ref{fig:maps_T_olr_clouds_alb} (panels a, b and c) illustrates the vertically integrated water cloud formation patterns across the three cases. 
The driest scenario (Case~2A) remains essentially cloud-free, showing the correlation between the cloud content and the water reservoir.
In Case~2C, clouds mainly form in the nightside, with a higher cloud density in the western hemisphere, similar to the cloud distribution observed in Case~2B (described in Section~\ref{sub:climate_results:refcase:water}).
The overall cloud content in Case~2C exceeds that of Case~2B, corresponding to its higher water vapor concentration.

\subsection{Effect of the atmospheric composition on the climate} 
\label{sub:climate_results:atmosphere_variation}

In this section, we discuss the impact of varying atmospheric compositions on the climate. 
The simulation setups presented here are Cases~2, 4 and~6 (defined in Table~\ref{tab:atmospheres}).
They correspond to atmospheres of \ce{N2} (Case~2), \ce{N2 + 376ppm CO2} (Case~4) and \ce{CO2} (Case~6).
All three cases assume synchronous rotation and 1~bar of surface pressure. They were simulated under dry and wet conditions, and the wet scenarios include a global water content of 8.3~cm~GEL (thus Case~2 with wet conditions corresponds to Case~2B).
We provide a summarized description of our results here.
The maps of surface temperatures, water vapor, clouds and TOA albedo corresponding to these results are available in Appendix~\ref{appendix:maps_atmospheres}.

\ce{CO2} has a warming effect on climate: in dry and wet scenarios, temperatures tend to increase across the surface with increasing partial pressures of \ce{CO2} in the atmosphere.
\ce{CO2} is a more effective ISR (i.e., incoming stellar radiation) absorber for a planet around an M~dwarf star than around a Sun-like star, due to its strong absorption lines in the near-infrared.
Despite this warming effect, the surface temperatures in the three setups under dry conditions remain below the freezing point of water in the coldest nightside regions.
Further simulations would be useful to investigate the minimum \ce{CO2} content needed to prevent the formation of these cold traps where ice could accumulate, as explored in \citet{turbet2016} and \citet{Taniguchi2026}.

In the wet scenarios, the three atmospheric setups show similar climate features: (1) they maintain temperatures above the freezing point of water across the surface, (2) they exhibit concentrations of atmospheric water vapor similar to Case~2B (due to similar water inventory), with most of the available water remaining gaseous within the atmosphere, and (3) they show similar cloud distributions, with clouds forming preferentially on the nightside.
In particular, both cases with \ce{N2}-dominated atmospheres (Cases~2 and~4) in wet conditions show comparable surface temperatures, with temperature conditions allowing condensation of water into liquid form within the two surface cold traps. 
In contrast, the \ce{CO2}-dominated atmosphere scenario (Case~6) shows higher surface temperatures, preventing the condensation of liquid water on the surface.
Therefore, Case~4 (in wet conditions) is analogous to Case~2B (i.e., our transient state in which most of the water is vaporized in the atmosphere, but some surface liquid water is sustained within the nightside cold traps), and Case~6 (in wet conditions) aligns with Case~2C (runaway state in which all available water remains in vapor form in the atmosphere).

\subsection{Effect of the atmospheric surface pressure on the climate}
\label{sub:climate_results:pressure_variation}

This section investigates the effects of varying mean surface pressures on the climate for \ce{N2} and \ce{CO2} atmospheres. 
We focus on Cases~1, 2 and~3, corresponding to \ce{N2} atmospheres with surface pressures of 0.1, 1 and 10~bars, respectively, and Cases~5 and~6, corresponding to \ce{CO2} atmospheres with surface pressures of 0.1 and 1~bar, respectively (as detailed in Table~\ref{tab:atmospheres}). All cases assume synchronous rotation.
As in the previous section, we provide a summarized description of our results here.
A more detailed description of the results is available in Appendix~\ref{appendix:maps_pressures}, with maps of surface temperatures, water vapor, clouds, TOA albedo, \ce{H2O} and \ce{CO2} ice surface densities.

\subsubsection{\ce{N2}-dominated atmospheres} 

For \ce{N2} atmospheres, we conducted simulations under dry and wet conditions. 
The scenarios in wet conditions include a water inventory of 8.3~cm~GEL.
In both dry and wet regimes, increasing surface pressures tend to increase temperatures across the surface and reduce the day/night temperature contrast, showing stronger heat redistribution.
Among the wet scenarios, the case with a thin atmosphere (0.1~bar) shows the highest temperature contrast, with surface temperatures on the nightside dropping below the freezing point of water.
This results in the condensation of water into ice on the surface nightside, with a higher ice surface density in the cold traps near the western terminator.
Therefore, Case~1 in wet conditions exhibits a climate regime analogous to Case~2A, with a water-depleted atmosphere and the presence of surface ice on the nightside.
In contrast, the temperatures for the case with a dense atmosphere (10~bars) under wet conditions are high enough to maintain water fully vaporized in the atmosphere. Thus, the climate regime of Case~3 in wet conditions is similar to Case~2C.

\subsubsection{\ce{CO2}-dominated atmospheres}

We conducted the simulations for Cases~5 and 6 (\ce{CO2} atmospheres) with a water inventory equivalent to 0.83~cm~GEL. 
In the case of a thin atmosphere (0.1~bar) and low water content (0.83~cm~GEL), the surface temperatures on the nightside fall below the water freezing point and the \ce{CO2} condensation temperature at 0.1~bar.
This enables water vapor and \ce{CO2} in the atmosphere to condense into ice in the cold traps at the surface.
As \ce{CO2} is the main constituent of the atmosphere in this case, this leads to the collapse of the atmosphere.

\begin{figure}[!htb]
   \centering
   \includegraphics[width=\hsize]{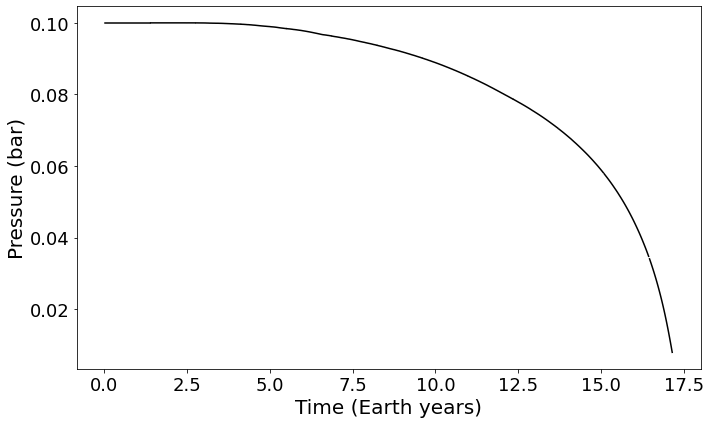}
    \caption{Temporal evolution of the mean atmospheric surface pressure for the collapsing case of a \ce{CO2}-dominated atmosphere initialized with a surface pressure of 0.1~bar (Case~5) and a global water content corresponding to 0.83~cm~GEL.}
    \label{fig:psurf_collapse}
\end{figure}

The temporal evolution of the mean surface pressure for Case~5 illustrated in Figure~\ref{fig:psurf_collapse} shows that the initial mean surface pressure of 0.1~bar decreases over time to around 0.01~bar, indicating the atmospheric collapse. The collapse of \ce{CO2} reduces the efficiency of the heat redistribution, resulting in a higher day/night thermal contrast that facilitates the condensation of atmospheric \ce{CO2} within the colder regions. This positive feedback is described in \cite{turbet2016}. This case highlights the importance of efficient heat redistribution in \ce{CO2}-dominated atmospheres, driven by sufficient \ce{CO2} abundance, to prevent atmospheric collapse and maintain the stability of the atmosphere \citep{turbet2016, maurel2025}.

\subsection{Comparison synchronous/asynchronous rotations}
\label{sub:climate_results:rotation_variation}

In this section, we analyze the impact of a scenario with an asynchronous rotation (5:2 spin-orbit resonance) and high obliquity (79°) on the climate of Ross~128~b.
Here we compare two setups assuming (1)~1:1 resonance with 0° obliquity and (2)~5:2 resonance with 79° obliquity.
We assume 1~bar of \ce{N2} and \ce{CO2} atmospheres (Cases~2 and 6, respectively) and 8.3~cm~GEL of water inventory.
As in the previous sections, we provide a summarized description of our results here (the corresponding maps of surface temperatures, water vapor, clouds and TOA albedo are available in Appendix~\ref{appendix:maps_rotations}).

In both \ce{N2} and \ce{CO2} atmospheres, the surface temperatures, water vapor and clouds are more uniformly distributed across the planet (especially along the equator) in the case of a 5:2 resonance.
The higher spin-orbit resonance (corresponding to a higher rotation rate of $1.8\times10^{-5}$) induces a more effective longitudinal transport of heat and water vapor across the atmosphere, and a significant reduction of the temperature contrasts. 
The high obliquity induces slightly higher temperatures at the poles, but the temperature contrasts between the warmer poles and the colder equator remain significantly lower compared to the day/night temperature contrasts in the synchronous configurations.
This homogenization of the temperature distribution (1)~enables cloud formation uniformly around the planet and (2)~prevents the condensation of water into liquid phase on the surface, a condition that occurs under synchronous rotation and for an \ce{N2} atmosphere (see Case~2B described in Section~\ref{sub:climate_results:refcase}).
Therefore, in the cases assuming 5:2 resonance, all available water is vaporized in the atmosphere, which is consistent with the climate regime represented by Case~2C.

\subsection{Overview of the results and link with the planetary albedo}
\label{sub:climate_results:overview}

\begin{figure*}[!htb]
    \centering
    \includegraphics[width=\hsize]{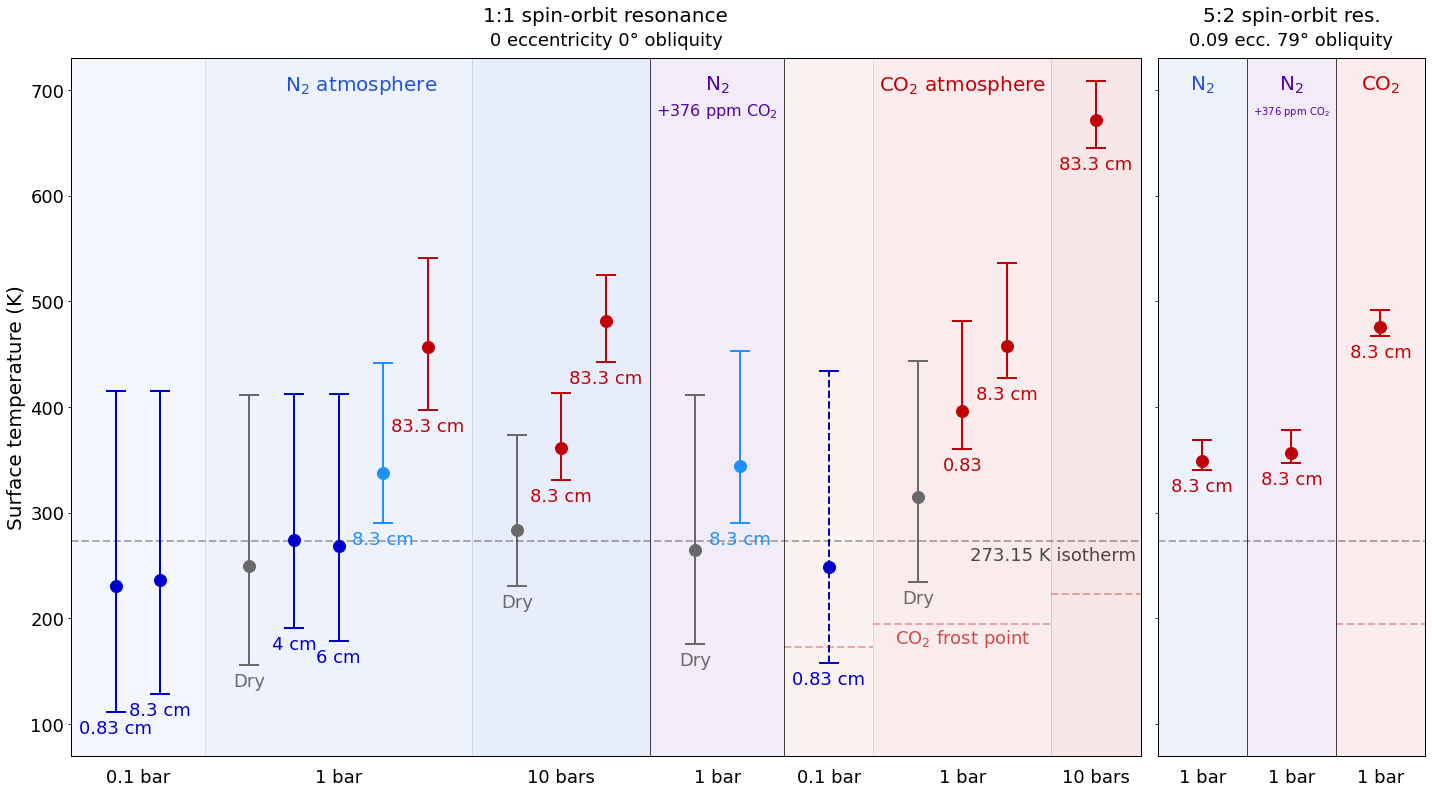}
    \caption{Mean surface temperatures for all simulated cases; markers indicate the global mean, while error bars extend from the minimum to the maximum surface temperatures. Gray markers represent the dry cases; dark blue markers represent the collapsed states (where water collapses into ice, Case~2A); light blue markers represent the transient states with trapped surface liquid water (Case~2B); and red markers represent the runaway states (where all water is vaporized, Case~2C). The dashed bar represents the case where atmospheric \ce{CO2} collapses into surface ice (atmospheric collapse). The left panel presents cases with synchronous rotations, and the right panel cases with asynchronous rotations, using spin-orbit resonance, eccentricity and obliquity values from \cite{valente2022}. The 273.15~K isotherm below which water freezes and the isotherms below which \ce{CO2} condenses at 0.1, 1 and 10~bars are represented by the gray and red dashed lines, respectively.}
    \label{fig:tsurf_summary}
\end{figure*}

The results are summarized in Figure~\ref{fig:tsurf_summary}, which shows the mean surface temperatures across all simulated cases, including their respective minimum and maximum values. The distribution of mean surface temperatures shows significant variability, ranging from approximately 230~K in scenarios characterized by tenuous \ce{N2}-dominated atmospheres with limited water reservoirs, to 671~K in cases involving dense \ce{CO2}-rich atmospheres with high water contents. This temperature range reflects diverse climate regimes, highlighting the high sensitivity of the planetary climate to a broad range of parameters, including total water inventory, surface pressure, atmospheric composition and the planetary rotation state.

The atmospheric water vapor concentration is determined by the total water reservoir of the planet. In runaway scenarios, characterized by high global temperatures, efficient heat redistribution and minimal day/night temperature contrasts, a linear correlation exists between the water vapor content and the surface temperature. We find that planets are in a runaway scenario even with a relatively small water content (transition around 8~cm~GEL). Moreover, scenarios with dense atmospheres or atmospheres enriched with other greenhouse gases such as \ce{CO2} tend to the runaway state.
In contrast, colder climate regimes, marked by high day/night thermal contrasts due to less effective heat redistribution, manifest in scenarios with very low global water contents ($\mathrm{<8~cm~GEL}$), thin atmospheres, or atmospheres poor in other greenhouse gases. Under such conditions, a substantial fraction of the atmospheric water vapor can condense into ice on the surface (see Figure~\ref{fig:maps_water}). However, the condensation occurs exclusively on the nightside, where surface temperatures allow phase transition, whereas the dayside remains free of condensation due to higher temperatures. This pattern is primarily governed by the atmospheric circulation regimes characteristic of synchronously rotating planets. Furthermore, while condensed ice scenarios tend to produce low cloud contents, water clouds preferentially form on the nightside in the warmer scenarios, in agreement with \citet{Kodama2019}, \citet{turbet2021} and \citet{chaverot2023}. 

In the context of assessing the planetary albedo, the main quantities potentially affecting the dayside reflectivity include the atmospheric water vapor concentration, the surface water condensation (into ice or liquid) and the cloud distribution, all of which modulated by the prevailing climate regime. In scenarios exhibiting runaway conditions or presence of surface liquid water, we expect high water vapor concentrations to decrease the planetary reflectivity due to the high water vapor absorption efficiency, particularly within the near-infrared. Moreover, the cloud formation, mainly occurring in the nightside, is not expected to significantly impact the planetary albedo. 
In collapsed scenarios, the surface water condensation mainly occurring in the nightside and the limited cloud coverage are not expected to have a substantial influence on the reflectivity. The subsequent section aims to quantitatively check these hypotheses.

\section{Detectability of Ross 128 b in reflected light}
\label{sec:observability_results}

This section presents the detectability analysis of Ross~128~b in reflected light, using it as a representative analog for the population of warm, short-orbit exoplanets that are overrepresented due to observational biases. With a maximum angular separation of 14.7~mas, Ross~128~b currently lies outside the detection limits of RISTRETTO and ANDES (see Figure~\ref{fig:contrast_vs_sep}); however, it remains a potential target for observation with PCS, which aims to also cover the visible and could reach the contrast and separation required for direct characterization. Given that Ross~128~b has the smallest angular separation and highest stellar irradiation among the targets of the ANDES golden sample, it can serve as a representative example for highly irradiated planets, which are more likely to exhibit a runaway climate if they host water. This work can then help to prepare the upcoming observations in reflected light with the ELT by providing a more accurate assessment of the spectral reflectivity of these targets. In this section, we discuss the Bond (\ref{sub:observability_results:bond}) and geometric (\ref{sub:observability_results:geometric}) albedos computed for all simulated cases explored in the previous section. The results are shown in Figures~\ref{fig:spectra} and \ref{fig:albedo_summary} and summarized in Table~\ref{tab:albedo}.

\subsection{Bond albedo}
\label{sub:observability_results:bond}

In this section, we calculate the Bond albedo (i.e., the reflectivity at the top of the atmosphere integrated over all wavelengths) of Ross~128~b assuming the different climates obtained in the GCM simulations described in Section~\ref{sec:climate_results}. The values are shown in Figure~\ref{fig:albedo_summary}, and compared to the albedo of the rocky surface represented by the dashed gray line ($A_{\rm surf}=0.2$). 

\subsubsection{Effect of atmospheric absorbers}

The Bond albedo for the dry, \ce{N2} atmosphere scenario at 1~bar of surface pressure is 0.21 (Case~2, as specified in Table~\ref{tab:atmospheres}). This value closely approximates the surface albedo, indicating a radiatively transparent atmosphere. 
In Case~2A (water vapor-depleted scenario), the Bond albedo is slightly decreased to 0.19. 
For fully vaporized atmospheres (Cases~2B and~2C) with mean water vapor mixing ratios of 83 and 833~kg\,m$^{-2}$, respectively, the Bond albedo further decreases to 0.11 and 0.05. This trend indicates that, for planets orbiting M~dwarf stars whose spectral emission peaks in the near-infrared, atmospheric water vapor reduces the Bond albedo by increasing absorption in the water spectral bands that coincide with significative stellar flux. For runaway climate states, increased water vapor concentrations would then enhance the planetary mean absorptivity, potentially limiting the detection of the total reflected flux of the planet. 
Moreover, water clouds are mainly located on the nightside \citep{Kodama2019, turbet2021, chaverot2023} and thus do not contribute to the reflectivity of the planet. 

The dry scenario at 1~bar of surface pressure simulated with an Earth-like composition (Case~4) exhibits a Bond albedo of 0.2, similar to the pure \ce{N2} atmosphere case. In contrast, a pure \ce{CO2} atmosphere as in Case~6 results in a lower Bond albedo of 0.15. This indicates that the Bond albedo reduces with increasing partial pressure of \ce{CO2}. 

Across all scenarios explored, Figure~\ref{fig:albedo_summary} shows that the Bond albedo ranges between 0.05 and 0.24, rarely exceeding the surface albedo value of 0.2. 
The mean planetary absorptivity tends to be higher, showing a significative sensitivity to the presence of efficient near-infrared absorbers in the atmosphere (water vapor and \ce{CO2} here). 
Therefore, photometric measurements could provide initial constraints on the presence of strong atmospheric absorbers, but high-resolution spectroscopy is needed to identify them.

\subsubsection{Effect of the atmospheric surface pressure}

In the dry, \ce{N2} atmosphere scenario with a surface pressure of 10~bars (Case~3), corresponding to a higher atmospheric density, the Bond albedo reaches a maximum value of 0.24, higher than the 0.21 value obtained at 1~bar. This enhancement is driven by the Rayleigh scattering of \ce{N2}, which is more efficient in denser atmospheres, thereby increasing the planetary albedo in the visible. Therefore, increased surface pressures can amplify the planetary reflectivity through Rayleigh scattering. However, multiple parameters, such as dayside cloud coverage, atmospheric hazes or surface ice, can lead to similar increases in the total reflected flux of the planet. These degeneracies limit the ability to disentangle individual contributions through photometric observations only.

\subsection{Spectral geometric albedo}
\label{sub:observability_results:geometric}

\begin{figure*}[!htb]
    \centering
    \includegraphics[width=\hsize]{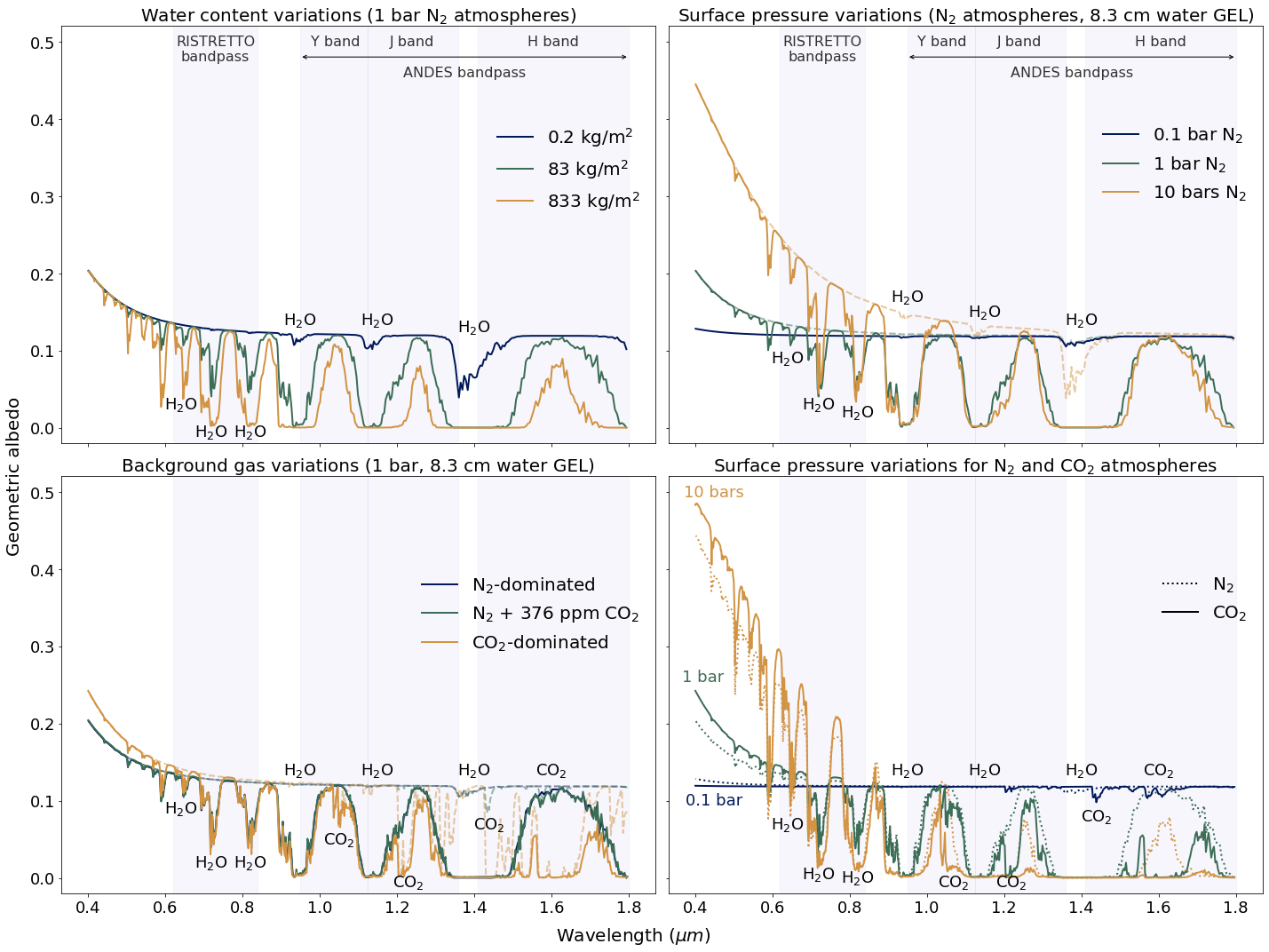}
    \caption{Geometric albedo spectra computed between 0.4 and 1.8~$\mu$m. The spectral coverage regions of RISTRETTO and ANDES are depicted by the shaded blue areas. All cases assume a synchronous rotation. \textit{Top left:} Spectra corresponding to Cases~2A, 2B and~2C, representative of the three climate states described in Section~\ref{sec:climate_results} (collapsed, transient and runaway states, corresponding to mean water vapor mixing ratios of 0.2, 83 and 833~kg\,m$^{-2}$, respectively). Each case assumes 1~bar of \ce{N2} atmosphere. \textit{Top right:} Spectra for Cases~1, 2B and~3, assuming an \ce{N2} atmosphere and varying surface pressures from 0.1 to 10~bars. The dashed spectra represent the dry scenarios, while the solid lines represent cases with a water inventory fixed to 8.3~cm~GEL. \textit{Bottom left:} Spectra for Cases~2, 4 and~6, corresponding to distinct atmospheric compositions and a given surface pressure of 1~bar. The dashed spectra represent the dry scenarios, while the solid lines represent cases with a water inventory fixed to 8.3~cm~GEL. \textit{Bottom right:} Spectra corresponding to cases with different surface pressures (0.1, 1 and 10~bars), for \ce{N2} (dashed lines, Cases~1, 2 and~3) and \ce{CO2} (solid lines, Cases~5, 6 and~7) atmospheres. The global water content corresponds to 0.83~cm~GEL for 0.1~bar; 8.3~cm~GEL for 1~bar; and 83.3~cm~GEL for 10~bars.}
    \label{fig:spectra}
\end{figure*}

\begin{figure*}[!htb]
    \centering
    \includegraphics[width=\hsize]{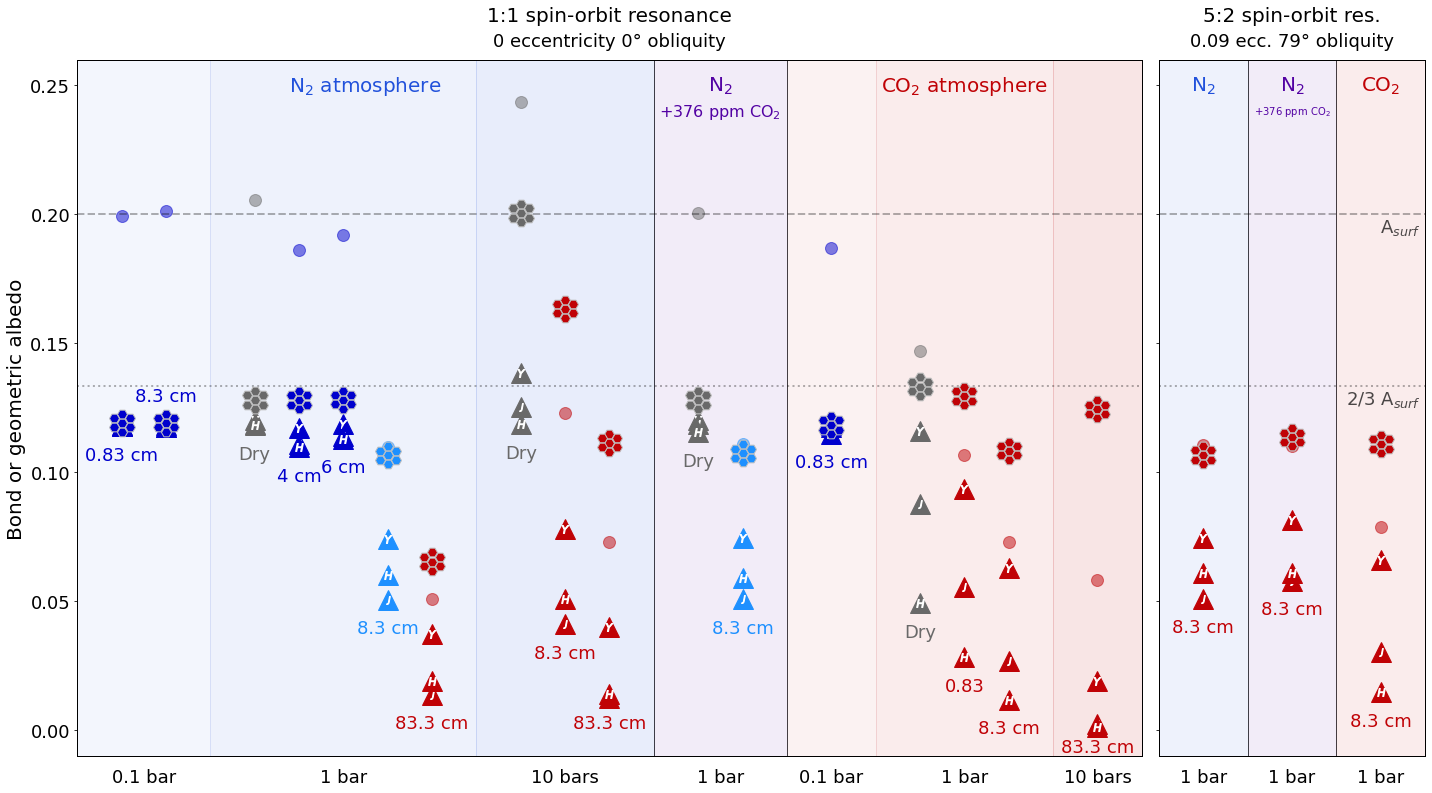}
    \caption{Mean Bond and geometric albedos for all simulated cases (the values can be found in Table~\ref{tab:albedo}). The Bond albedo values are represented by transparent circles. The mean geometric albedo values are computed for the RISTRETTO bandpass (hexagons), and the Y, J and H bands (triangles) as described in Table~\ref{tab:bands}. The dashed gray horizontal line indicates the rocky surface albedo $A_{\rm surf}$ considered in all GCM simulations. The dotted gray line delineates the geometric albedo of a Lambertian spherical surface with a surface albedo $A_{\rm surf}$ of 0.2, which is equal to $2/3$ of $A_{\rm surf}$ (the $2/3$ factor comes from the projection of a flat Lambertian disk on a spherical surface).}
    \label{fig:albedo_summary}
\end{figure*}

Figure~\ref{fig:spectra} presents the geometric albedo spectra computed with Pytmosph3R for all scenarios.
The mean geometric albedo values, derived over the spectral ranges covered by RISTRETTO and ANDES, are shown in Figure~\ref{fig:albedo_summary}.

\subsubsection{Variation of total water inventory}

The top panel of Figure~\ref{fig:spectra} shows the geometric albedo spectra corresponding to Cases~2A, 2B and 2C (representative of the three climate regimes described in Section~\ref{sec:climate_results}).
All scenarios exhibit spectral features associated with water vapor absorption bands in the near-infrared. 
The collapsed case, characterized by atmospheric water depletion (with a mean water vapor mixing ratio of 0.2~kg\,m$^{-2}$), shows relatively weak and narrow absorption features. In contrast, the transient and runaway scenarios, which involve higher atmospheric water vapor concentrations, are characterized by broader and deeper absorption bands, rapidly reaching zero flux levels within these spectral regions. This indicates a high optical thickness and efficient absorption of the stellar incident flux due to the presence of abundant atmospheric water vapor. This leads to the decrease of the overall reflected flux, which is consistent with the initial Bond albedo analyses. 

The spectral disparities between the collapsed and runaway scenarios shown in Figure~\ref{fig:spectra}, as described also in \cite{leconte2013aa}, suggest that high-resolution spectroscopy, given sufficient total flux, could allow differentiating these two different climate regimes. Further analyses are required to provide more quantitative assessments to determine how water vapor absorption decreases the reflected flux and impacts the detection sensitivity, by exploring more water vapor abundances scenarios between the two boundary values described in this study.

For RISTRETTO and ANDES, we can assess preliminary constraints on the detectability of a planet with 1~bar of \ce{N2} atmosphere and various water reservoirs. We can already infer from the analysis of the three spectra presented in Figure~\ref{fig:spectra} that the geometric albedo within the RISTRETTO bandpass is higher relative to that within the ANDES YJH bands.
The mean geometric albedo values illustrated in Figure~\ref{fig:albedo_summary} confirms this assessment.
The collapsed case (dark blue) exhibits the highest mean geometric albedo among the three cases for both RISTRETTO and ANDES, with values ranging from 0.11 to 0.13. These values closely correspond to the geometric albedo of a Lambertian sphere with a surface albedo of 0.2, which means that the atmosphere is relatively radiatively transparent. In both instruments, the geometric albedo shows a decreasing trend with increasing atmospheric water vapor concentration, reaching values from 0.01 to 0.07, which aligns with the initial Bond albedo estimate. This behavior is due to the sensitivity of both instrument bandpasses to water vapor absorption features, as shown in Figure~\ref{fig:spectra}. Therefore, the efficient absorption of the stellar incident radiation in these spectral regions limits the detection of the planetary reflected light. The detection sensitivity is particularly impacted within the Y, J and H bands covered by ANDES, which coincide with the main water vapor absorption features in the near-infrared (at 0.9, 1.1 and 1.4~$\mu$m). This results in a more pronounced disparity in geometric albedo between the two instruments as atmospheric water vapor abundance increases.

\subsubsection{Variation of surface pressure}

The top right panel of Figure~\ref{fig:spectra} shows the computed geometric albedo spectra for Cases~1, 2 and 3 (synchronous rotation, \ce{N2} atmosphere) with different surface pressures (0.1 to 10~bars). 
They include scenarios simulated under dry conditions\footnote{In the dry scenarios, the correlated-k tables used for the radiative transfer computations incorporate a residual water content. Therefore, minor spectral features associated with water vapor absorption remain observable in the spectra.} (dashed lines) and with a water inventory of 8.3~cm~GEL (solid lines).

The case at 0.1~bar represents a collapsed climate state where all available water is condensed into surface ice. This is evidenced by the spectrum, which exhibits a broad shape without spectral features, with a geometric albedo converging to $2/3$ of $A_{\rm surf}$ at all wavelengths. 
This case approximates a bare rock scenario without atmosphere, which would also be plausible given the proximity of the planet to its host star \citep{Kreidberg2025}. Under such conditions, the spectrum would be flatter and with fewer features. 

The 1~bar and 10~bars cases present two main spectral features that can provide hints on the atmospheric water vapor content and the atmospheric density: (1)~in the near-infrared, both cases show pronounced and saturated water absorption bands, indicating the presence of water vapor with comparable mixing ratios ($\sim$83~kg\,m$^{-2}$), as described in the previous part and (2)~both cases exhibit a significant increase of the flux in the visible, with the geometric albedo reaching 0.2 for the 1~bar case and 0.45 for the 10~bars case (for both dry and water-rich spectra). 
This increase in reflected flux is due to Rayleigh scattering and indicates a concentration of atmospheric molecules consistent with higher atmospheric surface pressure and density. 

Increased backscattering by the atmosphere boosts the reflected flux, thereby potentially improving the detectability. These trends are also observed in Figure~\ref{fig:albedo_summary}, which shows that the geometric albedo of an \ce{N2} atmosphere within the RISTRETTO bandpass exhibits an increase from 1 to 10~bars for both dry and water-rich scenarios (reaching 0.2 for the dry case). In contrast, the differences between 1 and 10~bars are negligible for the ANDES spectral bands, where the spectrum is dominated by water absorption. This discrepancy occurs because the RISTRETTO spectral range presents a higher sensitivity to the albedo boost driven by the Rayleigh scattering, as this spectral region coincides with the characteristic Rayleigh scattering peak, seen between 0.4 and $\sim$0.9~$\mu$m in the spectra (Figure~\ref{fig:spectra}). 
Moreover, a quantitative assessment of the contribution of Rayleigh scattering to the observed reflected flux is constrained by parameter degeneracies, including the surface albedo, as well as dayside cloud and haze coverage (if the climate allows it). However, in these scenarios, our climate simulations indicate that the cloud formation mainly occurs in the nightside, resulting in a negligible impact on the planetary spectral reflectivity. Consequently, clouds can be excluded from the degeneracy parameters.

\subsubsection{Variation of atmospheric composition}

The bottom left panel of Figure~\ref{fig:spectra} shows the synthetic spectra for Cases~2, 4 and 6 (synchronous rotation, 1~bar of surface pressure) with varying atmospheric gas compositions (\ce{N2}, Earth-like and \ce{CO2}). They include dry scenarios (dashed lines) and scenarios with water inventory of 8.3~cm~GEL (solid lines). 

The spectra of simulations assuming \ce{376 ppm} of \ce{CO2} (dry and water vapor-rich) show additional absorption features corresponding to the \ce{CO2} absorption bands at 1.1, 1.2, 1.4 and 1.6~$\mu$m. These spectral signatures are more prominent in \ce{CO2}-dominated atmospheres, which nearly absorb all incident flux within the ANDES YJH spectral bands. Consequently, for equivalent water vapor content, the mean geometric albedo within the YJH bands decreases significantly for \ce{CO2}-rich atmospheres, reaching values as low as 0.05 in dry conditions and 0.01 in the water-rich scenario for the H band (see Figure~\ref{fig:albedo_summary}). The high absorption efficiency of \ce{CO2}-rich atmospheres poses challenges for detection. 

In contrast, for a fixed water vapor amount, the mean geometric albedo values are similar in the spectral range of RISTRETTO across all three atmospheric compositions (0.13 and 0.11 for dry and water-rich scenarios respectively, see Figure~\ref{fig:albedo_summary}). Thus, the RISTRETTO bandpass is not sensitive to decreases in albedo caused by \ce{CO2} absorption, unlike the sensitivity to water vapor variations. 

Moreover, the spectra corresponding to \ce{CO2}-dominated atmospheres (both dry and water-rich) show a slight increase in the geometric albedo, reaching up to 0.25 at 0.4~$\mu$m (slightly higher than \ce{N2}-dominated atmospheres, reaching 0.2 at the same wavelength, see Figure~\ref{fig:spectra}). 
This difference indicates a higher Rayleigh scattering efficiency in \ce{CO2}-rich atmospheres for a given surface pressure. 
However, this does not affect the geometric albedo in the RISTRETTO bandpass, which remains similar for the three atmospheric compositions. This spectral feature becomes more significative in atmospheres with higher surface pressures (e.g., 10~bars). As shown in the bottom right panel of Figure~\ref{fig:spectra}, which compares spectra of \ce{N2} and \ce{CO2} atmospheres at various surface pressures, the geometric albedo in a 10~bar-\ce{CO2} environment increases to 0.49 at 0.4~$\mu$m, surpassing the 0.45 albedo observed in the \ce{N2} atmosphere at the same wavelength. Moreover, Figure~\ref{fig:albedo_summary} indicates a slight increase in the average geometric albedo for RISTRETTO from 0.11 to 0.12 for a 10~bar-\ce{CO2}-dominated, water-rich atmosphere (compared to \ce{N2}), due to the more efficient Rayleigh scattering by \ce{CO2} in denser atmospheres. Thus, the sensitivity of RISTRETTO to \ce{CO2} Rayleigh scattering requires high-pressure conditions to detect a discernable flux increase. However, even at high surface pressures, water absorption reduces the detectability of atmospheric signatures by decreasing the mean geometric albedo in the RISTRETTO bandpass.

\subsection{Overview of the results}

Figure~\ref{fig:albedo_summary} shows the mean Bond and geometric albedo values for all the cases we investigated, highlighting the influence of total water inventory, surface pressure, atmospheric composition and planetary rotation on the planetary spectral reflectivity. All values are given in Table~\ref{tab:albedo}.
Figure~\ref{fig:albedo_summary} reveals clear correlations between the planetary climate regime and the Bond albedo. Specifically, the transition between different climate states (cold and runaway) shows two distinct trends associated with each regime: cold states maintain a nearly constant Bond albedo (as seen for \ce{N2} atmospheres), indicating that variations in water content, surface pressure or atmospheric composition do not significantly impact the albedo. Conversely, runaway states show a linear relationship between the Bond albedo and the water reservoir, with higher water vapor contents leading to lower albedo values. 

Our results highlight the strong dependency of the albedo to the complexity of climate systems and several key parameters. More specifically, we find that the mean geometric albedo remains consistently low across all scenarios investigated in this study. Figure~\ref{fig:albedo_summary} shows that it ranges from 0.07 to 0.2 within the RISTRETTO bandpass and from 0 to 0.14 within the ANDES YJH bands. Consequently, assuming a constant albedo for different atmospheric properties can lead to significant inaccuracies in the assessment of the observability of potential targets. Moreover, based on our geometric albedo estimates (see Figure~\ref{fig:albedo_summary} and Table~\ref{tab:albedo}), adopting an Earth-like albedo of 0.3 in both visible and infrared likely overestimates the actual reflectivity of hot rocky exoplanets orbiting closer than the inner edge of the HZ of M~dwarf stars. 
This overestimation may lead to unrealistic detectability predictions. 
Therefore, for this specific planet population, our results suggest that conservative geometric albedo estimates, such as the ones we produced, should be considered when evaluating the detection limits of these targets and considering such scenarios. 

In our simulations, clouds do not have a significant effect on the planetary reflectivity. They are mostly on the nightside \citep[as in][]{turbet2021,chaverot2023}, in contrast to previous studies showing the reflective effect of dayside cloud formation patterns for slightly less irradiated planets (e.g., \citealt{yang2013}). 

To conclude, in the RISTRETTO bandpass, the albedo is mainly determined by Rayleigh scattering with thicker atmospheres reflecting more light even for higher water contents (see Figure~\ref{fig:albedo_summary}, 1~bar \ce{CO2} and 10~bars \ce{CO2}). 
In the ANDES bandpasses, the water content is the main parameter affecting the planetary albedo, due to the absorptivity of atmospheric water vapor (see Figure~\ref{fig:albedo_summary}, 1~bar \ce{N2} and 1~bar \ce{CO2}).
Therefore, these results highlight the need of a synergy between visible (e.g., RISTRETTO) and near-infrared wavelengths (e.g., ANDES) or an instrument which will combine both (e.g., PCS), as each wavelength ranges present sensitivities to different physical processes, which can improve the interpretation of the upcoming observations.

\section{Discussion}
\label{sec:discussion}

This study of the potential climate regimes and reflected-light observability of Ross~128~b relies on several assumptions and involves uncertainties that cannot be definitively excluded. These factors may significantly influence (to varying degrees) the planetary climate and reflectivity. Such factors include the obliquity, eccentricity, spin-orbit resonance and rotation rate; the orbital inclination and planetary mass; the planetary composition, radius and gravity; the surface composition and topography; the atmospheric escape rate \citep[e.g.,][]{Yamashiki2019}; the molecular composition, scale height and scattering properties of a potential atmosphere; the distribution, composition and scattering properties of potential clouds or hazes.

The presence of atmospheric hazes, which are not included in our current simulations, could increase the planetary albedo and amplify the reflected flux signal, thereby potentially improving the detectability of close-in, rocky planets. Previous studies have incorporated hazes into their climate simulations to model exoplanets such as Trappist-1~b \citep{maurel2025} or Trappist-1~e (including photochemistry, e.g., \citealt{Wolf2025}). Other works have included photochemistry to simulate other photochemical processes on environments like early Mars (Kuzucan et al. in prep.) or Trappist-1~c \citep{Jaziri2026}. Because photochemical hazes can be critical to assess the planetary climate, laboratory experiments aiming at determining the refractive indices of photochemical hazes can be applied on exoplanets \citep{Drant2025}. Including these factors into our simulations represents a potential area for future improvements, and should be considered as a next step in this study. 

Previous studies have conducted exoplanet climate modeling by implementing in their models Venus-like atmosphere conditions \citep{Quirino2023}, characterized by thick, highly reflective acid sulphuric cloud and haze layers, and a dense \ce{CO2}-dominated atmosphere (around 90~bars, significantly higher to our 10~bars \ce{CO2} atmosphere case). Our results show that our dense \ce{CO2} atmosphere cases present very low reflectivity due to the high water vapor absorptivity, however, the presence of cloud and haze layers, which are located in higher altitude levels, are expected to increase the planetary reflectivity. 
We tried to perform simulations assuming Venus-like atmospheric conditions, but numerical instability issues (especially in the upper atmosphere) prevented the model from converging.
These studies can also be particularly relevant to define observing strategies for the detection and characterization of Venus-like exoplanets, defined as post-runaway greenhouse planets \citep{Kane2026}. 

Our findings suggest that hazeless planets with optically thin atmospheres (i.e., low water contents, radiatively inactive gases, low surface pressures) are likely to appear dark. However, these results are highly sensitive to the assumed surface albedo. In this work, we adopted a low-albedo value of 0.2, as justified in Section~\ref{sub:method:climate}. However, it is important to note that surface types with higher albedos, such as ultramafic (low-silica), feldspathic, granitoid or clay surfaces \citep{Hu2012jun}, could increase the spectral reflectivity of scenarios with thin atmospheres. Further observations and analyses of planetary surfaces are required to determine their likely compositions and conclude on this question.

Questions regarding water inventories on rocky exoplanets are still open. In this study, we focus on planets with relatively low amounts of water, because 
(1)~the presence of large water reservoirs might be limited by water loss occurring during the long pre-main-sequence of M dwarf stars, during which the high stellar energy radiation might drive water dissociation and hydrogen escape \citep{Luger2015, Bolmont2017} and
(2)~the climate of highly irradiated planets with large water reservoirs tend to reach the runaway greenhouse state \citep{leconte2013nat, kopparapu2017, chaverot2022, chaverot2023, turbet2023}.
It is the case of Ross~128~b which is located interior to the HZ inner limit \citep{turbet2023} as calculated by \citet{kopparapu2017}.
In this extreme climate state, all available water is vaporized in the atmosphere and the lower atmosphere becomes optically thick, preventing the thermal radiation from the surface and the lower atmospheric layers to escape into space.
In this scenario, we expect the planetary albedo to be extremely low due to the high absorptivity of the moist atmosphere.
This is consistent with previous studies using 1D cloud-free models \citep{Goldblatt2013} and 3D GCMs \citep{chaverot2023}.

An exhaustive investigation of the climate and atmospheres of close-in, rocky exoplanets is essential for developing a robust database of synthetic reflectance spectra, thereby facilitating the interpretation of future observations. This would require to expand the parameter space to cover a broader range of atmospheric compositions and orbital configurations beyond those considered in this study. Defining representative reference or boundary cases help define the limits of the relevant parameter space, thereby providing constraints that refine the science cases and improve the interpretation of upcoming data. While processes such as atmospheric hazes can increase the planetary albedo and improve the prospects for detection, it is important to consider the scenarios derived from this work, which suggest a high probability that these warm exoplanets will exhibit low albedo and therefore appear optically dark.

The synthetic spectra derived from the simulated scenarios in this study are generated at spectral resolutions of $R=300$ or $R=500$, depending on the correlated-k tables used for each gas mixture. These resolutions are significantly lower than the targeted performances of RISTRETTO, ANDES and PCS, which aim for spectral resolutions of $R\geq100\,000$. To optimize the preparation for high-resolution observations, we need to increase the spectral resolution of the synthetic spectra. 

A next step will also consist in computing phase curves over the orbital period, considering the instrumental and optical constraints. This can help further refine this study by including potential spatial heterogeneities in the atmosphere and cloud coverage. 
Dayside clouds or ice on the dayside surface could break the east/west symmetry. 
However, our results suggest an absence of ice and cloud coverage on the dayside.
Therefore, we do not expect significant variations between orbital phases at $+90^\circ$ and $-90^\circ$.

\section{Conclusions}
\label{sec:ccl}

In this study, we first investigated the potential climate regimes of Ross~128~b, prototype of small, rocky exoplanets located close or interior to the inner edge of the Habitable Zone of M~dwarf stars, for different orbital (synchronous and asynchronous rotations) and atmospheric setups, with various atmospheric gases (\ce{N2} and \ce{CO2}), mean atmospheric surface pressures, and water inventories. Our results show that the climate of Ross~128~b, and by extension similar close-in, rocky exoplanets, is strongly sensitive to these parameters. This variability can induce a broad range of mean surface temperatures, from $\sim$250 to 650~K, and can drive the planet into distinct climate regimes with different implications for the observability. For most of the scenarios explored, the planet can either reach a collapsed state, characterized by a water-depleted atmosphere and nightside surface ice deposits, or a runaway state, which is a warmer state with fully vaporized water in the atmosphere. These two climate states are characteristic of the moist bistability observed in close-in, land rocky planets, as described in previous studies \citep[e.g.,][]{leconte2013aa}. Furthermore, we show that even a planet largely located outside the Habitable Zone, as defined in \citet{kopparapu2017}, may still sustain localized liquid water within the surface cold traps corresponding to the Rossby-wave gyres for a specific water inventory. This highlights the complexity of planetary climate systems and suggests that climate responses are not simply a linear function of the stellar irradiation (i.e., the orbital distance). Therefore, producing accurate synthetic observables requires GCM simulations across a broad parameter space.

Then, we generated synthetic geometric albedo spectra across the RISTRETTO and ANDES instrument bandpasses for all simulated scenarios. This enables to evaluate the detection and atmospheric characterization prospects via the combined approach of high-contrast imaging and high-resolution spectroscopy. Specifically, our synthetic spectra can provide more accurate albedo estimates, which are useful to determine the instrument detection limits (like planet-to-star contrast ratios) for targets within this growing exoplanet population. Our results indicate that, in most cases, the geometric albedo remains relatively low, ranging from 0 to 0.2, which is below the commonly assumed value of 0.3 used to calculate the detection limits. Therefore, assuming the albedo of the Earth (in the visible, which is not covered by ANDES) may be overly inaccurate for the planets of this population. This suggests a need to revisit the albedo assumptions based on this work to optimize the detectability predictions in future studies. 

\begin{acknowledgements}
    This work was carried out within the framework of the NCCR PlanetS, supported by the Swiss National Science Foundation under grants 51NF40\_182901 and 51NF40\_205606.
    E.B. acknowledges the financial support of the SNSF (grant number: 200021\_197176 and 200020\_215760). 
    GC acknowledges the financial support of the SNSF (grant number: P500PT\_217840).
    This research has made use of the Astrophysics Data System, funded by NASA under Cooperative Agreement 80NSSC25M7105.
\end{acknowledgements}

\section*{Data availability}

The GCM outputs corresponding to each simulation setups are available here: \url{https://doi.org/10.5281/zenodo.19208230}. Additional data can be shared on request to the corresponding author. The Generic-PCM (including documentation) can be downloaded from the SVN repository \url{https://svn.lmd.jussieu.fr/Planeto/trunk/LMDZ.GENERIC/}. The Pytmosph3R documentation can be found in: \url{https://perso.astrophy.u-bordeaux.fr/~jleconte/pytmosph3r-doc/index.html}.

\bibliography{biblio.bib}

@ARTICLE{anglada_escude2016,
       author = {{Anglada-Escud{\'e}}, Guillem and {Amado}, Pedro J. and {Barnes}, John and {Berdi{\~n}as}, Zaira M. and {Butler}, R. Paul and {Coleman}, Gavin A.~L. and {de La Cueva}, Ignacio and {Dreizler}, Stefan and {Endl}, Michael and {Giesers}, Benjamin and {Jeffers}, Sandra V. and {Jenkins}, James S. and {Jones}, Hugh R.~A. and {Kiraga}, Marcin and {K{\"u}rster}, Martin and {L{\'o}pez-Gonz{\'a}lez}, Mar{\'\i}a J. and {Marvin}, Christopher J. and {Morales}, Nicol{\'a}s and {Morin}, Julien and {Nelson}, Richard P. and {Ortiz}, Jos{\'e} L. and {Ofir}, Aviv and {Paardekooper}, Sijme-Jan and {Reiners}, Ansgar and {Rodr{\'\i}guez}, Eloy and {Rodr{\'\i}guez-L{\'o}pez}, Cristina and {Sarmiento}, Luis F. and {Strachan}, John P. and {Tsapras}, Yiannis and {Tuomi}, Mikko and {Zechmeister}, Mathias},
        title = "{A terrestrial planet candidate in a temperate orbit around Proxima Centauri}",
      journal = {\nat},
         year = 2016,
        month = aug,
       volume = {536},
       number = {7617},
        pages = {437-440},
          doi = {10.1038/nature19106},
archivePrefix = {arXiv},
       eprint = {1609.03449},
 primaryClass = {astro-ph.EP},
       adsurl = {https://ui.adsabs.harvard.edu/abs/2016Natur.536..437A}
}

@ARTICLE{gillon2016,
       author = {{Gillon}, Micha{\"e}l and {Jehin}, Emmanu{\"e}l and {Lederer}, Susan M. and {Delrez}, Laetitia and {de Wit}, Julien and {Burdanov}, Artem and {Van Grootel}, Val{\'e}rie and {Burgasser}, Adam J. and {Triaud}, Amaury H.~M.~J. and {Opitom}, Cyrielle and {Demory}, Brice-Olivier and {Sahu}, Devendra K. and {Bardalez Gagliuffi}, Daniella and {Magain}, Pierre and {Queloz}, Didier},
        title = "{Temperate Earth-sized planets transiting a nearby ultracool dwarf star}",
      journal = {\nat},
         year = 2016,
        month = may,
       volume = {533},
       number = {7602},
        pages = {221-224},
          doi = {10.1038/nature17448},
archivePrefix = {arXiv},
       eprint = {1605.07211},
 primaryClass = {astro-ph.EP},
       adsurl = {https://ui.adsabs.harvard.edu/abs/2016Natur.533..221G}
}

@article{gillon2017,
  title={Seven temperate terrestrial planets around the nearby ultracool dwarf star TRAPPIST-1},
  author={Gillon, Micha{\"e}l and Triaud, Amaury HMJ and Demory, Brice-Olivier and Jehin, Emmanu{\"e}l and Agol, Eric and Deck, Katherine M and Lederer, Susan M and De Wit, Julien and Burdanov, Artem and Ingalls, James G and others},
  journal={Nature},
  volume={542},
  number={7642},
  pages={456--460},
  year={2017},
  publisher={Nature Publishing Group}
}

@ARTICLE{snellen2015,
       author = {{Snellen}, I. and {de Kok}, R. and {Birkby}, J.~L. and {Brandl}, B. and {Brogi}, M. and {Keller}, C. and {Kenworthy}, M. and {Schwarz}, H. and {Stuik}, R.},
        title = "{Combining high-dispersion spectroscopy with high contrast imaging: Probing rocky planets around our nearest neighbors}",
      journal = {\aap},
         year = 2015,
        month = apr,
       volume = {576},
          eid = {A59},
        pages = {A59},
          doi = {10.1051/0004-6361/201425018},
archivePrefix = {arXiv},
       eprint = {1503.01136},
 primaryClass = {astro-ph.EP},
       adsurl = {https://ui.adsabs.harvard.edu/abs/2015A&A...576A..59S}
}

@ARTICLE{greene2023,
       author = {{Greene}, Thomas P. and {Bell}, Taylor J. and {Ducrot}, Elsa and {Dyrek}, Achr{\`e}ne and {Lagage}, Pierre-Olivier and {Fortney}, Jonathan J.},
        title = "{Thermal emission from the Earth-sized exoplanet TRAPPIST-1 b using JWST}",
      journal = {\nat},
         year = 2023,
        month = jun,
       volume = {618},
       number = {7963},
        pages = {39-42},
          doi = {10.1038/s41586-023-05951-7},
archivePrefix = {arXiv},
       eprint = {2303.14849},
 primaryClass = {astro-ph.EP},
       adsurl = {https://ui.adsabs.harvard.edu/abs/2023Natur.618...39G}
}

@ARTICLE{zieba2023,
       author = {{Zieba}, Sebastian and {Kreidberg}, Laura and {Ducrot}, Elsa and {Gillon}, Micha{\"e}l and {Morley}, Caroline and {Schaefer}, Laura and {Tamburo}, Patrick and {Koll}, Daniel D.~B. and {Lyu}, Xintong and {Acu{\~n}a}, Lorena and {Agol}, Eric and {Iyer}, Aishwarya R. and {Hu}, Renyu and {Lincowski}, Andrew P. and {Meadows}, Victoria S. and {Selsis}, Franck and {Bolmont}, Emeline and {Mandell}, Avi M. and {Suissa}, Gabrielle},
        title = "{No thick carbon dioxide atmosphere on the rocky exoplanet TRAPPIST-1 c}",
      journal = {\nat},
         year = 2023,
        month = aug,
       volume = {620},
       number = {7975},
        pages = {746-749},
          doi = {10.1038/s41586-023-06232-z},
archivePrefix = {arXiv},
       eprint = {2306.10150},
 primaryClass = {astro-ph.EP},
       adsurl = {https://ui.adsabs.harvard.edu/abs/2023Natur.620..746Z}
}

@ARTICLE{lovis2017,
       author = {{Lovis}, C. and {Snellen}, I. and {Mouillet}, D. and {Pepe}, F. and {Wildi}, F. and {Astudillo-Defru}, N. and {Beuzit}, J. -L. and {Bonfils}, X. and {Cheetham}, A. and {Conod}, U. and {Delfosse}, X. and {Ehrenreich}, D. and {Figueira}, P. and {Forveille}, T. and {Martins}, J.~H.~C. and {Quanz}, S.~P. and {Santos}, N.~C. and {Schmid}, H. -M. and {S{\'e}gransan}, D. and {Udry}, S.},
        title = "{Atmospheric characterization of Proxima b by coupling the SPHERE high-contrast imager to the ESPRESSO spectrograph}",
      journal = {\aap},
         year = 2017,
        month = mar,
       volume = {599},
          eid = {A16},
        pages = {A16},
          doi = {10.1051/0004-6361/201629682},
archivePrefix = {arXiv},
       eprint = {1609.03082},
 primaryClass = {astro-ph.EP},
       adsurl = {https://ui.adsabs.harvard.edu/abs/2017A&A...599A..16L}
}

@ARTICLE{bonfils2018,
       author = {{Bonfils}, X. and {Astudillo-Defru}, N. and {D{\'\i}az}, R. and {Almenara}, J. -M. and {Forveille}, T. and {Bouchy}, F. and {Delfosse}, X. and {Lovis}, C. and {Mayor}, M. and {Murgas}, F. and {Pepe}, F. and {Santos}, N.~C. and {S{\'e}gransan}, D. and {Udry}, S. and {W{\"u}nsche}, A.},
        title = "{A temperate exo-Earth around a quiet M dwarf at 3.4 parsec}",
      journal = {\aap},
         year = 2018,
        month = may,
       volume = {613},
          eid = {A25},
        pages = {A25},
          doi = {10.1051/0004-6361/201731973},
archivePrefix = {arXiv},
       eprint = {1711.06177},
 primaryClass = {astro-ph.EP},
       adsurl = {https://ui.adsabs.harvard.edu/abs/2018A&A...613A..25B}
}

@ARTICLE{turbet2016,
       author = {{Turbet}, Martin and {Leconte}, J{\'e}r{\'e}my and {Selsis}, Franck and {Bolmont}, Emeline and {Forget}, Fran{\c{c}}ois and {Ribas}, Ignasi and {Raymond}, Sean N. and {Anglada-Escud{\'e}}, Guillem},
        title = "{The habitability of Proxima Centauri b. II. Possible climates and observability}",
      journal = {\aap},
         year = 2016,
        month = dec,
       volume = {596},
          eid = {A112},
        pages = {A112},
          doi = {10.1051/0004-6361/201629577},
archivePrefix = {arXiv},
       eprint = {1608.06827},
 primaryClass = {astro-ph.EP},
       adsurl = {https://ui.adsabs.harvard.edu/abs/2016A&A...596A.112T}
}

@INPROCEEDINGS{lovis2024,
       author = {{Lovis}, Christophe and {Blind}, Nicolas and {Chazelas}, Bruno and {Shinde}, Muskan and {Bugatti}, Maddalena and {Restori}, Nathana{\"e}l. and {Dinis}, Isaac and {Genolet}, Ludovic and {Hughes}, Ian and {Sordet}, Micha{\"e}l. and {Schnell}, Robin and {Rihs}, Samuel and {Crausaz}, Adrien and {Turbet}, Martin and {Billot}, Nicolas and {Fusco}, Thierry and {Neichel}, Beno{\^\i}t and {Sauvage}, Jean-Fran{\c{c}}ois and {Santos Diaz}, Pablo and {Houelle}, Mathilde and {Blackman}, Joshua and {Lanotte}, Audrey and {K{\"u}hn}, Jonas and {Hagelberg}, Janis and {Guyon}, Olivier and {Martinez}, Patrice and {Spang}, Alain and {Mordasini}, Christoph and {Ehrenreich}, David and {Demory}, Brice-Olivier and {Bolmont}, Emeline},
        title = "{RISTRETTO: reflected-light exoplanet spectroscopy at the diffraction limit of the VLT}",
    booktitle = {Ground-based and Airborne Instrumentation for Astronomy X},
         year = 2024,
       editor = {{Bryant}, Julia J. and {Motohara}, Kentaro and {Vernet}, Jo{\"e}l. R.~D.},
       series = {Society of Photo-Optical Instrumentation Engineers (SPIE) Conference Series},
       volume = {13096},
        month = jul,
          eid = {130961I},
        pages = {130961I},
          doi = {10.1117/12.3020142},
       adsurl = {https://ui.adsabs.harvard.edu/abs/2024SPIE13096E..1IL}
}

@ARTICLE{hourdin2006,
       author = {{Hourdin}, Fr{\'e}d{\'e}ric and {Musat}, Ionela and {Bony}, Sandrine and {Braconnot}, Pascale and {Codron}, Francis and {Dufresne}, Jean-Louis and {Fairhead}, Laurent and {Filiberti}, Marie-Ang{\`e}le and {Friedlingstein}, Pierre and {Grandpeix}, Jean-Yves and {Krinner}, Gerhard and {Levan}, Phu and {Li}, Zhao-Xin and {Lott}, Fran{\c{c}}ois},
        title = "{The LMDZ4 general circulation model: climate performance and sensitivity to parametrized physics with emphasis on tropical convection}",
      journal = {Climate Dynamics},
         year = 2006,
        month = dec,
       volume = {27},
       number = {7-8},
        pages = {787-813},
          doi = {10.1007/s00382-006-0158-0},
       adsurl = {https://ui.adsabs.harvard.edu/abs/2006ClDy...27..787H}
}

@ARTICLE{wordsworth2010,
       author = {{Wordsworth}, R.~D. and {Forget}, F. and {Selsis}, F. and {Madeleine}, J. -B. and {Millour}, E. and {Eymet}, V.},
        title = "{Is Gliese 581d habitable? Some constraints from radiative-convective climate modeling}",
      journal = {\aap},
         year = 2010,
        month = nov,
       volume = {522},
          eid = {A22},
        pages = {A22},
          doi = {10.1051/0004-6361/201015053},
archivePrefix = {arXiv},
       eprint = {1005.5098},
 primaryClass = {astro-ph.EP},
       adsurl = {https://ui.adsabs.harvard.edu/abs/2010A&A...522A..22W}
}

@ARTICLE{wordsworth2011,
       author = {{Wordsworth}, Robin D. and {Forget}, Fran{\c{c}}ois and {Selsis}, Franck and {Millour}, Ehouarn and {Charnay}, Benjamin and {Madeleine}, Jean-Baptiste},
        title = "{Gliese 581d is the First Discovered Terrestrial-mass Exoplanet in the Habitable Zone}",
      journal = {\apjl},
         year = 2011,
        month = jun,
       volume = {733},
       number = {2},
          eid = {L48},
        pages = {L48},
          doi = {10.1088/2041-8205/733/2/L48},
archivePrefix = {arXiv},
       eprint = {1105.1031},
 primaryClass = {astro-ph.EP},
       adsurl = {https://ui.adsabs.harvard.edu/abs/2011ApJ...733L..48W}
}

@ARTICLE{selsis2011,
       author = {{Selsis}, F. and {Wordsworth}, R.~D. and {Forget}, F.},
        title = "{Thermal phase curves of nontransiting terrestrial exoplanets. I. Characterizing atmospheres}",
      journal = {\aap},
         year = 2011,
        month = aug,
       volume = {532},
          eid = {A1},
        pages = {A1},
          doi = {10.1051/0004-6361/201116654},
archivePrefix = {arXiv},
       eprint = {1104.4763},
 primaryClass = {astro-ph.EP},
       adsurl = {https://ui.adsabs.harvard.edu/abs/2011A&A...532A...1S}
}

@ARTICLE{bolmont2016,
       author = {{Bolmont}, Emeline and {Libert}, Anne-Sophie and {Leconte}, Jeremy and {Selsis}, Franck},
        title = "{Habitability of planets on eccentric orbits: Limits of the mean flux approximation}",
      journal = {\aap},
         year = 2016,
        month = jun,
       volume = {591},
          eid = {A106},
        pages = {A106},
          doi = {10.1051/0004-6361/201628073},
archivePrefix = {arXiv},
       eprint = {1604.06091},
 primaryClass = {astro-ph.EP},
       adsurl = {https://ui.adsabs.harvard.edu/abs/2016A&A...591A.106B}
}

@ARTICLE{zeng2019,
       author = {{Zeng}, Li and {Jacobsen}, Stein B. and {Sasselov}, Dimitar D. and {Petaev}, Michail I. and {Vanderburg}, Andrew and {Lopez-Morales}, Mercedes and {Perez-Mercader}, Juan and {Mattsson}, Thomas R. and {Li}, Gongjie and {Heising}, Matthew Z. and {Bonomo}, Aldo S. and {Damasso}, Mario and {Berger}, Travis A. and {Cao}, Hao and {Levi}, Amit and {Wordsworth}, Robin D.},
        title = "{Growth model interpretation of planet size distribution}",
      journal = {Proceedings of the National Academy of Science},
         year = 2019,
        month = may,
       volume = {116},
       number = {20},
        pages = {9723-9728},
          doi = {10.1073/pnas.1812905116},
archivePrefix = {arXiv},
       eprint = {1906.04253},
 primaryClass = {astro-ph.EP},
       adsurl = {https://ui.adsabs.harvard.edu/abs/2019PNAS..116.9723Z}
}

@ARTICLE{leconte2013aa,
       author = {{Leconte}, J. and {Forget}, F. and {Charnay}, B. and {Wordsworth}, R. and {Selsis}, F. and {Millour}, E. and {Spiga}, A.},
        title = "{3D climate modeling of close-in land planets: Circulation patterns, climate moist bistability, and habitability}",
      journal = {\aap},
         year = 2013,
        month = jun,
       volume = {554},
          eid = {A69},
        pages = {A69},
          doi = {10.1051/0004-6361/201321042},
archivePrefix = {arXiv},
       eprint = {1303.7079},
 primaryClass = {astro-ph.EP},
       adsurl = {https://ui.adsabs.harvard.edu/abs/2013A&A...554A..69L}
}

@ARTICLE{turbet2021,
       author = {{Turbet}, Martin and {Bolmont}, Emeline and {Chaverot}, Guillaume and {Ehrenreich}, David and {Leconte}, J{\'e}r{\'e}my and {Marcq}, Emmanuel},
        title = "{Day-night cloud asymmetry prevents early oceans on Venus but not on Earth}",
      journal = {\nat},
         year = 2021,
        month = oct,
       volume = {598},
       number = {7880},
        pages = {276-280},
          doi = {10.1038/s41586-021-03873-w},
archivePrefix = {arXiv},
       eprint = {2110.08801},
 primaryClass = {astro-ph.EP},
       adsurl = {https://ui.adsabs.harvard.edu/abs/2021Natur.598..276T}
}

@ARTICLE{chaverot2023,
       author = {{Chaverot}, Guillaume and {Bolmont}, Emeline and {Turbet}, Martin},
        title = "{First exploration of the runaway greenhouse transition with a 3D General Circulation Model}",
      journal = {\aap},
         year = 2023,
        month = dec,
       volume = {680},
          eid = {A103},
        pages = {A103},
          doi = {10.1051/0004-6361/202346936},
archivePrefix = {arXiv},
       eprint = {2309.05449},
 primaryClass = {astro-ph.EP},
       adsurl = {https://ui.adsabs.harvard.edu/abs/2023A&A...680A.103C}
}

@INCOLLECTION{showman2013,
       author = {{Showman}, A.~P. and {Wordsworth}, R.~D. and {Merlis}, T.~M. and {Kaspi}, Y.},
        title = "{Atmospheric Circulation of Terrestrial Exoplanets}",
    booktitle = {Comparative Climatology of Terrestrial Planets},
         year = 2013,
       editor = {{Mackwell}, Stephen J. and {Simon-Miller}, Amy A. and {Harder}, Jerald W. and {Bullock}, Mark A.},
        pages = {277-327},
          doi = {10.2458/azu_uapress_9780816530595-ch012},
       adsurl = {https://ui.adsabs.harvard.edu/abs/2013cctp.book..277S}
}

@ARTICLE{matsuno1966,
       author = {{Matsuno}, Taroh},
        title = "{赤道近くでの準地衡風的運動赤道近くでの準地衡風的運動Quasi-Geostrophic Motions in the Equatorial Area}",
      journal = {Journal of the Meteorological Society of Japan},
         year = 1966,
        month = jan,
       volume = {44},
       number = {1},
        pages = {25-43},
          doi = {10.2151/jmsj1965.44.1_25},
       adsurl = {https://ui.adsabs.harvard.edu/abs/1966JMeSJ..44...25M}
}

@ARTICLE{gill1980,
       author = {{Gill}, A.~E.},
        title = "{Some simple solutions for heat-induced tropical circulation}",
      journal = {Quarterly Journal of the Royal Meteorological Society},
         year = 1980,
        month = jul,
       volume = {106},
       number = {449},
        pages = {447-462},
          doi = {10.1002/qj.49710644905},
       adsurl = {https://ui.adsabs.harvard.edu/abs/1980QJRMS.106..447G}
}

@ARTICLE{showman2011,
       author = {{Showman}, Adam P. and {Polvani}, Lorenzo M.},
        title = "{Equatorial Superrotation on Tidally Locked Exoplanets}",
      journal = {\apj},
         year = 2011,
        month = sep,
       volume = {738},
       number = {1},
          eid = {71},
        pages = {71},
          doi = {10.1088/0004-637X/738/1/71},
archivePrefix = {arXiv},
       eprint = {1103.3101},
 primaryClass = {astro-ph.EP},
       adsurl = {https://ui.adsabs.harvard.edu/abs/2011ApJ...738...71S}
}

@ARTICLE{valente2022,
       author = {{Valente}, Ema F.~S. and {Correia}, Alexandre C.~M.},
        title = "{Tidal excitation of the obliquity of Earth-like planets in the habitable zone of M-dwarf stars}",
      journal = {\aap},
         year = 2022,
        month = sep,
       volume = {665},
          eid = {A130},
        pages = {A130},
          doi = {10.1051/0004-6361/202244010},
archivePrefix = {arXiv},
       eprint = {2307.08770},
 primaryClass = {astro-ph.EP},
       adsurl = {https://ui.adsabs.harvard.edu/abs/2022A&A...665A.130V}
}

@ARTICLE{gillon2025,
       author = {{Gillon}, Micha{\"e}l and {Ducrot}, Elsa and {Bell}, Taylor J. and {Huang}, Ziyu and {Lincowski}, Andrew and {Lyu}, Xintong and {Maurel}, Alice and {Revol}, Alexandre and {Agol}, Eric and {Bolmont}, Emeline and {Dong}, Chuanfei and {Fauchez}, Thomas J. and {Koll}, Daniel D.~B. and {Leconte}, J{\'e}r{\'e}my and {Meadows}, Victoria S. and {Selsis}, Franck and {Turbet}, Martin and {Charnay}, Benjamin and {Delre}, Laetita and {Demory}, Brice-Olivier and {Householder}, Aaron and {Zieba}, Sebastian and {Berardo}, David and {Dyrek}, Achr{\`e}ne and {Edwards}, Billy and {de Wit}, Julien and {Greene}, Thomas P. and {Hu}, Renyu and {Iro}, Nicolas and {Kreidberg}, Laura and {Lagage}, Pierre-Olivier and {Lustig-Yaeger}, Jacob and {Iyer}, Aishwarya},
        title = "{First JWST thermal phase curves of temperate terrestrial exoplanets reveal no thick atmosphere around TRAPPIST-1 b and c}",
      journal = {Nature Astronomy},
         year = 2025,
        month = sep,
          eid = {arXiv:2509.02128},
        pages = {arXiv:2509.02128},
          doi = {10.48550/arXiv.2509.02128},
archivePrefix = {arXiv},
       eprint = {2509.02128},
 primaryClass = {astro-ph.EP},
       adsurl = {https://ui.adsabs.harvard.edu/abs/2025arXiv250902128G}
}

@ARTICLE{bugatti2025,
       author = {{Bugatti}, Maddalena and {Lovis}, Christophe and {Billot}, Nicolas and {Blind}, Nicolas and {Lavie}, Baptiste and {Turbet}, Martin and {Chazelas}, Bruno and {Pepe}, Francesco},
        title = "{Simulating RISTRETTO: Proxima b detectability in reflected light}",
      journal = {\aap},
         year = 2025,
        month = oct,
       volume = {702},
          eid = {A230},
        pages = {A230},
          doi = {10.1051/0004-6361/202556398},
archivePrefix = {arXiv},
       eprint = {2509.07644},
 primaryClass = {astro-ph.EP},
       adsurl = {https://ui.adsabs.harvard.edu/abs/2025A&A...702A.230B}
}

@INPROCEEDINGS{marconi2022,
       author = {{Marconi}, A. and {Abreu}, M. and {Adibekyan}, V. and {Alberti}, V. and {Albrecht}, S. and {Alcaniz}, J. and {Aliverti}, M. and {Allende Prieto}, C. and {Alvarado G{\'o}mez}, J.~D. and {Amado}, P.~J. and {Amate}, M. and {Andersen}, M.~I. and {Artigau}, E. and {Baker}, C. and {Baldini}, V. and {Balestra}, A. and {Barnes}, S.~A. and {Baron}, F. and {Barros}, S.~C.~C. and {Bauer}, S.~M. and {Beaulieu}, M. and {Bellido-Tirado}, O. and {Benneke}, B. and {Bensby}, T. and {Bergin}, E.~A. and {Biazzo}, K. and {Bik}, A. and {Birkby}, J.~L. and {Blind}, N. and {Boisse}, I. and {Bolmont}, E. and {Bonaglia}, M. and {Bonfils}, X. and {Borsa}, F. and {Brandeker}, A. and {Brandner}, W. and {Broeg}, C.~H. and {Brogi}, M. and {Brousseau}, D. and {Brucalassi}, A. and {Brynnel}, J. and {Buchhave}, L.~A. and {Buscher}, D.~F. and {Cabral}, A. and {Calderone}, G. and {Calvo-Ortega}, R. and {Canto Martins}, B.~L. and {Cantalloube}, F. and {Carbonaro}, L. and {Chauvin}, G. and {Chazelas}, B. and {Cheffot}, A.-L. and {Cheng}, Y.~S. and {Chiavassa}, A. and {Christensen}, L. and {Cirami}, R. and {Cook}, N.~J. and {Cooke}, R.~J. and {Coretti}, I. and {Covino}, S. and {Cowan}, N. and {Cresci}, G. and {Cristiani}, S. and {Cunha Parro}, V. and {Cupani}, G. and {D'Odorico}, V. and {de Castro Le{\~a}o}, I. and {De Cia}, A. and {De Medeiros}, J.~R. and {Debras}, F. and {Debus}, M. and {Demangeon}, O. and {Dessauges-Zavadsky}, M. and {Di Marcantonio}, P. and {Dionies}, F. and {Doyon}, R. and {Dunn}, J. and {Ehrenreich}, D. and {Faria}, J.~P. and {Feruglio}, C. and {Fisher}, M. and {Fontana}, A. and {Fumagalli}, M. and {Fusco}, T. and {Fynbo}, J. and {Gabella}, O. and {Gaessler}, W. and {Gallo}, E. and {Gao}, X. and {Genolet}, L. and {Genoni}, M. and {Giacobbe}, P. and {Giro}, E. and {Gon{\c{c}}alves}, R.~S. and {Gonzalez}, O. and {Gonz{\'a}lez Hern{\'a}ndez}, J.~I. and {Gracia T{\'e}mich}, F. and {Haehnelt}, M.~G. and {Haniff}, C. and {Hatzes}, A. and {Helled}, R. and {Hoeijmakers}, H.~J. and {Huke}, P. and {J{\"a}rvinen}, S. and {J{\"a}rvinen}, A. and {Kaminski}, A. and {Korn}, A. and {Kouach}, D. and {Kowzan}, G. and {Kreidberg}, L. and {Landoni}, M. and {Lanotte}, A. and {Lavail}, A. and {Li}, J. and {Liske}, J. and {Lovis}, C. and {Lucatello}, S. and {Lunney}, D. and {MacIntosh}, M. and {Madhusudhan}, N. and {Magrini}, L. and {Maiolino}, R. and {Malo}, L. and {Man}, A. and {Marquart}, T. and {Marques}, E.~L. and {Martins}, A.~M. and {Martins}, C.~J.~A.~P. and {Maslowski}, P. and {Mason}, C. and {Mason}, E. and {McCracken}, R.~A. and {Mergo}, P. and {Micela}, G. and {Mitchell}, T. and {Molli{\`e}re}, P. and {Monteiro}, M. and {Montgomery}, D. and {Mordasini}, C. and {Morin}, J. and {Mucciarelli}, A. and {Murphy}, M.~T. and {N'Diaye}, M. and {Neichel}, B. and {Niedzielski}, A.~T. and {Niemczura}, E. and {Nortmann}, L. and {Noterdaeme}, P. and {Nunes}, N. and {Oggioni}, L. and {Oliva}, E. and {{\"O}nel}, H. and {Origlia}, L. and {{\"O}stlin}, G. and {Palle}, E. and {Papaderos}, P. and {Pariani}, G. and {Pe{\~n}ate Castro}, J. and {Pepe}, F. and {Perreault Levasseur}, L. and {Petit}, P. and {Pino}, L. and {Piqueras}, J. and {Pollo}, A. and {Poppenhaeger}, K. and {Quirrenbach}, A. and {Rauscher}, E. and {Rebolo}, R. and {Redaelli}, E.~M.~A. and {Reffert}, S. and {Reid}, D.~T. and {Reiners}, A. and {Richter}, P. and {Riva}, M. and {Rivoire}, S. and {Rodr{\'\i}guez-L{\'o}pez}, C. and {Roederer}, I.~U. and {Romano}, D. and {Rousseau}, S. and {Rowe}, J. and {Salvadori}, S. and {Santos}, N. and {Santos Diaz}, P. and {Sanz-Forcada}, J. and {Sarajlic}, M. and {Sauvage}, J.-F. and {Sch{\"a}fer}, S. and {Schiavon}, R.~P. and {Schmidt}, T.~M. and {Selmi}, C. and {Sivanandam}, S. and {Sordet}, M. and {Sordo}, R. and {Sortino}, F. and {Sosnowska}, D. and {Sousa}, S.~G. and {Stempels}, E. and {Strassmeier}, K.~G. and {Su{\'a}rez Mascare{\~n}o}, A. and {Sulich}, A.},
        title = "{ANDES, the high resolution spectrograph for the ELT: science case, baseline design and path to construction}",
    booktitle = {Ground-based and Airborne Instrumentation for Astronomy IX},
         year = 2022,
       editor = {{Evans}, Christopher J. and {Bryant}, Julia J. and {Motohara}, Kentaro},
       series = {Society of Photo-Optical Instrumentation Engineers (SPIE) Conference Series},
       volume = {12184},
        month = aug,
          eid = {1218424},
        pages = {1218424},
          doi = {10.1117/12.2628689},
       adsurl = {https://ui.adsabs.harvard.edu/abs/2022SPIE12184E..24M}
}

@ARTICLE{palle2025,
       author = {{Palle}, Enric and {Biazzo}, Katia and {Bolmont}, Emeline and {Molli{\`e}re}, Paul and {Poppenhaeger}, Katja and {Birkby}, Jayne and {Brogi}, Matteo and {Chauvin}, Gael and {Chiavassa}, Andrea and {Hoeijmakers}, Jens and {Lellouch}, Emmanuel and {Lovis}, Christophe and {Maiolino}, Roberto and {Nortmann}, Lisa and {Parviainen}, Hannu and {Pino}, Lorenzo and {Turbet}, Martin and {Weder}, Jesse and {Albrecht}, Simon and {Antoniucci}, Simone and {Barros}, Susana C. and {Beaudoin}, Andre and {Benneke}, Bjorn and {Boisse}, Isabelle and {Bonomo}, Aldo S. and {Borsa}, Francesco and {Brandeker}, Alexis and {Brandner}, Wolfgang and {Buchhave}, Lars A. and {Cheffot}, Anne-Laure and {Deborde}, Robin and {Debras}, Florian and {Doyon}, Rene and {Di Marcantonio}, Paolo and {Giacobbe}, Paolo and {Gonz{\'a}lez Hern{\'a}ndez}, Jonay I. and {Helled}, Ravit and {Kreidberg}, Laura and {Machado}, Pedro and {Maldonado}, Jesus and {Marconi}, Alessandro and {Martins}, B.~L. Canto and {Miceli}, Adriano and {Mordasini}, Christoph and {N'Diaye}, Mamadou and {Niedzielski}, Andrzej and {Nisini}, Brunella and {Origlia}, Livia and {Peroux}, Celine and {Pietrow}, Alexander G.~M. and {Pinna}, Enrico and {Rauscher}, Emily and {Reffert}, Sabine and {Rodr{\'\i}guez-L{\'o}pez}, Cristina and {Rousselot}, Philippe and {Sanna}, Nicoletta and {Santos}, Nuno C. and {Simonnin}, Adrien and {Su{\'a}rez Mascare{\~n}o}, Alejandro and {Zanutta}, Alessio and {Zapatero-Osorio}, Maria Rosa and {Zechmeister}, Mathias},
        title = "{Ground-breaking exoplanet science with the ANDES spectrograph at the ELT}",
      journal = {Experimental Astronomy},
         year = 2025,
        month = jun,
       volume = {59},
       number = {3},
          eid = {29},
        pages = {29},
          doi = {10.1007/s10686-025-10000-4},
archivePrefix = {arXiv},
       eprint = {2311.17075},
 primaryClass = {astro-ph.IM},
       adsurl = {https://ui.adsabs.harvard.edu/abs/2025ExA....59...29P}
}

@ARTICLE{kasper2021,
       author = {{Kasper}, M. and {Cerpa Urra}, N. and {Pathak}, P. and {Bonse}, M. and {Nousiainen}, J. and {Engler}, B. and {Heritier}, C.~T. and {Kammerer}, J. and {Leveratto}, S. and {Rajani}, C. and {Bristow}, P. and {Le Louarn}, M. and {Madec}, P.-Y. and {Str{\"o}bele}, S. and {Verinaud}, C. and {Glauser}, A. and {Quanz}, S.~P. and {Helin}, T. and {Keller}, C. and {Snik}, F. and {Boccaletti}, A. and {Chauvin}, G. and {Mouillet}, D. and {Kulcs{\'a}r}, C. and {Raynaud}, H.-F.},
        title = "{PCS {\textemdash} A Roadmap for Exoearth Imaging with the ELT}",
      journal = {The Messenger},
         year = 2021,
        month = mar,
       volume = {182},
        pages = {38-43},
          doi = {10.18727/0722-6691/5221},
archivePrefix = {arXiv},
       eprint = {2103.11196},
 primaryClass = {astro-ph.IM},
       adsurl = {https://ui.adsabs.harvard.edu/abs/2021Msngr.182...38K}
}

@ARTICLE{astudillo_defru2017,
       author = {{Astudillo-Defru}, N. and {Forveille}, T. and {Bonfils}, X. and {S{\'e}gransan}, D. and {Bouchy}, F. and {Delfosse}, X. and {Lovis}, C. and {Mayor}, M. and {Murgas}, F. and {Pepe}, F. and {Santos}, N.~C. and {Udry}, S. and {W{\"u}nsche}, A.},
        title = "{The HARPS search for southern extra-solar planets. XLI. A dozen planets around the M dwarfs GJ 3138, GJ 3323, GJ 273, GJ 628, and GJ 3293}",
      journal = {\aap},
         year = 2017,
        month = jun,
       volume = {602},
          eid = {A88},
        pages = {A88},
          doi = {10.1051/0004-6361/201630153},
archivePrefix = {arXiv},
       eprint = {1703.05386},
 primaryClass = {astro-ph.EP},
       adsurl = {https://ui.adsabs.harvard.edu/abs/2017A&A...602A..88A}
}

@ARTICLE{wright2016,
       author = {{Wright}, D.~J. and {Wittenmyer}, R.~A. and {Tinney}, C.~G. and {Bentley}, J.~S. and {Zhao}, Jinglin},
        title = "{Three Planets Orbiting Wolf 1061}",
      journal = {\apjl},
         year = 2016,
        month = feb,
       volume = {817},
       number = {2},
          eid = {L20},
        pages = {L20},
          doi = {10.3847/2041-8205/817/2/L20},
archivePrefix = {arXiv},
       eprint = {1512.05154},
 primaryClass = {astro-ph.EP},
       adsurl = {https://ui.adsabs.harvard.edu/abs/2016ApJ...817L..20W}
}

@INPROCEEDINGS{lovis2022,
       author = {{Lovis}, Christophe and {Blind}, Nicolas and {Chazelas}, Bruno and {K{\"u}hn}, Jonas G. and {Genolet}, Ludovic and {Hughes}, Ian and {Sordet}, Micha{\"e}l. and {Schnell}, Robin and {Turbet}, Martin and {Fusco}, Thierry and {Sauvage}, Jean-Fran{\c{c}}ois and {Bugatti}, Maddalena and {Billot}, Nicolas and {Hagelberg}, Janis and {Hocini}, Eddy and {Guyon}, Olivier},
        title = "{RISTRETTO: high-resolution spectroscopy at the diffraction limit of the VLT}",
    booktitle = {Ground-based and Airborne Instrumentation for Astronomy IX},
         year = 2022,
       editor = {{Evans}, Christopher J. and {Bryant}, Julia J. and {Motohara}, Kentaro},
       series = {Society of Photo-Optical Instrumentation Engineers (SPIE) Conference Series},
       volume = {12184},
        month = aug,
          eid = {121841Q},
        pages = {121841Q},
          doi = {10.1117/12.2627923},
archivePrefix = {arXiv},
       eprint = {2208.14838},
 primaryClass = {astro-ph.EP},
       adsurl = {https://ui.adsabs.harvard.edu/abs/2022SPIE12184E..1QL}
}

@ARTICLE{ribas2017,
       author = {{Ribas}, Ignasi and {Gregg}, Michael D. and {Boyajian}, Tabetha S. and {Bolmont}, Emeline},
        title = "{The full spectral radiative properties of Proxima Centauri}",
      journal = {\aap},
         year = 2017,
        month = jul,
       volume = {603},
          eid = {A58},
        pages = {A58},
          doi = {10.1051/0004-6361/201730582},
archivePrefix = {arXiv},
       eprint = {1704.08449},
 primaryClass = {astro-ph.SR},
       adsurl = {https://ui.adsabs.harvard.edu/abs/2017A&A...603A..58R}
}

@article{dressing2015,
	author = {{Dressing}, Courtney D. and {Charbonneau}, David},
	journal = {\apj},
	month = Jul,
	pages = {45},
	title = {{The Occurrence of Potentially Habitable Planets Orbiting M Dwarfs Estimated from the Full Kepler Dataset and an Empirical Measurement of the Detection Sensitivity}},
	volume = {807},
	year = 2015}

@article{mignon2025,
	author = {{Mignon}, L. and {Delfosse}, X. and {Meunier}, N. and {Chaverot}, G. and {Burn}, R. and {Bonfils}, X. and {Bouchy}, F. and {Astudillo-Defru}, N. and {Lo Curto}, G. and {Gaisne}, G. and {Udry}, S. and {Forveille}, T. and {Segransan}, D. and {Lovis}, C. and {Santos}, N.~C. and {Mayor}, M.},
	journal = {\aap},
	month = aug,
	pages = {A146},
	title = {{Radial velocity homogeneous analysis of M dwarfs observed with HARPS: II. Detection limits and planetary occurrence statistics}},
	volume = {700},
	year = 2025}

@article{rackham_transit_2019,
	title = {The {Transit} {Light} {Source} {Effect} {II}: {The} {Impact} of {Stellar} {Heterogeneity} on {Transmission} {Spectra} of {Planets} {Orbiting} {Broadly} {Sun}-like {Stars}},
	volume = {157},
	issn = {1538-3881},
	shorttitle = {The {Transit} {Light} {Source} {Effect} {II}},
	url = {http://arxiv.org/abs/1812.06184},
	doi = {10.3847/1538-3881/aaf892},
	number = {3},
	urldate = {2021-06-15},
	journal = {The Astronomical Journal},
	author = {Rackham, Benjamin V. and Apai, Dániel and Giampapa, Mark S.},
	month = feb,
	year = {2019},
	note = {arXiv: 1812.06184},
	pages = {96},
}

@ARTICLE{suarez_mascareno2025,
       author = {{Su{\'a}rez Mascare{\~n}o}, Alejandro and {Artigau}, {\'E}tienne and {Mignon}, Lucile and {Delfosse}, Xavier and {Cook}, Neil J. and {Bouchy}, Fran{\c{c}}ois and {Doyon}, Ren{\'e} and {Gonz{\'a}lez Hern{\'a}ndez}, Jonay I. and {Vandal}, Thomas and {de Castro Le{\~a}o}, Izan and {Stefanov}, Atanas K. and {Faria}, Jo{\~a}o and {Cadieux}, Charles and {Lamontagne}, Pierrot and {Baron}, Fr{\'e}d{\'e}rique and {Barros}, Susana C.~C. and {Benneke}, Bj{\"o}rn and {Bonfils}, Xavier and {Bryan}, Marta and {Canto Martins}, Bruno L. and {Cloutier}, Ryan and {Cowan}, Nicolas B. and {de Freitas}, Daniel Brito and {De Medeiros}, Jose Renan and {Delgado-Mena}, Elisa and {Figueira}, Pedro and {Dumusque}, Xavier and {Ehrenreich}, David and {Lafreni{\`e}re}, David and {Lovis}, Christophe and {Malo}, Lison and {Melo}, Claudio and {Mordasini}, Christoph and {Pepe}, Francesco and {Rebolo}, Rafael and {Rowe}, Jason and {Santos}, Nuno C. and {S{\'e}gransan}, Damien and {Udry}, St{\'e}phane and {Valencia}, Diana and {Wade}, Gregg and {Abreu}, Manuel and {Aguiar}, Jos{\'e} L.~A. and {Al Moulla}, Khaled and {Allain}, Guillaume and {Allart}, Romain and {Arial}, Tomy and {Auger}, Hugues and {Bazinet}, Luc and {Blind}, Nicolas and {Bohlender}, David and {Boisse}, Isabelle and {Boucher}, Anne and {Bourrier}, Vincent and {Bovay}, S{\'e}bastien and {Broeg}, Christopher and {Brousseau}, Denis and {Cabral}, Alexandre and {Carmona}, Andres and {Carteret}, Yann and {Challita}, Zalpha and {Chazelas}, Bruno and {Coelho}, Jo{\~a}o and {Cointepas}, Marion and {Conod}, Uriel and {Cristo}, Eduardo and {Costa Silva}, Ana Rita and {Darveau-Bernier}, Antoine and {Dauplaise}, Laurie and {Delisle}, Jean-Baptiste and {de Lima Gomes}, Roseane and {Forveille}, Thierry and {Frensch}, Yolanda G.~C. and {Gracia T{\'e}mich}, F{\'e}lix and {Fontinele}, Dasaev O. and {Gagn{\'e}}, Jonathan and {Genest}, Fr{\'e}d{\'e}ric and {Genolet}, Ludovic and {Gomes da Silva}, Jo{\~a}o and {Grieves}, Nolan and {Hernandez}, Olivier and {Hobson}, Melissa J. and {Hoeijmakers}, H. Jens and {Hubin}, Norbert and {Jahandar}, Farbod and {Jayawardhana}, Ray and {K{\"a}ufl}, Hans-Ulrich and {Kerley}, Dan and {Kolb}, Johann and {Krishnamurthy}, Vigneshwaran and {Kung}, Benjamin and {L'Heureux}, Alexandrine and {Larue}, Pierre and {Leath}, Henry and {Lim}, Olivia and {Lo Curto}, Gaspare and {Martins}, Allan M. and {Matthews}, Jaymie and {Mayer}, Jean-S{\'e}bastien and {Messias}, Yuri S. and {Metchev}, Stan and {Moranta}, Leslie and {Mounzer}, Dany and {Nari}, Nicola and {Nielsen}, Louise D. and {Osborn}, Ares and {Ouellet}, Mathieu and {Otegi}, Jon and {Parc}, L{\'e}na and {Pasquini}, Luca and {Passegger}, Vera M. and {Pelletier}, Stefan and {Peroux}, C{\'e}line and {Piaulet-Ghorayeb}, Caroline and {Plotnykov}, Mykhaylo and {Pompei}, Emanuela and {Poulin-Girard}, Anne-Sophie and {Rasilla}, Jos{\'e} Luis and {Reshetov}, Vladimir and {Saint-Antoine}, Jonathan and {Sarajlic}, Mirsad and {Saviane}, Ivo and {Schnell}, Robin and {Segovia}, Alex and {Seidel}, Julia and {Silber}, Armin and {Sinclair}, Peter and {Sordet}, Michael and {Sosnowska}, Danuta and {Srivastava}, Avidaan and {Teixeira}, M{\'a}rcio A. and {Thibault}, Simon and {Vall{\'e}e}, Philippe and {Vaulato}, Valentina and {Wardenier}, Joost P. and {Wehbe}, Bachar and {Weisserman}, Drew and {Wevers}, Ivan and {Wildi}, Fran{\c{c}}ois and {Yariv}, Vincent and {Zins}, G{\'e}rard},
        title = "{Diving into the planetary system of Proxima with NIRPS: Breaking the metre per second barrier in the infrared}",
      journal = {\aap},
         year = 2025,
        month = jul,
       volume = {700},
          eid = {A11},
        pages = {A11},
          doi = {10.1051/0004-6361/202553728},
archivePrefix = {arXiv},
       eprint = {2507.21751},
 primaryClass = {astro-ph.EP},
       adsurl = {https://ui.adsabs.harvard.edu/abs/2025A&A...700A..11S}
}

@ARTICLE{mann2015,
       author = {{Mann}, Andrew W. and {Feiden}, Gregory A. and {Gaidos}, Eric and {Boyajian}, Tabetha and {von Braun}, Kaspar},
        title = "{How to Constrain Your M Dwarf: Measuring Effective Temperature, Bolometric Luminosity, Mass, and Radius}",
      journal = {\apj},
         year = 2015,
        month = may,
       volume = {804},
       number = {1},
          eid = {64},
        pages = {64},
          doi = {10.1088/0004-637X/804/1/64},
archivePrefix = {arXiv},
       eprint = {1501.01635},
 primaryClass = {astro-ph.SR},
       adsurl = {https://ui.adsabs.harvard.edu/abs/2015ApJ...804...64M}
}

@ARTICLE{kasting1993,
       author = {{Kasting}, James F. and {Whitmire}, Daniel P. and {Reynolds}, Ray T.},
        title = "{Habitable Zones around Main Sequence Stars}",
      journal = {\icarus},
         year = 1993,
        month = jan,
       volume = {101},
       number = {1},
        pages = {108-128},
          doi = {10.1006/icar.1993.1010},
       adsurl = {https://ui.adsabs.harvard.edu/abs/1993Icar..101..108K}
}

@ARTICLE{rajpurohit2013,
       author = {{Rajpurohit}, A.~S. and {Reyl{\'e}}, C. and {Allard}, F. and {Homeier}, D. and {Schultheis}, M. and {Bessell}, M.~S. and {Robin}, A.~C.},
        title = "{The effective temperature scale of M dwarfs}",
      journal = {\aap},
         year = 2013,
        month = aug,
       volume = {556},
          eid = {A15},
        pages = {A15},
          doi = {10.1051/0004-6361/201321346},
archivePrefix = {arXiv},
       eprint = {1304.4072},
 primaryClass = {astro-ph.SR},
       adsurl = {https://ui.adsabs.harvard.edu/abs/2013A&A...556A..15R}
}

@ARTICLE{batalha2013,
       author = {{Batalha}, Natalie M. and {Rowe}, Jason F. and {Bryson}, Stephen T. and {Barclay}, Thomas and {Burke}, Christopher J. and {Caldwell}, Douglas A. and {Christiansen}, Jessie L. and {Mullally}, Fergal and {Thompson}, Susan E. and {Brown}, Timothy M. and {Dupree}, Andrea K. and {Fabrycky}, Daniel C. and {Ford}, Eric B. and {Fortney}, Jonathan J. and {Gilliland}, Ronald L. and {Isaacson}, Howard and {Latham}, David W. and {Marcy}, Geoffrey W. and {Quinn}, Samuel N. and {Ragozzine}, Darin and {Shporer}, Avi and {Borucki}, William J. and {Ciardi}, David R. and {Gautier}, III, Thomas N. and {Haas}, Michael R. and {Jenkins}, Jon M. and {Koch}, David G. and {Lissauer}, Jack J. and {Rapin}, William and {Basri}, Gibor S. and {Boss}, Alan P. and {Buchhave}, Lars A. and {Carter}, Joshua A. and {Charbonneau}, David and {Christensen-Dalsgaard}, Joergen and {Clarke}, Bruce D. and {Cochran}, William D. and {Demory}, Brice-Olivier and {Desert}, Jean-Michel and {Devore}, Edna and {Doyle}, Laurance R. and {Esquerdo}, Gilbert A. and {Everett}, Mark and {Fressin}, Francois and {Geary}, John C. and {Girouard}, Forrest R. and {Gould}, Alan and {Hall}, Jennifer R. and {Holman}, Matthew J. and {Howard}, Andrew W. and {Howell}, Steve B. and {Ibrahim}, Khadeejah A. and {Kinemuchi}, Karen and {Kjeldsen}, Hans and {Klaus}, Todd C. and {Li}, Jie and {Lucas}, Philip W. and {Meibom}, S{\o}ren and {Morris}, Robert L. and {Pr{\v{s}}a}, Andrej and {Quintana}, Elisa and {Sanderfer}, Dwight T. and {Sasselov}, Dimitar and {Seader}, Shawn E. and {Smith}, Jeffrey C. and {Steffen}, Jason H. and {Still}, Martin and {Stumpe}, Martin C. and {Tarter}, Jill C. and {Tenenbaum}, Peter and {Torres}, Guillermo and {Twicken}, Joseph D. and {Uddin}, Kamal and {Van Cleve}, Jeffrey and {Walkowicz}, Lucianne and {Welsh}, William F.},
        title = "{Planetary Candidates Observed by Kepler. III. Analysis of the First 16 Months of Data}",
      journal = {\apjs},
         year = 2013,
        month = feb,
       volume = {204},
       number = {2},
          eid = {24},
        pages = {24},
          doi = {10.1088/0067-0049/204/2/24},
archivePrefix = {arXiv},
       eprint = {1202.5852},
 primaryClass = {astro-ph.EP},
       adsurl = {https://ui.adsabs.harvard.edu/abs/2013ApJS..204...24B}
}

@ARTICLE{petigura2013,
       author = {{Petigura}, Erik A. and {Howard}, Andrew W. and {Marcy}, Geoffrey W.},
        title = "{Prevalence of Earth-size planets orbiting Sun-like stars}",
      journal = {Proceedings of the National Academy of Science},
         year = 2013,
        month = nov,
       volume = {110},
       number = {48},
        pages = {19273-19278},
          doi = {10.1073/pnas.1319909110},
archivePrefix = {arXiv},
       eprint = {1311.6806},
 primaryClass = {astro-ph.EP},
       adsurl = {https://ui.adsabs.harvard.edu/abs/2013PNAS..11019273P}
}

@ARTICLE{fulton2018,
       author = {{Fulton}, Benjamin J. and {Petigura}, Erik A.},
        title = "{The California-Kepler Survey. VII. Precise Planet Radii Leveraging Gaia DR2 Reveal the Stellar Mass Dependence of the Planet Radius Gap}",
      journal = {\aj},
         year = 2018,
        month = dec,
       volume = {156},
       number = {6},
          eid = {264},
        pages = {264},
          doi = {10.3847/1538-3881/aae828},
archivePrefix = {arXiv},
       eprint = {1805.01453},
 primaryClass = {astro-ph.EP},
       adsurl = {https://ui.adsabs.harvard.edu/abs/2018AJ....156..264F}
}

@ARTICLE{he2019,
       author = {{He}, Matthias Y. and {Ford}, Eric B. and {Ragozzine}, Darin},
        title = "{Architectures of exoplanetary systems - I. A clustered forward model for exoplanetary systems around Kepler's FGK stars}",
      journal = {\mnras},
         year = 2019,
        month = dec,
       volume = {490},
       number = {4},
        pages = {4575-4605},
          doi = {10.1093/mnras/stz2869},
archivePrefix = {arXiv},
       eprint = {1907.07773},
 primaryClass = {astro-ph.EP},
       adsurl = {https://ui.adsabs.harvard.edu/abs/2019MNRAS.490.4575H}
}

@ARTICLE{espinoza2025,
       author = {{Espinoza}, N{\'e}stor and {Allen}, Natalie H. and {Glidden}, Ana and {Lewis}, Nikole K. and {Seager}, Sara and {Ca{\~n}as}, Caleb I. and {Grant}, David and {Gressier}, Am{\'e}lie and {Courreges}, Shelby and {Stevenson}, Kevin B. and {Ranjan}, Sukrit and {Col{\'o}n}, Knicole and {Morris}, Brett M. and {MacDonald}, Ryan J. and {Long}, Douglas and {Wakeford}, Hannah R. and {Valenti}, Jeff A. and {Alderson}, Lili and {Batalha}, Natasha E. and {Challener}, Ryan C. and {Huang}, Jingcheng and {Lin}, Zifan and {Louie}, Dana R. and {Mullens}, Elijah and {Valentine}, Daniel and {Mountain}, C. Matt and {Pueyo}, Laurent and {Perrin}, Marshall D. and {Bellini}, Andrea and {Kammerer}, Jens and {Libralato}, Mattia and {Rebollido}, Isabel and {Rickman}, Emily and {Sohn}, Sangmo Tony and {van der Marel}, Roeland P.},
        title = "{JWST-TST DREAMS: NIRSpec/PRISM Transmission Spectroscopy of the Habitable Zone Planet TRAPPIST-1 e}",
      journal = {\apjl},
         year = 2025,
        month = sep,
       volume = {990},
       number = {2},
          eid = {L52},
        pages = {L52},
          doi = {10.3847/2041-8213/adf42e},
archivePrefix = {arXiv},
       eprint = {2509.05414},
 primaryClass = {astro-ph.EP},
       adsurl = {https://ui.adsabs.harvard.edu/abs/2025ApJ...990L..52E}
}

@ARTICLE{glidden2025,
       author = {{Glidden}, Ana and {Ranjan}, Sukrit and {Seager}, Sara and {Espinoza}, N{\'e}stor and {MacDonald}, Ryan J. and {Allen}, Natalie H. and {Ca{\~n}as}, Caleb I. and {Grant}, David and {Gressier}, Am{\'e}lie and {Stevenson}, Kevin B. and {Batalha}, Natasha E. and {Lewis}, Nikole K. and {Long}, Douglas and {Wakeford}, Hannah R. and {Alderson}, Lili and {Challener}, Ryan C. and {Col{\'o}n}, Knicole and {Huang}, Jingcheng and {Lin}, Zifan and {Louie}, Dana R. and {Mullens}, Elijah and {Sotzen}, Kristin S. and {Valenti}, Jeff A. and {Valentine}, Daniel and {Clampin}, Mark and {Mountain}, C. Matt and {Perrin}, Marshall and {van der Marel}, Roeland P.},
        title = "{JWST-TST DREAMS: Secondary Atmosphere Constraints for the Habitable Zone Planet TRAPPIST-1 e}",
      journal = {\apjl},
         year = 2025,
        month = sep,
       volume = {990},
       number = {2},
          eid = {L53},
        pages = {L53},
          doi = {10.3847/2041-8213/adf62e},
archivePrefix = {arXiv},
       eprint = {2509.05407},
 primaryClass = {astro-ph.EP},
       adsurl = {https://ui.adsabs.harvard.edu/abs/2025ApJ...990L..53G}
}

@ARTICLE{piaulet_Ghorayeb2025,
       author = {{Piaulet-Ghorayeb}, Caroline and {Benneke}, Bj{\"o}rn and {Turbet}, Martin and {Moore}, Keavin and {Roy}, Pierre-Alexis and {Lim}, Olivia and {Doyon}, Ren{\'e} and {Fauchez}, Thomas J. and {Albert}, Lo{\"\i}c and {Radica}, Michael and {Coulombe}, Louis-Philippe and {Lafreni{\`e}re}, David and {Cowan}, Nicolas B. and {Belzile}, Danika and {Musfirat}, Kamrul and {Kaur}, Mehramat and {L'Heureux}, Alexandrine and {Johnstone}, Doug and {MacDonald}, Ryan J. and {Allart}, Romain and {Dang}, Lisa and {Kaltenegger}, Lisa and {Pelletier}, Stefan and {Rowe}, Jason F. and {Taylor}, Jake and {Turner}, Jake D.},
        title = "{Strict Limits on Potential Secondary Atmospheres on the Temperate Rocky Exo-Earth TRAPPIST-1 d}",
      journal = {\apj},
         year = 2025,
        month = aug,
       volume = {989},
       number = {2},
          eid = {181},
        pages = {181},
          doi = {10.3847/1538-4357/adf207},
archivePrefix = {arXiv},
       eprint = {2508.08416},
 primaryClass = {astro-ph.EP},
       adsurl = {https://ui.adsabs.harvard.edu/abs/2025ApJ...989..181P}
}

@ARTICLE{fauchez2025,
       author = {{Fauchez}, Thomas J. and {Ducrot}, Elsa and {Rackham}, Benjamin V. and {Stevenson}, Kevin B. and {Mayorga}, L.~C. and {de Wit}, Julien},
        title = "{Stellar Models Also Limit Exoplanet Atmosphere Studies in Emission}",
      journal = {\apj},
         year = 2025,
        month = aug,
       volume = {989},
       number = {2},
          eid = {170},
        pages = {170},
          doi = {10.3847/1538-4357/adf068},
archivePrefix = {arXiv},
       eprint = {2502.19585},
 primaryClass = {astro-ph.EP},
       adsurl = {https://ui.adsabs.harvard.edu/abs/2025ApJ...989..170F}
}

@ARTICLE{del_genio2019,
       author = {{Del Genio}, Anthony D. and {Way}, Michael J. and {Amundsen}, David S. and {Aleinov}, Igor and {Kelley}, Maxwell and {Kiang}, Nancy Y. and {Clune}, Thomas L.},
        title = "{Habitable Climate Scenarios for Proxima Centauri b with a Dynamic Ocean}",
      journal = {Astrobiology},
         year = 2019,
        month = jan,
       volume = {19},
       number = {1},
        pages = {99-125},
          doi = {10.1089/ast.2017.1760},
       adsurl = {https://ui.adsabs.harvard.edu/abs/2019AsBio..19...99D}
}

@ARTICLE{turbet2018,
       author = {{Turbet}, Martin and {Bolmont}, Emeline and {Leconte}, Jeremy and {Forget}, Fran{\c{c}}ois and {Selsis}, Franck and {Tobie}, Gabriel and {Caldas}, Anthony and {Naar}, Joseph and {Gillon}, Micha{\"e}l},
        title = "{Modeling climate diversity, tidal dynamics and the fate of volatiles on TRAPPIST-1 planets}",
      journal = {\aap},
         year = 2018,
        month = may,
       volume = {612},
          eid = {A86},
        pages = {A86},
          doi = {10.1051/0004-6361/201731620},
archivePrefix = {arXiv},
       eprint = {1707.06927},
 primaryClass = {astro-ph.EP},
       adsurl = {https://ui.adsabs.harvard.edu/abs/2018A&A...612A..86T}
}

@ARTICLE{souto2018,
       author = {{Souto}, Diogo and {Unterborn}, Cayman T. and {Smith}, Verne V. and {Cunha}, Katia and {Teske}, Johanna and {Covey}, Kevin and {Rojas-Ayala}, B{\'a}rbara and {Garc{\'\i}a-Hern{\'a}ndez}, D.~A. and {Stassun}, Keivan and {Zamora}, Olga and {Masseron}, Thomas and {Johnson}, J.~A. and {Majewski}, Steven R. and {J{\"o}nsson}, Henrik and {Gilhool}, Steven and {Blake}, Cullen and {Santana}, Felipe},
        title = "{Stellar and Planetary Characterization of the Ross 128 Exoplanetary System from APOGEE Spectra}",
      journal = {\apjl},
         year = 2018,
        month = jun,
       volume = {860},
       number = {1},
          eid = {L15},
        pages = {L15},
          doi = {10.3847/2041-8213/aac896},
archivePrefix = {arXiv},
       eprint = {1805.11633},
 primaryClass = {astro-ph.SR},
       adsurl = {https://ui.adsabs.harvard.edu/abs/2018ApJ...860L..15S}
}

@ARTICLE{newton2016,
       author = {{Newton}, Elisabeth R. and {Irwin}, Jonathan and {Charbonneau}, David and {Berta-Thompson}, Zachory K. and {Dittmann}, Jason A. and {West}, Andrew A.},
        title = "{The Rotation and Galactic Kinematics of Mid M Dwarfs in the Solar Neighborhood}",
      journal = {\apj},
         year = 2016,
        month = apr,
       volume = {821},
       number = {2},
          eid = {93},
        pages = {93},
          doi = {10.3847/0004-637X/821/2/93},
archivePrefix = {arXiv},
       eprint = {1511.00957},
 primaryClass = {astro-ph.SR},
       adsurl = {https://ui.adsabs.harvard.edu/abs/2016ApJ...821...93N}
}

@ARTICLE{fauchez2019,
       author = {{Fauchez}, Thomas J. and {Turbet}, Martin and {Villanueva}, Geronimo L. and {Wolf}, Eric T. and {Arney}, Giada and {Kopparapu}, Ravi K. and {Lincowski}, Andrew and {Mandell}, Avi and {de Wit}, Julien and {Pidhorodetska}, Daria and {Domagal-Goldman}, Shawn D. and {Stevenson}, Kevin B.},
        title = "{Impact of Clouds and Hazes on the Simulated JWST Transmission Spectra of Habitable Zone Planets in the TRAPPIST-1 System}",
      journal = {\apj},
         year = 2019,
        month = dec,
       volume = {887},
       number = {2},
          eid = {194},
        pages = {194},
          doi = {10.3847/1538-4357/ab5862},
archivePrefix = {arXiv},
       eprint = {1911.08596},
 primaryClass = {astro-ph.EP},
       adsurl = {https://ui.adsabs.harvard.edu/abs/2019ApJ...887..194F}
}

@ARTICLE{bower2025,
       author = {{Bower}, Dan J. and {Thompson}, Maggie A. and {Hakim}, Kaustubh and {Tian}, Meng and {Sossi}, Paolo A.},
        title = "{Diversity of Low-mass Planet Atmospheres in the C─H─O─N─S─Cl System with Interior Dissolution, Nonideality, and Condensation: Application to TRAPPIST-1e and Sub-Neptunes}",
      journal = {\apj},
         year = 2025,
        month = dec,
       volume = {995},
       number = {1},
          eid = {59},
        pages = {59},
          doi = {10.3847/1538-4357/ae1479},
archivePrefix = {arXiv},
       eprint = {2507.00499},
 primaryClass = {astro-ph.EP},
       adsurl = {https://ui.adsabs.harvard.edu/abs/2025ApJ...995...59B}
}

@ARTICLE{chaverot2022,
       author = {{Chaverot}, Guillaume and {Turbet}, Martin and {Bolmont}, Emeline and {Leconte}, J{\'e}r{\'e}my},
        title = "{How does the background atmosphere affect the onset of the runaway greenhouse?}",
      journal = {\aap},
         year = 2022,
        month = feb,
       volume = {658},
          eid = {A40},
        pages = {A40},
          doi = {10.1051/0004-6361/202142286},
archivePrefix = {arXiv},
       eprint = {2111.07662},
 primaryClass = {astro-ph.EP},
       adsurl = {https://ui.adsabs.harvard.edu/abs/2022A&A...658A..40C}
}

@ARTICLE{leconte2013nat,
       author = {{Leconte}, J{\'e}r{\'e}my and {Forget}, Francois and {Charnay}, Benjamin and {Wordsworth}, Robin and {Pottier}, Aliz{\'e}e},
        title = "{Increased insolation threshold for runaway greenhouse processes on Earth-like planets}",
      journal = {\nat},
         year = 2013,
        month = dec,
       volume = {504},
       number = {7479},
        pages = {268-271},
          doi = {10.1038/nature12827},
archivePrefix = {arXiv},
       eprint = {1312.3337},
 primaryClass = {astro-ph.EP},
       adsurl = {https://ui.adsabs.harvard.edu/abs/2013Natur.504..268L}
}

@ARTICLE{turbet2019,
       author = {{Turbet}, Martin and {Tran}, Ha and {Pirali}, Olivier and {Forget}, Fran{\c{c}}ois and {Boulet}, Christian and {Hartmann}, Jean-Michel},
        title = "{Far infrared measurements of absorptions by CH$_{4}$ + CO$_{2}$ and H$_{2}$ + CO$_{2}$ mixtures and implications for greenhouse warming on early Mars}",
      journal = {\icarus},
         year = 2019,
        month = mar,
       volume = {321},
        pages = {189-199},
          doi = {10.1016/j.icarus.2018.11.021},
archivePrefix = {arXiv},
       eprint = {1805.02595},
 primaryClass = {astro-ph.EP},
       adsurl = {https://ui.adsabs.harvard.edu/abs/2019Icar..321..189T}
}

@ARTICLE{karman2019,
       author = {{Karman}, Tijs and {Gordon}, Iouli E. and {van der Avoird}, Ad and {Baranov}, Yury I. and {Boulet}, Christian and {Drouin}, Brian J. and {Groenenboom}, Gerrit C. and {Gustafsson}, Magnus and {Hartmann}, Jean-Michel and {Kurucz}, Robert L. and {Rothman}, Laurence S. and {Sun}, Kang and {Sung}, Keeyoon and {Thalman}, Ryan and {Tran}, Ha and {Wishnow}, Edward H. and {Wordsworth}, Robin and {Vigasin}, Andrey A. and {Volkamer}, Rainer and {van der Zande}, Wim J.},
        title = "{Update of the HITRAN collision-induced absorption section}",
      journal = {\icarus},
         year = 2019,
        month = aug,
       volume = {328},
        pages = {160-175},
          doi = {10.1016/j.icarus.2019.02.034},
       adsurl = {https://ui.adsabs.harvard.edu/abs/2019Icar..328..160K}
}

@ARTICLE{mlawer2012,
       author = {{Mlawer}, E.~J. and {Payne}, V.~H. and {Moncet}, J.-L. and {Delamere}, J.~S. and {Alvarado}, M.~J. and {Tobin}, D.~C.},
        title = "{Development and recent evaluation of the MT\_CKD model of continuum absorption}",
      journal = {Philosophical Transactions of the Royal Society of London Series A},
         year = 2012,
        month = jun,
       volume = {370},
       number = {1968},
        pages = {2520-2556},
          doi = {10.1098/rsta.2011.0295},
       adsurl = {https://ui.adsabs.harvard.edu/abs/2012RSPTA.370.2520M}
}

@ARTICLE{parc2024,
       author = {{Parc}, L{\'e}na and {Bouchy}, Fran{\c{c}}ois and {Venturini}, Julia and {Dorn}, Caroline and {Helled}, Ravit},
        title = "{From super-Earths to sub-Neptunes: Observational constraints and connections to theoretical models}",
      journal = {\aap},
         year = 2024,
        month = aug,
       volume = {688},
          eid = {A59},
        pages = {A59},
          doi = {10.1051/0004-6361/202449911},
archivePrefix = {arXiv},
       eprint = {2406.04311},
 primaryClass = {astro-ph.EP},
       adsurl = {https://ui.adsabs.harvard.edu/abs/2024A&A...688A..59P}
}

@ARTICLE{wolf2019,
       author = {{Wolf}, E.~T. and {Kopparapu}, R.~K. and {Haqq-Misra}, J.},
        title = "{Simulated Phase-dependent Spectra of Terrestrial Aquaplanets in M Dwarf Systems}",
      journal = {\apj},
         year = 2019,
        month = may,
       volume = {877},
       number = {1},
          eid = {35},
        pages = {35},
          doi = {10.3847/1538-4357/ab184a},
archivePrefix = {arXiv},
       eprint = {1906.02697},
 primaryClass = {astro-ph.EP},
       adsurl = {https://ui.adsabs.harvard.edu/abs/2019ApJ...877...35W}
}

@ARTICLE{fauchez2025b,
       author = {{Fauchez}, Thomas J. and {Villanueva}, Geronimo L. and {Kofman}, Vincent and {Suissa}, Gabriella and {Kopparapu}, Ravi K.},
        title = "{From global climate models (GCMs) to exoplanet spectra with the Global Emission Spectra (GlobES)}",
      journal = {Astronomy and Computing},
         year = 2025,
        month = oct,
       volume = {53},
          eid = {100982},
        pages = {100982},
          doi = {10.1016/j.ascom.2025.100982},
archivePrefix = {arXiv},
       eprint = {2507.09048},
 primaryClass = {astro-ph.EP},
       adsurl = {https://ui.adsabs.harvard.edu/abs/2025A&C....5300982F}
}

@ARTICLE{kopparapu2013,
       author = {{Kopparapu}, Ravi Kumar and {Ramirez}, Ramses and {Kasting}, James F. and {Eymet}, Vincent and {Robinson}, Tyler D. and {Mahadevan}, Suvrath and {Terrien}, Ryan C. and {Domagal-Goldman}, Shawn and {Meadows}, Victoria and {Deshpande}, Rohit},
        title = "{Habitable Zones around Main-sequence Stars: New Estimates}",
      journal = {\apj},
         year = 2013,
        month = mar,
       volume = {765},
       number = {2},
          eid = {131},
        pages = {131},
          doi = {10.1088/0004-637X/765/2/131},
archivePrefix = {arXiv},
       eprint = {1301.6674},
 primaryClass = {astro-ph.EP},
       adsurl = {https://ui.adsabs.harvard.edu/abs/2013ApJ...765..131K}
}

@ARTICLE{kopparapu2014,
       author = {{Kopparapu}, Ravi Kumar and {Ramirez}, Ramses M. and {SchottelKotte}, James and {Kasting}, James F. and {Domagal-Goldman}, Shawn and {Eymet}, Vincent},
        title = "{Habitable Zones around Main-sequence Stars: Dependence on Planetary Mass}",
      journal = {\apjl},
         year = 2014,
        month = jun,
       volume = {787},
       number = {2},
          eid = {L29},
        pages = {L29},
          doi = {10.1088/2041-8205/787/2/L29},
archivePrefix = {arXiv},
       eprint = {1404.5292},
 primaryClass = {astro-ph.EP},
       adsurl = {https://ui.adsabs.harvard.edu/abs/2014ApJ...787L..29K}
}

@ARTICLE{kopparapu2016,
       author = {{Kopparapu}, Ravi kumar and {Wolf}, Eric T. and {Haqq-Misra}, Jacob and {Yang}, Jun and {Kasting}, James F. and {Meadows}, Victoria and {Terrien}, Ryan and {Mahadevan}, Suvrath},
        title = "{The Inner Edge of the Habitable Zone for Synchronously Rotating Planets around Low-mass Stars Using General Circulation Models}",
      journal = {\apj},
         year = 2016,
        month = mar,
       volume = {819},
       number = {1},
          eid = {84},
        pages = {84},
          doi = {10.3847/0004-637X/819/1/84},
archivePrefix = {arXiv},
       eprint = {1602.05176},
 primaryClass = {astro-ph.EP},
       adsurl = {https://ui.adsabs.harvard.edu/abs/2016ApJ...819...84K}
}

@ARTICLE{yang2013,
       author = {{Yang}, Jun and {Cowan}, Nicolas B. and {Abbot}, Dorian S.},
        title = "{Stabilizing Cloud Feedback Dramatically Expands the Habitable Zone of Tidally Locked Planets}",
      journal = {\apjl},
         year = 2013,
        month = jul,
       volume = {771},
       number = {2},
          eid = {L45},
        pages = {L45},
          doi = {10.1088/2041-8205/771/2/L45},
archivePrefix = {arXiv},
       eprint = {1307.0515},
 primaryClass = {astro-ph.EP},
       adsurl = {https://ui.adsabs.harvard.edu/abs/2013ApJ...771L..45Y}
}

@ARTICLE{maurel2025,
       author = {{Maurel}, Alice and {Turbet}, Martin and {Ducrot}, Elsa and {Leconte}, J{\'e}r{\'e}my and {Chaverot}, Guillaume and {Milcareck}, Gwenael and {Revol}, Alexandre and {Charnay}, Benjamin and {Fauchez}, Thomas J. and {Gillon}, Micha{\"e}l and {Mechineau}, Alexandre and {Bolmont}, Emeline and {Millour}, Ehouarn and {Selsis}, Franck and {Beaulieu}, Jean-Philippe and {Drossart}, Pierre},
        title = "{Constraints on the possible atmospheres on TRAPPIST-1 b: insights from 3D climate modeling}",
      journal = {\aap},
         year = 2025,
        month = sep,
       volume = {701},
          eid = {A193},
        pages = {A193},
          doi = {10.1051/0004-6361/202554243},
archivePrefix = {arXiv},
       eprint = {2509.02120},
 primaryClass = {astro-ph.EP},
       adsurl = {https://ui.adsabs.harvard.edu/abs/2025A&A...701A.193M}
}

@ARTICLE{Toon1989,
       author = {{Toon}, Owen B. and {McKay}, C.~P. and {Ackerman}, T.~P. and {Santhanam}, K.},
        title = "{Rapid calculation of radiative heating rates and photodissociation rates in inhomogeneous multiple scattering atmospheres}",
      journal = {\jgr},
         year = 1989,
        month = nov,
       volume = {94},
        pages = {16287-16301},
          doi = {10.1029/JD094iD13p16287},
       adsurl = {https://ui.adsabs.harvard.edu/abs/1989JGR....9416287T}
}

@ARTICLE{Caldas2019,
       author = {{Caldas}, A. and {Leconte}, J. and {Selsis}, F. and {Waldmann}, I.~P. and {Bord{\'e}}, P. and {Rocchetto}, M. and {Charnay}, B.},
        title = "{Effects of a fully 3D atmospheric structure on exoplanet transmission spectra: retrieval biases due to day-night temperature gradients}",
      journal = {\aap},
         year = 2019,
        month = mar,
       volume = {623},
          eid = {A161},
        pages = {A161},
          doi = {10.1051/0004-6361/201834384},
archivePrefix = {arXiv},
       eprint = {1901.09932},
 primaryClass = {astro-ph.EP},
       adsurl = {https://ui.adsabs.harvard.edu/abs/2019A&A...623A.161C}
}

@ARTICLE{Falco2022,
       author = {{Falco}, Aur{\'e}lien and {Zingales}, Tiziano and {Pluriel}, William and {Leconte}, J{\'e}r{\'e}my},
        title = "{Toward a multidimensional analysis of transmission spectroscopy. I. Computation of transmission spectra using a 1D, 2D, or 3D atmosphere structure}",
      journal = {\aap},
         year = 2022,
        month = feb,
       volume = {658},
          eid = {A41},
        pages = {A41},
          doi = {10.1051/0004-6361/202141940},
archivePrefix = {arXiv},
       eprint = {2110.11799},
 primaryClass = {astro-ph.EP},
       adsurl = {https://ui.adsabs.harvard.edu/abs/2022A&A...658A..41F}
}

@ARTICLE{Meador1980,
       author = {{Meador}, W.~E. and {Weaver}, W.~R.},
        title = "{Two-stream approximations to radiative transfer in planetary atmospheres: a unified description of existing methods and a new improvement.}",
      journal = {Journal of the Atmospheric Sciences},
         year = 1980,
        month = mar,
       volume = {37},
        pages = {630-643},
          doi = {10.1175/1520-0469(1980)037<0630:TSATRT>2.0.CO;2},
       adsurl = {https://ui.adsabs.harvard.edu/abs/1980JAtS...37..630M}
}

@ARTICLE{Leconte2021,
       author = {{Leconte}, J{\'e}r{\'e}my},
        title = "{Spectral binning of precomputed correlated-k coefficients}",
      journal = {\aap},
         year = 2021,
        month = jan,
       volume = {645},
          eid = {A20},
        pages = {A20},
          doi = {10.1051/0004-6361/202039040},
archivePrefix = {arXiv},
       eprint = {2012.01428},
 primaryClass = {astro-ph.EP},
       adsurl = {https://ui.adsabs.harvard.edu/abs/2021A&A...645A..20L}
}

@ARTICLE{kasting1988,
       author = {{Kasting}, J.~F.},
        title = "{Runaway and moist greenhouse atmospheres and the evolution of Earth and Venus}",
      journal = {\icarus},
         year = 1988,
        month = jun,
       volume = {74},
       number = {3},
        pages = {472-494},
          doi = {10.1016/0019-1035(88)90116-9},
       adsurl = {https://ui.adsabs.harvard.edu/abs/1988Icar...74..472K}
}

@ARTICLE{kopparapu2017,
       author = {{Kopparapu}, Ravi kumar and {Wolf}, Eric T. and {Arney}, Giada and {Batalha}, Natasha E. and {Haqq-Misra}, Jacob and {Grimm}, Simon L. and {Heng}, Kevin},
        title = "{Habitable Moist Atmospheres on Terrestrial Planets near the Inner Edge of the Habitable Zone around M Dwarfs}",
      journal = {\apj},
         year = 2017,
        month = aug,
       volume = {845},
       number = {1},
          eid = {5},
        pages = {5},
          doi = {10.3847/1538-4357/aa7cf9},
archivePrefix = {arXiv},
       eprint = {1705.10362},
 primaryClass = {astro-ph.EP},
       adsurl = {https://ui.adsabs.harvard.edu/abs/2017ApJ...845....5K}
}

@ARTICLE{way2016,
       author = {{Way}, M.~J. and {Del Genio}, Anthony D. and {Kiang}, Nancy Y. and {Sohl}, Linda E. and {Grinspoon}, David H. and {Aleinov}, Igor and {Kelley}, Maxwell and {Clune}, Thomas},
        title = "{Was Venus the first habitable world of our solar system?}",
      journal = {\grl},
         year = 2016,
        month = aug,
       volume = {43},
       number = {16},
        pages = {8376-8383},
          doi = {10.1002/2016GL069790},
archivePrefix = {arXiv},
       eprint = {1608.00706},
 primaryClass = {astro-ph.EP},
       adsurl = {https://ui.adsabs.harvard.edu/abs/2016GeoRL..43.8376W}
}

@ARTICLE{lim2023,
       author = {{Lim}, Olivia and {Benneke}, Bj{\"o}rn and {Doyon}, Ren{\'e} and {MacDonald}, Ryan J. and {Piaulet}, Caroline and {Artigau}, {\'E}tienne and {Coulombe}, Louis-Philippe and {Radica}, Michael and {L'Heureux}, Alexandrine and {Albert}, Lo{\"\i}c and {Rackham}, Benjamin V. and {de Wit}, Julien and {Salhi}, Salma and {Roy}, Pierre-Alexis and {Flagg}, Laura and {Fournier-Tondreau}, Marylou and {Taylor}, Jake and {Cook}, Neil J. and {Lafreni{\`e}re}, David and {Cowan}, Nicolas B. and {Kaltenegger}, Lisa and {Rowe}, Jason F. and {Espinoza}, N{\'e}stor and {Dang}, Lisa and {Darveau-Bernier}, Antoine},
        title = "{Atmospheric Reconnaissance of TRAPPIST-1 b with JWST/NIRISS: Evidence for Strong Stellar Contamination in the Transmission Spectra}",
      journal = {\apjl},
         year = 2023,
        month = sep,
       volume = {955},
       number = {1},
          eid = {L22},
        pages = {L22},
          doi = {10.3847/2041-8213/acf7c4},
archivePrefix = {arXiv},
       eprint = {2309.07047},
 primaryClass = {astro-ph.EP},
       adsurl = {https://ui.adsabs.harvard.edu/abs/2023ApJ...955L..22L}
}

@BOOK{pierrehumbert2010,
       author = {{Pierrehumbert}, Raymond T.},
        title = "{Principles of Planetary Climate}",
         year = 2010,
       adsurl = {https://ui.adsabs.harvard.edu/abs/2010ppc..book.....P}
}

@ARTICLE{yang2014,
       author = {{Yang}, Jun and {Bou{\'e}}, Gwena{\"e}l and {Fabrycky}, Daniel C. and {Abbot}, Dorian S.},
        title = "{Strong Dependence of the Inner Edge of the Habitable Zone on Planetary Rotation Rate}",
      journal = {\apjl},
         year = 2014,
        month = may,
       volume = {787},
       number = {1},
          eid = {L2},
        pages = {L2},
          doi = {10.1088/2041-8205/787/1/L2},
archivePrefix = {arXiv},
       eprint = {1404.4992},
 primaryClass = {astro-ph.EP},
       adsurl = {https://ui.adsabs.harvard.edu/abs/2014ApJ...787L...2Y}
}

@ARTICLE{forget2014,
       author = {{Forget}, F. and {Leconte}, J.},
        title = "{Possible climates on terrestrial exoplanets}",
      journal = {Philosophical Transactions of the Royal Society of London Series A},
         year = 2014,
        month = mar,
       volume = {372},
       number = {2014},
        pages = {20130084-20130084},
          doi = {10.1098/rsta.2013.0084},
archivePrefix = {arXiv},
       eprint = {1311.3101},
 primaryClass = {astro-ph.EP},
       adsurl = {https://ui.adsabs.harvard.edu/abs/2014RSPTA.37230084F}
}

@ARTICLE{maurice2024,
       author = {{Maurice}, M. and {Dasgupta}, R. and {Hassanzadeh}, P.},
        title = "{Volatile atmospheres of lava worlds}",
      journal = {\aap},
         year = 2024,
        month = aug,
       volume = {688},
          eid = {A47},
        pages = {A47},
          doi = {10.1051/0004-6361/202347749},
archivePrefix = {arXiv},
       eprint = {2405.09284},
 primaryClass = {astro-ph.EP},
       adsurl = {https://ui.adsabs.harvard.edu/abs/2024A&A...688A..47M}
}

@book{Milankovitch1941,
    author = {{Milankovitch}, M.},
    booktitle = {Koniglich Serbische Akademie},
    publisher = {Koniglich Serbische Akademie},
    title = {{Kanon der Erdebestrahlung und seine anwendung auf das eiszeitenproblem}},
    year = 1941}

@ARTICLE{Yamashiki2019,
       author = {{Yamashiki}, Yosuke A. and {Maehara}, Hiroyuki and {Airapetian}, Vladimir and {Notsu}, Yuta and {Sato}, Tatsuhiko and {Notsu}, Shota and {Kuroki}, Ryusuke and {Murashima}, Keiya and {Sato}, Hiroaki and {Namekata}, Kosuke and {Sasaki}, Takanori and {Scott}, Thomas B. and {Bando}, Hina and {Nashimoto}, Subaru and {Takagi}, Fuka and {Ling}, Cassandra and {Nogami}, Daisaku and {Shibata}, Kazunari},
        title = "{Impact of Stellar Superflares on Planetary Habitability}",
      journal = {\apj},
         year = 2019,
        month = aug,
       volume = {881},
       number = {2},
          eid = {114},
        pages = {114},
          doi = {10.3847/1538-4357/ab2a71},
archivePrefix = {arXiv},
       eprint = {1906.06797},
 primaryClass = {astro-ph.SR},
       adsurl = {https://ui.adsabs.harvard.edu/abs/2019ApJ...881..114Y}
}

@ARTICLE{Wolf2025,
       author = {{Wolf}, Eric T. and {Schwieterman}, Edward W. and {Haqq-Misra}, Jacob and {Fauchez}, Thomas J. and {Bastelberger}, Sandra T. and {Leung}, Michaela and {Peacock}, Sarah and {Villanueva}, Geronimo L. and {Kopparapu}, Ravi K.},
        title = "{Chemistry, Climate, and Transmission Spectra of TRAPPIST-1 e Explored with a Multimodel Sparse Sampled Ensemble}",
      journal = {The Planetary Science Journal},
         year = 2025,
        month = oct,
       volume = {6},
       number = {10},
          eid = {231},
        pages = {231},
          doi = {10.3847/PSJ/ae031e},
archivePrefix = {arXiv},
       eprint = {2510.18704},
 primaryClass = {astro-ph.EP},
       adsurl = {https://ui.adsabs.harvard.edu/abs/2025PSJ.....6..231W}
}

@ARTICLE{Jaziri2026,
       author = {{Jaziri}, Adam Y. and {Carrasco}, Nathalie and {Charnay}, Benjamin},
        title = "{Possible favored Great Oxidation Event scenario on exoplanets around M-Stars with the example of TRAPPIST-1e}",
      journal = {Nature Scientific Report},
         year = 2026,
        month = jan,
          eid = {arXiv:2601.18324},
        pages = {arXiv:2601.18324},
          doi = {10.48550/arXiv.2601.18324},
archivePrefix = {arXiv},
       eprint = {2601.18324},
 primaryClass = {astro-ph.EP},
       adsurl = {https://ui.adsabs.harvard.edu/abs/2026arXiv260118324J}
}

@ARTICLE{Drant2025,
       author = {{Drant}, T. and {Tian}, M. and {Carrasco}, N. and {Heng}, K.},
        title = "{Inferring the interior oxygen fugacity of rocky exoplanets from observations: Assessing biases by atmospheric chemistry}",
      journal = {\aap},
         year = 2025,
        month = jun,
       volume = {698},
          eid = {A76},
        pages = {A76},
          doi = {10.1051/0004-6361/202452016},
       adsurl = {https://ui.adsabs.harvard.edu/abs/2025A&A...698A..76D}
}

@ARTICLE{Quirino2023,
       author = {{Quirino}, Diogo and {Gilli}, Gabriella and {Kaltenegger}, Lisa and {Navarro}, Thomas and {Fauchez}, Thomas J. and {Turbet}, Martin and {Leconte}, J{\'e}r{\'e}my and {Lebonnois}, S{\'e}bastien and {Gonz{\'a}lez-Galindo}, Francisco},
        title = "{3D Global climate model of an exo-Venus: a modern Venus-like atmosphere for the nearby super-Earth LP 890-9 c}",
      journal = {\mnras},
         year = 2023,
        month = jul,
       volume = {523},
       number = {1},
        pages = {L86-L91},
          doi = {10.1093/mnrasl/slad045},
archivePrefix = {arXiv},
       eprint = {2311.04675},
 primaryClass = {astro-ph.EP},
       adsurl = {https://ui.adsabs.harvard.edu/abs/2023MNRAS.523L..86Q}
}

@ARTICLE{Kane2026,
       author = {{Kane}, Stephen R. and {Bott}, Kimberly M. and {Goodis Gordon}, Kenneth E. and {Miles}, Emma L. and {Ostberg}, Colby M. and {Byrne}, Paul K. and {Carone}, Ludmila and {Daylan}, Tansu and {Garc{\'\i}a Mu{\~n}oz}, Antonio and {Harada}, Caleb K. and {Hu}, Renyu and {Izenberg}, Noam. R. and {Kohler}, Erika and {Rice}, Malena and {Sagynbayeva}, Sabina and {Scherf}, Manuel and {Schwieterman}, Edward W. and {Woitke}, Peter},
        title = "{Imaging Venus-like Worlds: Spectral, Polarimetric, and UV Diagnostics for the Habitable Worlds Observatory}",
      journal = {\pasp},
         year = 2026,
        month = feb,
       volume = {138},
       number = {2},
          eid = {024404},
        pages = {024404},
          doi = {10.1088/1538-3873/ae417d},
archivePrefix = {arXiv},
       eprint = {2602.02728},
 primaryClass = {astro-ph.EP},
       adsurl = {https://ui.adsabs.harvard.edu/abs/2026PASP..138b4404K}
}

@ARTICLE{Madden2018,
       author = {{Madden}, J.~H. and {Kaltenegger}, Lisa},
        title = "{A Catalog of Spectra, Albedos, and Colors of Solar System Bodies for Exoplanet Comparison}",
      journal = {Astrobiology},
         year = 2018,
        month = dec,
       volume = {18},
       number = {12},
        pages = {1559-1573},
          doi = {10.1089/ast.2017.1763},
archivePrefix = {arXiv},
       eprint = {1807.11442},
 primaryClass = {astro-ph.EP},
       adsurl = {https://ui.adsabs.harvard.edu/abs/2018AsBio..18.1559M}
}

@ARTICLE{turbet2023,
       author = {{Turbet}, Martin and {Fauchez}, Thomas J. and {Leconte}, Jeremy and {Bolmont}, Emeline and {Chaverot}, Guillaume and {Forget}, Francois and {Millour}, Ehouarn and {Selsis}, Franck and {Charnay}, Benjamin and {Ducrot}, Elsa and {Gillon}, Micha{\"e}l and {Maurel}, Alice and {Villanueva}, Geronimo L.},
        title = "{Water condensation zones around main sequence stars}",
      journal = {\aap},
         year = 2023,
        month = nov,
       volume = {679},
          eid = {A126},
        pages = {A126},
          doi = {10.1051/0004-6361/202347539},
archivePrefix = {arXiv},
       eprint = {2308.15110},
 primaryClass = {astro-ph.EP},
       adsurl = {https://ui.adsabs.harvard.edu/abs/2023A&A...679A.126T}
}

@ARTICLE{Hammond2020,
       author = {{Hammond}, Mark and {Tsai}, Shang-Min and {Pierrehumbert}, Raymond T.},
        title = "{The Equatorial Jet Speed on Tidally Locked Planets. I. Terrestrial Planets}",
      journal = {\apj},
         year = 2020,
        month = sep,
       volume = {901},
       number = {1},
          eid = {78},
        pages = {78},
          doi = {10.3847/1538-4357/abb08b},
archivePrefix = {arXiv},
       eprint = {2009.00358},
 primaryClass = {astro-ph.EP},
       adsurl = {https://ui.adsabs.harvard.edu/abs/2020ApJ...901...78H}
}

@ARTICLE{Koll2016,
       author = {{Koll}, Daniel D.~B. and {Abbot}, Dorian S.},
        title = "{Temperature Structure and Atmospheric Circulation of Dry Tidally Locked Rocky Exoplanets}",
      journal = {\apj},
         year = 2016,
        month = jul,
       volume = {825},
       number = {2},
          eid = {99},
        pages = {99},
          doi = {10.3847/0004-637X/825/2/99},
archivePrefix = {arXiv},
       eprint = {1605.01066},
 primaryClass = {astro-ph.EP},
       adsurl = {https://ui.adsabs.harvard.edu/abs/2016ApJ...825...99K}
}

@ARTICLE{Yang2019,
       author = {{Yang}, Jun and {Leconte}, J{\'e}r{\'e}my and {Wolf}, Eric T. and {Merlis}, Timothy and {Koll}, Daniel D.~B. and {Forget}, Fran{\c{c}}ois and {Abbot}, Dorian S.},
        title = "{Simulations of Water Vapor and Clouds on Rapidly Rotating and Tidally Locked Planets: A 3D Model Intercomparison}",
      journal = {\apj},
         year = 2019,
        month = apr,
       volume = {875},
       number = {1},
          eid = {46},
        pages = {46},
          doi = {10.3847/1538-4357/ab09f1},
archivePrefix = {arXiv},
       eprint = {1912.11329},
 primaryClass = {astro-ph.EP},
       adsurl = {https://ui.adsabs.harvard.edu/abs/2019ApJ...875...46Y}
}

@ARTICLE{Taniguchi2026,
       author = {{Taniguchi}, Keigo and {Kodama}, Takanori and {Turbet}, Martin and {Chaverot}, Guillaume and {Millour}, Ehouarn and {Genda}, Hidenori},
        title = "{Atmospheric Collapse and Habitability on Tidally Locked Exoplanets}",
      journal = {\apj},
         year = 2026,
        month = mar,
       volume = {1000},
       number = {1},
          eid = {99},
        pages = {99},
          doi = {10.3847/1538-4357/ae47f2},
archivePrefix = {arXiv},
       eprint = {2603.09260},
 primaryClass = {astro-ph.EP},
       adsurl = {https://ui.adsabs.harvard.edu/abs/2026ApJ..1000...99T}
}

@ARTICLE{Goldblatt2013,
       author = {{Goldblatt}, Colin and {Robinson}, Tyler D. and {Zahnle}, Kevin J. and {Crisp}, David},
        title = "{Low simulated radiation limit for runaway greenhouse climates}",
      journal = {Nature Geoscience},
         year = 2013,
        month = aug,
       volume = {6},
       number = {8},
        pages = {661-667},
          doi = {10.1038/ngeo1892},
       adsurl = {https://ui.adsabs.harvard.edu/abs/2013NatGe...6..661G}
}

@ARTICLE{Hu2012jun,
       author = {{Hu}, Renyu and {Ehlmann}, Bethany L. and {Seager}, Sara},
        title = "{Theoretical Spectra of Terrestrial Exoplanet Surfaces}",
      journal = {\apj},
         year = 2012,
        month = jun,
       volume = {752},
       number = {1},
          eid = {7},
        pages = {7},
          doi = {10.1088/0004-637X/752/1/7},
archivePrefix = {arXiv},
       eprint = {1204.1544},
 primaryClass = {astro-ph.EP},
       adsurl = {https://ui.adsabs.harvard.edu/abs/2012ApJ...752....7H}
}

@ARTICLE{Hu2012dec,
       author = {{Hu}, Renyu and {Seager}, Sara and {Bains}, William},
        title = "{Photochemistry in Terrestrial Exoplanet Atmospheres. I. Photochemistry Model and Benchmark Cases}",
      journal = {\apj},
         year = 2012,
        month = dec,
       volume = {761},
       number = {2},
          eid = {166},
        pages = {166},
          doi = {10.1088/0004-637X/761/2/166},
archivePrefix = {arXiv},
       eprint = {1210.6885},
 primaryClass = {astro-ph.EP},
       adsurl = {https://ui.adsabs.harvard.edu/abs/2012ApJ...761..166H}
}

@ARTICLE{Roccetti2024,
       author = {{Roccetti}, Giulia and {Bugliaro}, Luca and {G{\"o}dde}, Felix and {Emde}, Claudia and {Hamann}, Ulrich and {Manev}, Mihail and {Sterzik}, Michael Fritz and {Wehrum}, Cedric},
        title = "{HAMSTER: Hyperspectral Albedo Maps dataset with high Spatial and TEmporal Resolution}",
      journal = {Atmospheric Measurement Techniques},
         year = 2024,
        month = oct,
       volume = {17},
       number = {20},
        pages = {6025-6046},
          doi = {10.5194/amt-17-6025-2024},
archivePrefix = {arXiv},
       eprint = {2407.18030},
 primaryClass = {astro-ph.EP},
       adsurl = {https://ui.adsabs.harvard.edu/abs/2024AMT....17.6025R}
}

@ARTICLE{Roccetti2025,
       author = {{Roccetti}, Giulia and {Emde}, Claudia and {Sterzik}, Michael F. and {Manev}, Mihail and {Seidel}, Julia V. and {Bagnulo}, Stefano},
        title = "{Planet Earth in reflected and polarized light: I. Three-dimensional radiative transfer simulations of realistic surface-atmosphere systems}",
      journal = {\aap},
         year = 2025,
        month = may,
       volume = {697},
          eid = {A170},
        pages = {A170},
          doi = {10.1051/0004-6361/202554167},
archivePrefix = {arXiv},
       eprint = {2504.02048},
 primaryClass = {astro-ph.EP},
       adsurl = {https://ui.adsabs.harvard.edu/abs/2025A&A...697A.170R}
}

@ARTICLE{Buratti1996,
       author = {{Buratti}, Bonnie J. and {Hillier}, John K. and {Wang}, Michael},
        title = "{The Lunar Opposition Surge: Observations by Clementine}",
      journal = {\icarus},
         year = 1996,
        month = dec,
       volume = {124},
       number = {2},
        pages = {490-499},
          doi = {10.1006/icar.1996.0225},
       adsurl = {https://ui.adsabs.harvard.edu/abs/1996Icar..124..490B}
}

@ARTICLE{Mallama2017,
       author = {{Mallama}, Anthony},
        title = "{The Spherical Bolometric Albedo of Planet Mercury}",
      journal = {arXiv e-prints},
         year = 2017,
        month = mar,
          eid = {arXiv:1703.02670},
        pages = {arXiv:1703.02670},
          doi = {10.48550/arXiv.1703.02670},
archivePrefix = {arXiv},
       eprint = {1703.02670},
 primaryClass = {astro-ph.EP},
       adsurl = {https://ui.adsabs.harvard.edu/abs/2017arXiv170302670M}
}

@ARTICLE{Zieba2026,
       author = {{Zieba}, Sebastian and {Kreidberg}, Laura and {Coy}, Brandon P. and {Bello-Arufe}, Aaron and {Paragas}, Kimberly and {Lyu}, Xintong and {Hu}, Renyu and {Iyer}, Aishwarya and {Kite}, Edwin S. and {Koll}, Daniel D.~B. and {Wohlfarth}, Kay and {Whittaker}, Emerson and {Knutson}, Heather and {Wordsworth}, Robin and {Morley}, Caroline and {Schaefer}, Laura},
        title = "{The dark and featureless surface of rocky exoplanet LHS 3844 b from JWST mid-infrared spectroscopy}",
      journal = {Nature Astronomy},
         year = 2026,
        month = may,
          doi = {10.1038/s41550-026-02860-3},
archivePrefix = {arXiv},
       eprint = {2605.00100},
 primaryClass = {astro-ph.EP},
       adsurl = {https://ui.adsabs.harvard.edu/abs/2026NatAs.tmp...97Z}
}

@ARTICLE{Cassidy1975,
       author = {{Cassidy}, W. and {Hapke}, B.},
        title = "{Effects of Darkening Processes on Surfaces of Airless Bodies}",
      journal = {\icarus},
         year = 1975,
        month = jul,
       volume = {25},
       number = {3},
        pages = {371-383},
          doi = {10.1016/0019-1035(75)90002-0},
       adsurl = {https://ui.adsabs.harvard.edu/abs/1975Icar...25..371C}
}

@ARTICLE{Pieters2016,
       author = {{Pieters}, Carle M. and {Noble}, Sarah K.},
        title = "{Space weathering on airless bodies}",
      journal = {Journal of Geophysical Research (Planets)},
         year = 2016,
        month = oct,
       volume = {121},
       number = {10},
        pages = {1865-1884},
          doi = {10.1002/2016JE005128},
       adsurl = {https://ui.adsabs.harvard.edu/abs/2016JGRE..121.1865P}
}

@ARTICLE{Luger2015,
       author = {{Luger}, R. and {Barnes}, R.},
        title = "{Extreme Water Loss and Abiotic O2Buildup on Planets Throughout the Habitable Zones of M Dwarfs}",
      journal = {Astrobiology},
         year = 2015,
        month = feb,
       volume = {15},
       number = {2},
        pages = {119-143},
          doi = {10.1089/ast.2014.1231},
archivePrefix = {arXiv},
       eprint = {1411.7412},
 primaryClass = {astro-ph.EP},
       adsurl = {https://ui.adsabs.harvard.edu/abs/2015AsBio..15..119L}
}

@ARTICLE{Bolmont2017,
       author = {{Bolmont}, E. and {Selsis}, F. and {Owen}, J.~E. and {Ribas}, I. and {Raymond}, S.~N. and {Leconte}, J. and {Gillon}, M.},
        title = "{Water loss from terrestrial planets orbiting ultracool dwarfs: implications for the planets of TRAPPIST-1}",
      journal = {\mnras},
         year = 2017,
        month = jan,
       volume = {464},
       number = {3},
        pages = {3728-3741},
          doi = {10.1093/mnras/stw2578},
archivePrefix = {arXiv},
       eprint = {1605.00616},
 primaryClass = {astro-ph.EP},
       adsurl = {https://ui.adsabs.harvard.edu/abs/2017MNRAS.464.3728B}
}

@ARTICLE{Kreidberg2025,
       author = {{Kreidberg}, Laura and {Stevenson}, Kevin B.},
        title = "{A first look at rocky exoplanets with JWST}",
      journal = {Proceedings of the National Academy of Science},
         year = 2025,
        month = sep,
       volume = {122},
       number = {39},
          eid = {e2416190122},
        pages = {e2416190122},
          doi = {10.1073/pnas.2416190122},
archivePrefix = {arXiv},
       eprint = {2507.00933},
 primaryClass = {astro-ph.EP},
       adsurl = {https://ui.adsabs.harvard.edu/abs/2025PNAS..12216190K}
}

@ARTICLE{Kodama2019,
       author = {{Kodama}, T. and {Genda}, H. and {O'ishi}, R. and {Abe-Ouchi}, A. and {Abe}, Y.},
        title = "{Inner Edge of Habitable Zones for Earth-Sized Planets With Various Surface Water Distributions}",
      journal = {Journal of Geophysical Research (Planets)},
         year = 2019,
        month = aug,
       volume = {124},
       number = {8},
        pages = {2306-2324},
          doi = {10.1029/2019JE006037},
archivePrefix = {arXiv},
       eprint = {1908.05909},
 primaryClass = {astro-ph.EP},
       adsurl = {https://ui.adsabs.harvard.edu/abs/2019JGRE..124.2306K}
}

@ARTICLE{Boutle2017,
       author = {{Boutle}, Ian A. and {Mayne}, Nathan J. and {Drummond}, Benjamin and {Manners}, James and {Goyal}, Jayesh and {Hugo Lambert}, F. and {Acreman}, David M. and {Earnshaw}, Paul D.},
        title = "{Exploring the climate of Proxima B with the Met Office Unified Model}",
      journal = {\aap},
         year = 2017,
        month = may,
       volume = {601},
          eid = {A120},
        pages = {A120},
          doi = {10.1051/0004-6361/201630020},
archivePrefix = {arXiv},
       eprint = {1702.08463},
 primaryClass = {astro-ph.EP},
       adsurl = {https://ui.adsabs.harvard.edu/abs/2017A&A...601A.120B}
}

@ARTICLE{Yang2020,
       author = {{Yang}, Jun and {Ji}, Weiwen and {Zeng}, Yaoxuan},
        title = "{Transition from eyeball to snowball driven by sea-ice drift on tidally locked terrestrial planets}",
      journal = {Nature Astronomy},
         year = 2020,
        month = jan,
       volume = {4},
        pages = {58-66},
          doi = {10.1038/s41550-019-0883-z},
archivePrefix = {arXiv},
       eprint = {1912.11377},
 primaryClass = {astro-ph.EP},
       adsurl = {https://ui.adsabs.harvard.edu/abs/2020NatAs...4...58Y}
}
\bibliographystyle{aa}



\begin{appendix} 

\onecolumn

\section{Calculations of the dimensionless Rossby deformation length}
\label{appendix:rossby}

Following \citet{leconte2013aa}, the dimensionless equatorial Rossby deformation length is defined as
\begin{equation}
    \frac{L_{\rm Ro}}{R_{\rm p}} = \sqrt{\frac{N \ H}{2 \ \Omega \ R_{\rm p}}},
    \label{equ:rossby_deformation_length}
\end{equation}
where $L_{\rm Ro}$ is the equatorial Rossby deformation radius, $R_{\rm p}$ is the planetary radius, $N$ is the Brunt-Väisälä frequency, $H$ is the pressure scale height and $\Omega$ is the planet rotation rate. The Brunt-Väisälä frequency describes the stratification stability of a fluid to vertical perturbations. 
In a dry stably stratified atmosphere, it is given by
\begin{equation}
    N^2 = \frac{g}{T} \left( \frac{g}{c_{\rm p}} + \frac{\mathrm{d}T}{\mathrm{d}z} \right),
    \label{equ:brunt_vaisala}
\end{equation}
where $g$ is the gravitational acceleration, T is the temperature, $c_{\rm p}$ is the heat capacity of the gas at constant pressure and $\mathrm{d}T / \mathrm{d}z$ is the lapse rate. 
The pressure scale height is given by
\begin{equation}
    H = \frac{k_{\rm B} \ T}{m_{\rm a} \ g}, 
    \label{equ:scale_height}
\end{equation}
where $k_{\rm B}$ is the Boltzmann constant and $m_{\rm a}$ is the mean molecular weight of the air, defined as $m_{\rm a} = M / N_{\rm A}$, where $M$ is the molar mass and $N_{\rm A}$ is the Avogadro constant. 
We assume an isothermal atmosphere, which gives the same expression as in \citet{leconte2013aa}
\begin{equation}
    \frac{L_{\rm Ro}}{R_{\rm p}} = \sqrt{\frac{k_{\rm B} \ T^{1/2}}{2 \ \Omega \ R_{\rm p} \ m_{\rm a} \ c_{\rm p}^{1/2}}}.
    \label{equ:isotherm_rossby_deformation_length}
\end{equation}

We use the equilibrium temperature of Ross~128~b for $T$, calculated assuming a Bond albedo of 0.3. Considering an \ce{N2}-dominated atmosphere, the molar mass is set to $M = 28 \times 10^{-3}$~$\mathrm{kg\,mol^{-1}}$ and the specific heat capacity is $c_{\rm p} = 1003.16$~$\mathrm{J\,K^{-1}\,kg^{-1}}$. Assuming a synchronous rotation, this gives $L_{\rm Ro} / R_{\rm p} = 1.2$. 
We find a similar result with $T = 230$~K, which is the lowest mean surface temperature derived from our GCM simulations. 
With $T = 482$~K (highest mean surface temperature from our \ce{N2} atmospheres simulations), the ratio increases to $L_{\rm Ro} / R_{\rm p} = 1.4$. 

For a \ce{CO2}-dominated atmosphere, with $M = 44 \times 10^{-3}$~$\mathrm{kg\,mol^{-1}}$ and $c_{\rm p} = 840$~$\mathrm{J\,K^{-1}\,kg^{-1}}$, we find $L_{\rm Ro} / R_{\rm p} = 1.0$ with the equilibrium temperature and $L_{\rm Ro} / R_{\rm p} = 1.3$ with the maximum mean surface temperature of 671~K derived from our GCM simulations. 
Our calculations are performed with specific assumptions and simplifications. More accurate calculations should be done, but it is out of scope of this study.

\section{Complementary simulation for Case~2B}
\label{appendix:water}

\begin{figure}[!htb]
   \sidecaption
   \includegraphics[width=7.5cm]{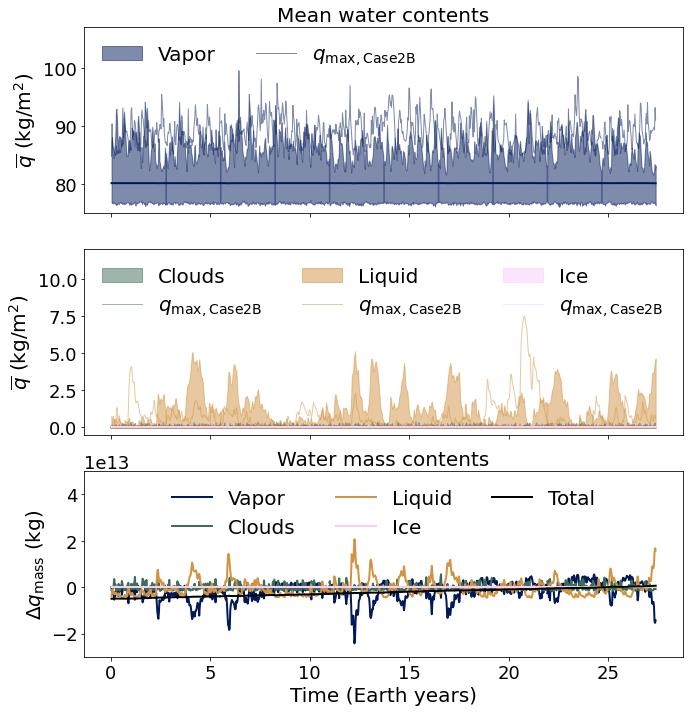}
      \caption{Temporal evolutions of the column-integrated water vapor (top panel), water clouds, surface liquid and ice water density (middle panel) for a simulation setup similar to Case~2B, but for a global water content equivalent to 8~cm~GEL initially in liquid phase at the surface. The mean values are represented by the solid lines, while the minimum and maximum values are defined by the filled areas in the first two panels. The thin lines show the maximum values of the corresponding variables of Case~2B. The lower panel represents the deviations of the total water mass content for each quantity from its respective initial value.}
         \label{fig:water_evolution_appendix}
   \end{figure}

The variables shown in Figure~\ref{fig:water_evolution_appendix} were computed for a global water content of 8~cm~GEL initially in liquid phase at the surface. As in Case~2B, this setup also shows temporal variations in the surface condensation of liquid water, correlated with the variations of atmospheric water vapor concentration (described in Section~\ref{sub:climate_results:refcase}).

\section{Complementary maps for Section \ref{sec:climate_results}}

\subsection{Effect of the atmospheric composition on the climate}
\label{appendix:maps_atmospheres}

\begin{figure*}[!htb]
    \centering
    \includegraphics[width=0.85\textwidth, clip]{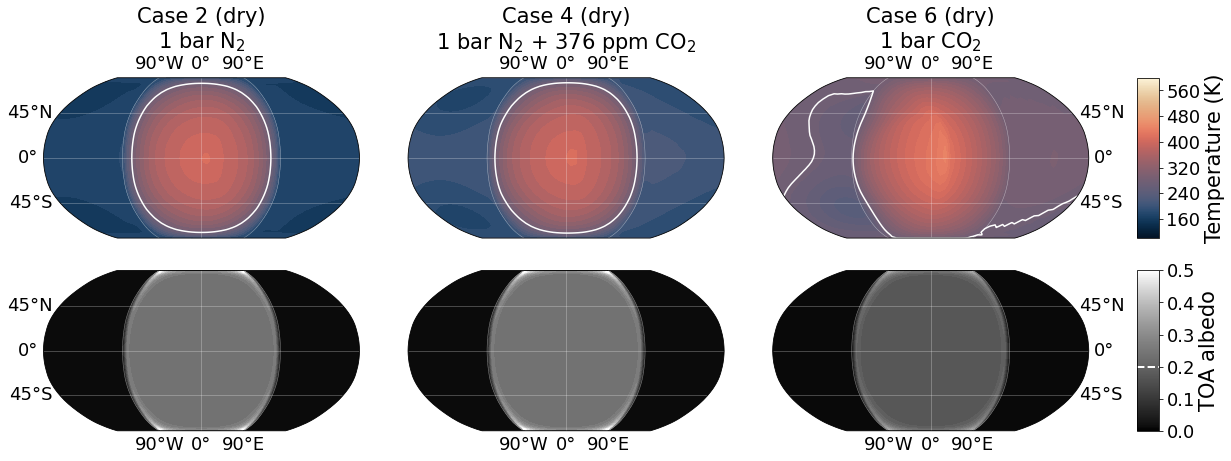}
    \caption{Latitude-longitude maps of surface temperature and TOA albedo for Cases~2, 4 and~6 under dry conditions. The variables are averaged over $\sim$50 orbits. The solid white line contour corresponds to the 273.15~K isotherm below which water (if present) freezes. The dashed white line on the albedo color bar corresponds to the surface albedo.}
    \label{fig:maps_varying_atmos_dry}
\end{figure*}

\begin{figure*}[!htb]
    \centering
    \includegraphics[width=0.85\textwidth, clip]{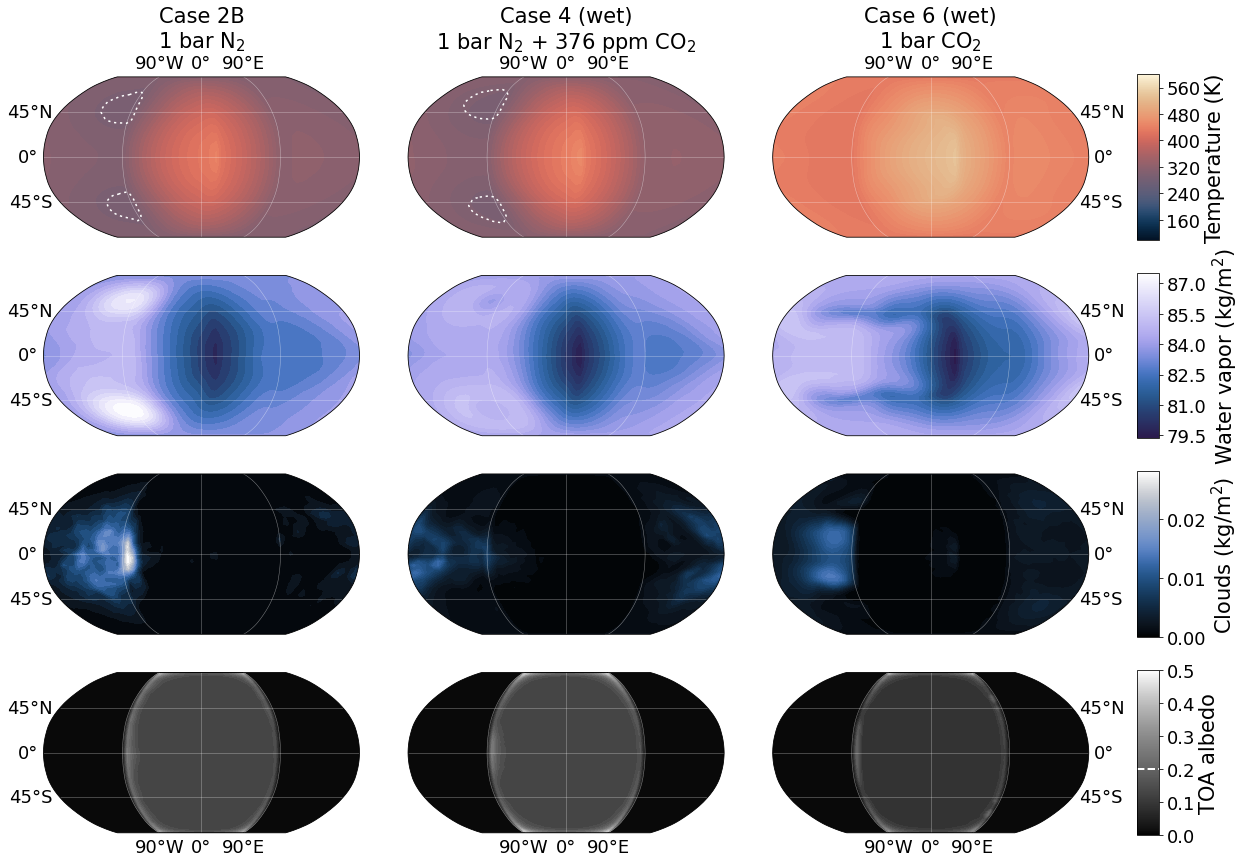}
    \caption{Latitude-longitude maps of surface temperature, column-integrated water vapor, water cloud column and TOA albedo for Cases~2, 4 and~6 under wet conditions (i.e., with a global water content of 8.3~cm~GEL). Here Case~2 corresponds to Case~2B. The variables are averaged over $\sim$50 orbits for Cases~2 and~4 and $\sim$26 orbits for Case~6. The dotted white line contour corresponds to the location of surface liquid water. The dashed white line on the albedo color bar corresponds to the surface albedo.}
    \label{fig:maps_varying_atmos_wet}
\end{figure*}

In this section, we present the maps of our GCM results corresponding to Cases~2, 4 and~6 under dry and wet conditions. 
Figure~\ref{fig:maps_varying_atmos_dry} presents the surface temperatures and TOA albedo for Cases~2, 4 and~6 under dry conditions. 
Figure~\ref{fig:maps_varying_atmos_wet} shows the surface temperatures, atmospheric water vapor, water clouds and TOA albedo for Cases~2, 4 and~6 under wet conditions (with 8.3~cm~GEL of water inventory). 

The case with 1~bar of \ce{CO2} atmosphere presents the highest surface temperatures among the three cases for both dry and wet conditions (Figures~\ref{fig:maps_varying_atmos_dry} and \ref{fig:maps_varying_atmos_wet}).
In dry conditions, Figure~\ref{fig:maps_varying_atmos_dry} shows for Case~6 a high temperature contrast between the dayside and the nightside of $\sim$209~K, about 100~K higher compared to the corresponding wet conditions.

\subsection{Effect of the atmospheric surface pressure on the climate}
\label{appendix:maps_pressures}

\begin{figure}[!htb]
   \sidecaption
   \includegraphics[width=11cm]{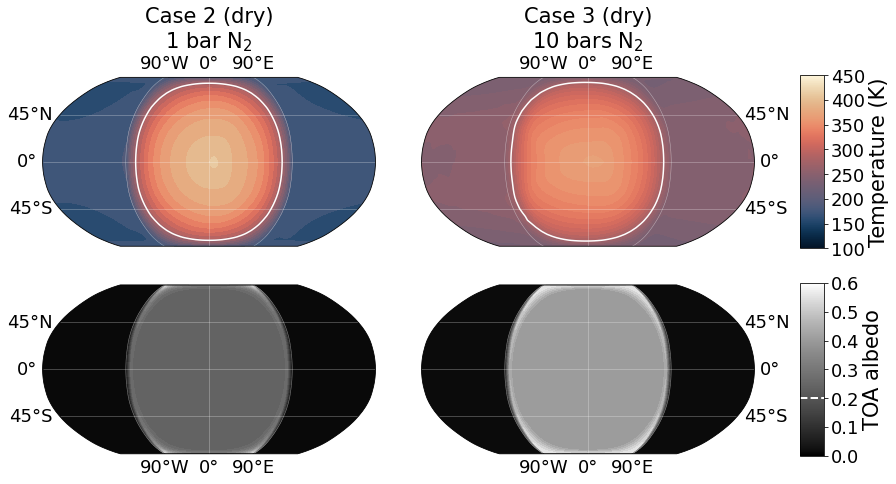}
      \caption{Latitude-longitude maps of surface temperature and TOA albedo for Cases~2 and 3 under dry conditions. The variables are averaged over $\sim$50 orbits. The solid white line contour corresponds to the 273.15~K isotherm below which water (if present) freezes. The dashed white line on the albedo color bar corresponds to the surface albedo.}
         \label{fig:maps_varying_psurf_n2_dry}
   \end{figure}

\begin{figure*}[!htb]
    \centering
    \includegraphics[width=0.85\textwidth, clip]{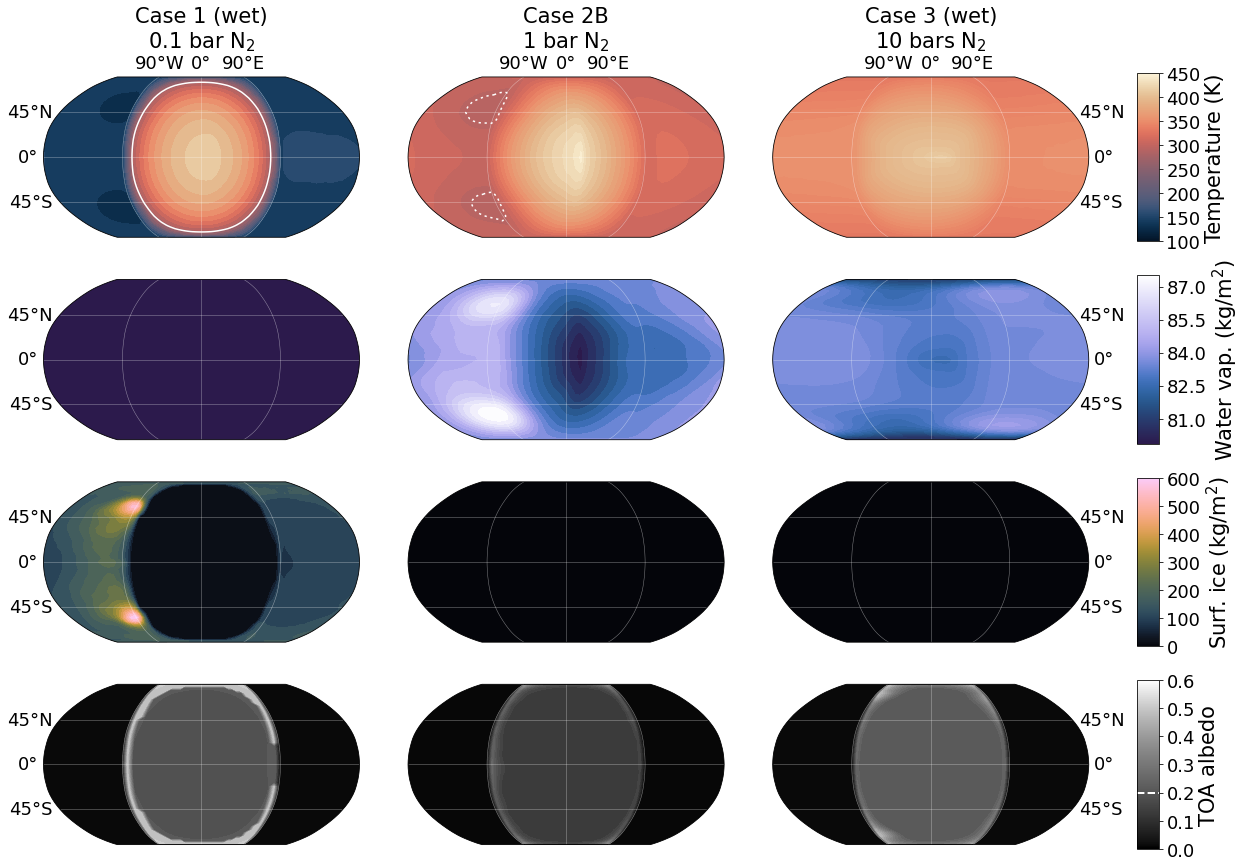}
    \caption{Latitude-longitude maps of surface temperature, column-integrated water vapor, water clouds, water ice surface density and TOA albedo for Cases~1, 2 and~3 under wet conditions (i.e., with a global water content of 8.3~cm~GEL). Here Case~2 corresponds to Case~2B. The variables are averaged over $\sim$50 orbits. The solid white line contour corresponds to the 273.15~K isotherm below which water (if present) freezes. The dotted white line contour corresponds to the location of surface liquid water. The dashed white line on the albedo color bar corresponds to the surface albedo.}
    \label{fig:maps_varying_psurf_n2_wet}
\end{figure*}

\begin{figure}[!htb]
   \sidecaption
   \includegraphics[width=11cm]{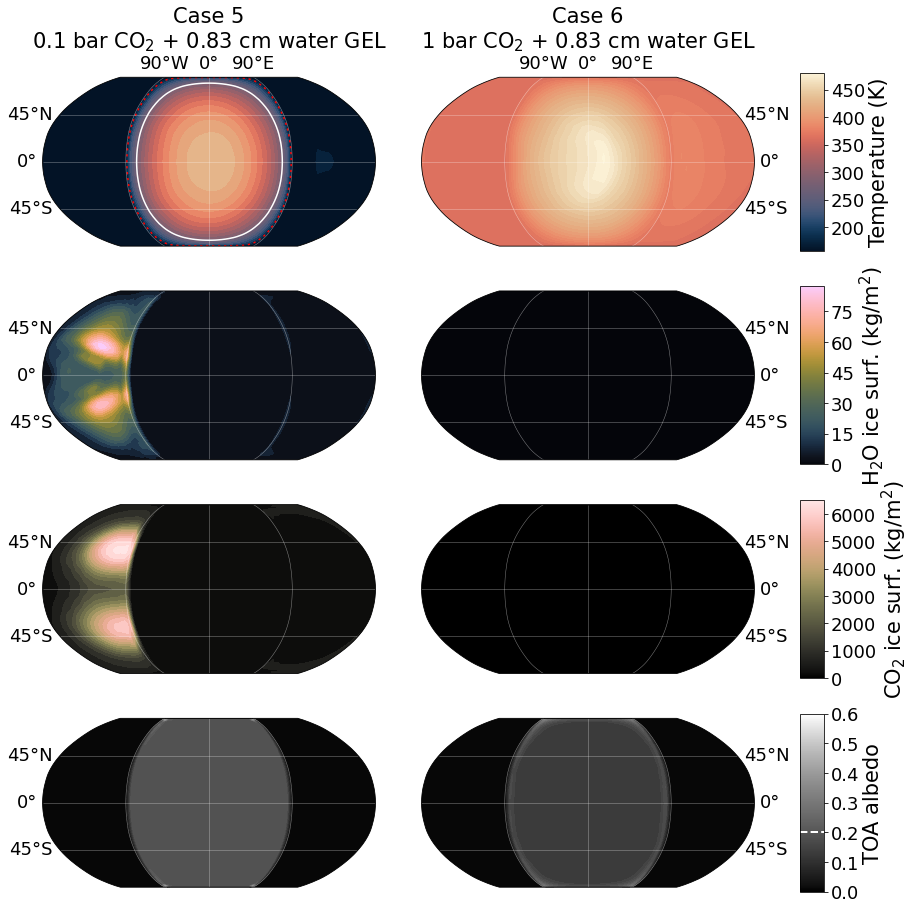}
      \caption{Latitude-longitude maps of surface temperature, water ice surface density, \ce{CO2} ice surface density and TOA albedo for Cases~5 and~6 under wet conditions (i.e., with a global water content of 0.83~cm~GEL). The variables are averaged over $\sim$26 orbits. The solid white line contour corresponds to the 273.15~K isotherm below which water (if present) freezes. The dotted red line contour corresponds to the 173.15~K isotherm below which \ce{CO2} condenses at 0.1~bar. The dashed white line on the albedo color bar corresponds to the surface albedo.}
         \label{fig:maps_varying_psurf_co2}
   \end{figure}

This section presents the maps of the GCM simulations corresponding to Cases~1, 2 and~3 (\ce{N2} atmospheres) and Cases~5 and 6 (\ce{CO2} atmospheres).
Figure~\ref{fig:maps_varying_psurf_n2_dry} presents the surface temperatures and TOA albedo for Cases~2 and~3 in dry conditions.
Figure~\ref{fig:maps_varying_psurf_n2_wet} shows the surface temperatures, water vapor, water clouds, water ice surface density and TOA albedo for Cases~1, 2 and~3 under wet conditions (with 8.3 cm GEL of water inventory). 
Figure~\ref{fig:maps_varying_psurf_co2} shows the surface temperatures, water ice surface density, \ce{CO2} ice surface density and TOA albedo for Cases~5 and~6 under wet conditions (with 0.83 cm GEL of water inventory). 

For \ce{N2}-dominated atmospheres, Figure~\ref{fig:maps_varying_psurf_n2_dry} suggests that increasing the surface pressure from 1 to 10~bars (under dry conditions) raises the mean surface temperature from approximately 250~K to 283~K. 
In Figure~\ref{fig:maps_varying_psurf_n2_wet}, the case of a thin atmosphere (0.1~bar) exhibits the highest temperature contrast among the wet scenarios with $\sim$286~K, which is about 200~K higher than in the case of a dense atmosphere (10~bars).
This behavior is due to the low efficiency of the atmospheric heat redistribution at low pressure. 
In the case at 10~bars of surface pressure under dry conditions, the nightside temperatures remain below the freezing point of water. However, they are high enough in the water-rich scenario to sustain water vapor within the atmosphere.
Therefore, decreasing the surface pressure from 1~bar to 0.1~bar in the transient climate state of Case~2B, assuming a similar initial water content, shifts the climate toward the collapsed state, and raising the surface pressure by a factor of 10 in Case~2B leads to a shift toward a runaway climate state, as represented by Case~2C.
Moreover, Figure~\ref{fig:maps_varying_psurf_n2_wet} shows that the surface temperatures and water vapor in Case~3 (in wet conditions) present more uniform distributions across the planet compared to Case~2C (see Figures~\ref{fig:maps_T_winds} and \ref{fig:maps_water}), showing an effective meridional heat redistribution in dense atmospheres.

In the case of \ce{CO2}-dominated atmospheres, Figure~\ref{fig:maps_varying_psurf_co2} reveals that the low-pressure scenario exhibits the lowest mean surface temperature at $\sim$249~K (roughly $\sim$150~K lower than the 1~bar case) and shows the highest day/night thermal contrast ($\sim$276~K).

\newpage

\subsection{Comparison synchronous/asynchronous rotations}
\label{appendix:maps_rotations}

\begin{figure*}[!htb]
    \centering
    \includegraphics[width=1\textwidth, clip]{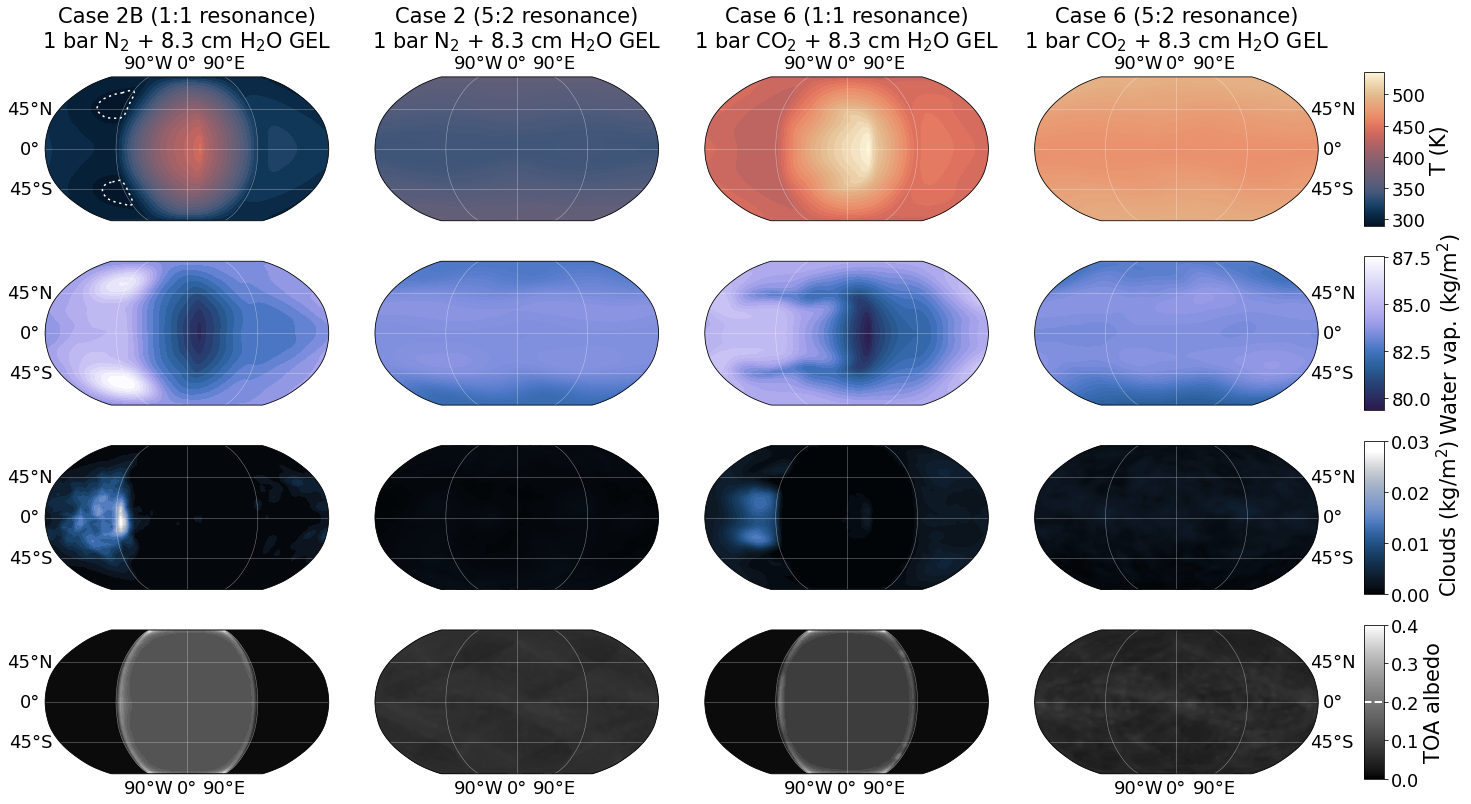}
    \caption{Latitude-longitude maps of surface temperature, column-integrated water vapor, water clouds and TOA albedo for Cases~2 and 6 (under wet conditions, i.e., with 8.3~cm~GEL of water inventory) for synchronous (1:1 resonance, 0° obliquity) and asynchronous (5:2 resonance, 79° obliquity) configurations. The variables are averaged over $\sim$50 (Case~2B), $\sim$20 (Case~2, 5:2 resonance), $\sim$26 (Case~6, 1:1 resonance) and $\sim$5 orbits (Case~6, 5:2 resonance). The dotted white line contour corresponds to the location of surface liquid water. The dashed white line on the albedo color bar corresponds to the surface albedo.}
    \label{fig:maps_asynchronous}
\end{figure*}

In this section, we present the maps corresponding the GCM results for Cases~2 and 6 (under wet conditions) for synchronous and asynchronous rotations.  
Here Case~2 assuming synchronous rotation corresponds to Case~2B.
Figure~\ref{fig:maps_asynchronous} shows the surface temperatures, water vapor, clouds and TOA albedo.

\newpage

\section{Bond and geometric albedo for all scenarios}

Table~\ref{tab:albedo} summarizes the mean Bond and geometric albedos. The mean geometric albedo are computed with Pytmosph3R within the spectral bandpasses of RISTRETTO and ANDES (including the Y, J and H bands covered with the SCAO mode of ANDES) for all configurations simulated with the Generic~PCM.

\begin{table}[h!]
	\caption{Albedo values computed for all scenarios.}
	\label{tab:albedo}      
	\centering                         
	\begin{tabular}{c c c c c c c}        
		\hline\hline                
		Scenario & GEL of water (cm) & $A_{\rm Bond}$\tablefootmark{*} & $A_{\rm g,~RISTRETTO}$\tablefootmark{**} & $A_{\rm g,~Y~band}$\tablefootmark{**} & $A_{\rm g,~J~band}$\tablefootmark{**} & $A_{\rm g,~H~band}$\tablefootmark{**} \\ 
		\hline
        \multicolumn{7}{c}{Synchronous rotations scenarios (1:1 resonance)} \\
        \hline
		\multirow{2}{*}{0.1 bar N$_{2}$} & 0.83 & 0.20 & 0.12 & 0.12 & 0.12 & 0.12 \\  
		& 8.3 & 0.20 & 0.12 & 0.12 & 0.12 & 0.12 \\
		\hline
		\multirow{5}{*}{1 bar N$_{2}$} & 0 & 0.21 & 0.13 & 0.12 & 0.12 & 0.12 \\
		& 4 & 0.19 & 0.13 & 0.12 & 0.11 & 0.11 \\
		& 6 & 0.19 & 0.13 & 0.12 & 0.11 & 0.11 \\
		& 8.3 & 0.11 & 0.11 & 0.07 & 0.05 & 0.06 \\
		& 83.3 & 0.05 & 0.07 & 0.04 & 0.01 & 0.02 \\
		\hline
		\multirow{3}{*}{10 bars N$_{2}$} & 0 & 0.24 & 0.20 & 0.14 & 0.13 & 0.12 \\
		& 8.3 & 0.12 & 0.16 & 0.08 & 0.04 & 0.05 \\
		& 83.3 & 0.07 & 0.11 & 0.04 & 0.01 & 0.01 \\
		\hline
		\multirow{2}{*}{1 bar \ce{N2 + 376ppm CO2}} & 0 & 0.20 & 0.13 & 0.12 & 0.12 & 0.12 \\  
		& 8.3 & 0.11 & 0.11 & 0.07 & 0.05 & 0.06 \\
		\hline
		0.1 bar CO$_{2}$ & 0.83 & 0.19 & 0.12 & 0.12 & 0.12 & 0.11 \\
		\hline
		\multirow{3}{*}{1 bar CO$_{2}$} & 0 & 0.15 & 0.13 & 0.12 & 0.09 & 0.05 \\
		& 0.83 & 0.11 & 0.13 & 0.09 & 0.06 & 0.03 \\
		& 8.3 & 0.07 & 0.11 & 0.06 & 0.03 & 0.01 \\
		\hline
		10 bars CO$_{2}$ & 83.3 & 0.06 & 0.12 & 0.02 & 0.00 & 0.00 \\
		\hline\hline
        \multicolumn{7}{c}{Asynchronous rotations scenarios (5:2 resonance)} \\
        \hline
		1 bar N$_{2}$ & 8.3 & 0.11 & 0.11 & 0.07 & 0.05 & 0.06 \\
		\hline
		1 bar \ce{N2 + 376ppm CO2} & 8.3 & 0.11 & 0.11 & 0.08 & 0.06 & 0.06 \\
		\hline
		1 bar CO$_{2}$ & 8.3 & 0.08 & 0.11 & 0.07 & 0.03 & 0.01 \\
		\hline
	\end{tabular}
    \tablefoot{
        \tablefoottext{*}{Mean Bond albedo computed from the Generic PCM data.} 
        \tablefoottext{**}{Mean geometric albedo computed with Pytmosph3R for the RISTRETTO bandpass and the YJH bands covered with the SCAO mode of ANDES.}
    }
\end{table}

\end{appendix}


\end{document}